\documentclass{jfm}

\usepackage{subfig}
\usepackage{xcolor}
\usepackage{amssymb}
\usepackage{comment}
\usepackage{graphicx}
\usepackage{newtxtext}
\usepackage{newtxmath}
\usepackage{natbib}
\usepackage{hyperref}
\hypersetup{
    colorlinks = true,
    urlcolor   = blue,
    citecolor  = black,
}

\newcommand{\RomanNumeralCaps}[1]
\linenumbers
\graphicspath{{./figures/}}

\newcommand{\SE}{SE}
\newcommand{\ST}{ST}
\newcommand{\E}{NS}
\newcommand{\ct}{{ct}}
\newcommand{\ce}{{ce}}
\newcommand{\ke}{{KE}}
\newcommand{\subfigthree}{0.27}
\newcommand{\subfigfour}{0.23}
\newcommand{\newTime}{{T}}
\title{Yield-Stress Fluid Mixing: Localization Mechanisms and Regime Transitions}

\author{Mohammad Reza Daneshvar Garmroodi\aff{1} and
  Ida Karimfazli\aff{1}\corresp{\email{Ida.karimfazli@concordia.ca}}}

\affiliation{\aff{1} Department of Mechanical, Industrial and Aerospace Engineering, Concordia University, 1515 St.Catherine W., Montreal, QC H3G 2W1, Canada}

\begin{document}
\maketitle

\begin{abstract}
We explore the mechanisms and regimes of mixing in yield-stress fluids by simulating the stirring of an infinite, two-dimensional domain filled with a Bingham fluid. A cylindrical stirrer moves along a circular path at constant speed to stir the fluid, with an initially quiescent domain marked by a passive dye in the lower half, facilitating the analysis of dye interface evolution and mixing dynamics. We first examine the mixing process in Newtonian fluids, identifying three key mechanisms: interface stretching and folding around the stirrer's path, diffusion across streamlines, and dye advection and interface stretching due to vortex shedding. Introducing yield stress into the system leads to notable localization effects in mixing, manifesting through three mechanisms: advection of vortices within a finite distance of the stirrer, vortex entrapment near the stirrer, and complete suppression of vortex shedding at high yield stresses. Based on these mechanisms, we classify three distinct mixing regimes in yield-stress fluids: (i) Regime \SE, where shed vortices escape the central region, (ii) Regime \ST, where shed vortices remain trapped near the stirrer, and (iii) Regime \E, where no vortex shedding occurs. These regimes are quantitatively distinguished through spectral analysis of energy oscillations, revealing transitions and the critical Bingham and Reynolds numbers. The transitions are captured through effective Reynolds numbers, supporting a hypothesis that mixing regime transitions in yield-stress fluids share fundamental characteristics with bluff-body flow dynamics. The findings provide a mechanistic framework for understanding and predicting mixing behaviors in yield-stress fluids, suggesting that the localization mechanisms and mixing regimes observed here are archetypal for stirred-tank applications.
\end{abstract}

\begin{keywords}
Authors should not enter keywords on the manuscript, as these must be chosen by the author during the online submission process and will then be added during the typesetting process (see \href{https://www.cambridge.org/core/journals/journal-of-fluid-mechanics/information/list-of-keywords}{Keyword PDF} for the full list).  Other classifications will be added at the same time.
\end{keywords}

%%%%%%%%%%%%%%%%%%%%%%%%%%%%%%%%%%%%%%%%%%

\section{Introduction}
Mixing is ubiquitous in both natural and industrial environments. Applications span a vast range of Reynolds numbers and length scales (see Figure 1 in \cite{ottino1990mixing} for illustration). From the coffee we drink, household cleaning products, and oil extraction to the human digestive system and pharmaceutical production, various materials undergo mixing processes daily. 

Despite its prevalence, mixing remains one of the more challenging paradigms in engineering to systematically define, frame, and understand \citep{spencer1951mixing,ottino1990mixing,villermaux2019mixing}. In the simplest context, mixing entails the homogenization of a passive tracer (\emph{Level-1}); however, it can be more intricately tied to flow dynamics, as when rheology depends on tracer concentration (\emph{Level-2}), or chemical reactions occur during the process (\emph{Level-3}) \citep{dimotakis2005turbulent}.

A significant body of literature focuses on Level-1 mixing, i.e., mixing of a passive dye in fluid. Even within this limited scope, the parameter space is extensive: mixing may be in-line, active, or passive, or take place in a stirred tank. Factors like domain geometry, impeller shape, size, position, stirring protocol, and speed all influence mixing behavior. Additionally, fluid rheology plays a critical role. Given the vast range of parameters and the complexity of the problem, most studies have focused on Newtonian fluids, and mixing remains an active research area (see \cite{warhaft2000passive,peltier2003mixing,wunsch2004vertical,caulfield2021layering} and references therein).

Many fluids in polymer processing, food engineering, bioengineering, physiology, and chemical engineering are non-Newtonian, with a subset exhibiting yield stress, such as polymeric gels, muds, paints, and cosmetics. Yield-stress fluids are highly viscous materials that flow only when the applied shear stress exceeds a threshold known as the yield stress \citep{balmforth2014yielding,coussot2014yield,bonn2017yield}. It was recognized early on that turbulent mixing in these fluids is economically and technologically impractical. \cite{spencer1951mixing} proposed \emph{streamline mixing}, achieved by continuously deforming the fluid to (a) increase the surface area of the interface and (b) distribute it throughout the material volume. With few exceptions \citep{derksen2013simulations,garmroodi2024mixing}, studies of mixing in yield-stress fluids focus on the mixing of passive dyes.

Initial efforts to understand mixing of non-Newtonian fluids were dedicated to establishing a relationship between impeller speed (in stirred tanks) and the fluid shear rate. According to \citep{metzner1957agitation}, mixing flows of non-Newtonian fluids were qualitatively understood at best in the 1950s, a sentiment that remained accurate for decades. One of the first studies on mixing yield-stress fluids was conducted by \cite{solomon1981cavern}, who experimentally identified the well-mixed regions in Xanthan gum and Carbopol solutions in stirred tanks. The coexistence of flowing and stagnant regions presented a challenge, as the latter remained unmixed. \cite{whitcomb1978rheology} introduced the term \emph{cavern} to describe the well-mixed region where the fluid was yielded.

Though debate remains regarding the existence of a true yield stress and optimal measurement methods \citep{barnes1999yield,divoux2011stress,dinkgreve2016different}, viscoplastic models are widely used to analyze and predict flows of yield-stress fluids \citep{mitsoulis2017numerical}. These models consider the fluid rigid below the yield stress and flowing with shear rate-dependent viscosity when exceeded. 

\cite{solomon1981cavern} used viscoplastic models and a yield criterion to estimate cavern size by assuming it to be spherical. Subsequent studies explored various impellers and developed cavern size estimates based on different simplified geometries; see, for example, \cite{galindo1996comparison,tanguy1996numerical,amanullah1998new,pakzad2013novel,sossa2015computational,ameur2020newly}. Generally, yield stress and shear-thinning viscosity reduce the mixing rate and extent, and cavern size. However, a mechanistic understanding of how flow dynamics and mixing are interlinked, especially with respect to rheological parameters, remains elusive.

Seminal works by \cite{Aref_1984} and \cite{ottino1989kinematics} demonstrated the role of chaotic flows in effective mixing, emphasizing the necessity of three-dimensional or transient two-dimensional flows. A parallel branch of research has since modeled mixing by analyzing dynamical systems, with a primary focus on chaotic mixing in two-dimensional time-periodic flows; for an overview of studies of Newtonian fluids, see \cite{aref2017frontiers}.

\cite{niederkorn1994chaotic} numerically investigated chaotic advection in journal bearing flows for shear-thinning fluids, examining tracer advection, periodic points, stretching along unstable manifolds, and stretching rate of fluid elements. They found that shear-thinning viscosity decreases the amount of stretching. \cite{fan2001tangential} studied advective mixing of viscoplastic fluids between eccentric cylinders, showing that tracer coverage may depend on initial tracer position, and chaotic advection can be achieved through alternating cylinder rotations. Their results highlight qualitative transitions but do not predict when such transitions will occur.

Experimental studies by \cite{wendell2013intermittent} explored mixing in a rotating tank filled with a yield-stress fluid stirred by cylindrical rods rotating with constant angular velocity in an eggbeater configuration. Higher period ratios and lower yield stresses enhanced mixing, although mixing efficiency decreased in resonance conditions. The intermittent yielding and unyielding near the tank wall was hypothesized to explain the decreased mixing of yield-stress fluids. Further experiments by \cite{boujlel2016rate} in the same setup measured mixing rate using dye concentration variance, showing that mixing consists of rapid stretching and folding followed by slower diffusion-dominated mixing. They concluded that mixing rate is proportional to the volume of highly sheared fluid during each rod rotation.

In summary, while the qualitative impact of yield stress on mixing is understood - yield stress limits cavern size and filament stretching, thus decreasing mixing rate - a mechanistic description connecting fluid dynamics to transitions in yield-stress fluid mixing remains absent. Decades after \cite{niederkorn1994chaotic}, design procedures still rely heavily on empiricism with limited fundamental understanding of fluid mechanics (see \cite{Paul2004,uhl2012mixing}).

The primary objective of this manuscript is to identify and elucidate the mechanisms behind different mixing regimes and localization in yield-stress fluids within a two-dimensional periodically stirred domain. We consider an infinite domain filled with a quiescent viscoplastic fluid stirred by a cylinder moving at constant speed along a circular path. By exploring a range of mixing speeds and yield stresses, we aim to characterize flow and mixing dynamics and establish a mechanistic link between them. The remainder of this paper is organized as follows: Section \ref{sec:formulation} presents the model problem, governing equations, and numerical methods. Section \ref{sec:results} discusses flow dynamics and mixing regimes using representative cases and maps out mixing mechanisms and flow regimes in the ($Re,Bn$) plane. Finally, Section \ref{sec:summary} summarizes our findings.

%--------------------------------------------------------------------------------------------------
\section{Problem setup}
\label{sec:formulation}

\subsection{Model problem}

\begin{figure}
	\centering
	\subfloat{
		\includegraphics[trim=0cm 0cm 0cm 0cm, clip=true,height=.35\textwidth]{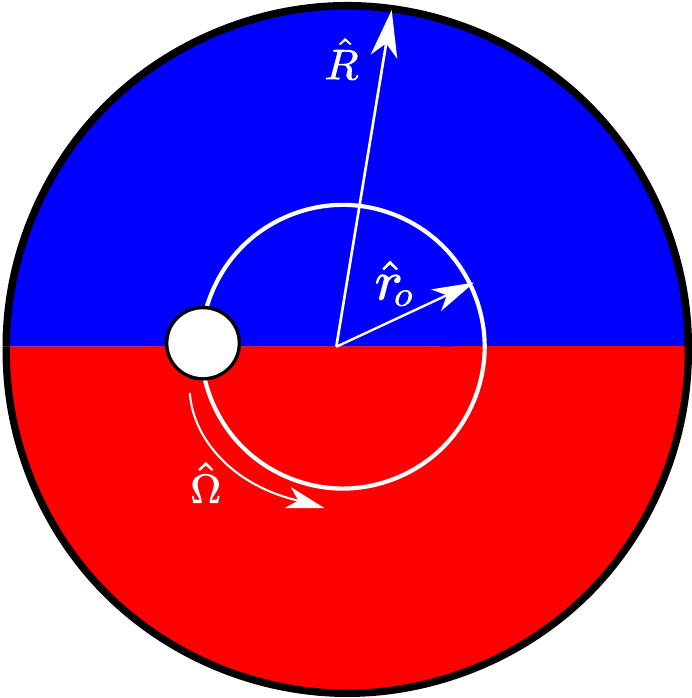}
	}
	\caption{Schematic of the domain geometry and initial conditions. The solid white line indicates the stirrer's path. The red and blue colors indicate the dyed and dye-free regions. Note that the figure is not to scale. 
	}
	\label{fig:geometry}
\end{figure}

We investigate the stirring of a viscoplastic fluid (VPF) using a circular stirrer of diameter $\hat{d}_s$, which moves at a constant speed ($\hat{r}_o \hat{\Omega}$) along a circular trajectory with radius $\hat{r}_o$ (solid white line in Figure \ref{fig:geometry}), where the ratio $\hat{r}_o/\hat{d}_s = c$. In this study, we set $c = 2$, and $\hat{\Omega}>0$ denotes the angular velocity of the stirrer. Dimensional quantities are denoted by a $\hat{.}$ symbol, while dimensionless quantities are written without it. To simulate mixing in an infinite domain, we employ a circular computational domain with a radius $\hat{R} \gg \hat{r}_o$, ensuring minimal boundary effects due to the domain's sufficiently large size. To monitor mixing, the fluid in the bottom half of the domain is marked with a passive dye (shown in red in Figure \ref{fig:geometry}), where $\alpha = 1$, while the rest of the fluid is dye-free (shown in blue), with $\alpha = 0$. The no-slip boundary condition is imposed on the walls of both the vessel and the stirrer.

The fluid is modeled using the Bingham model. The dimensionless form of the constitutive equation is given by:
\begin{align*}
\left\{
\begin{array}{lr}
\displaystyle \boldsymbol{\tau} = \left( \frac{Bn}{\dot{\gamma}} + 1 \right) \dot{\boldsymbol{\gamma}}
& \text{if} \quad \tau \ge Bn \\
\dot{\boldsymbol{\gamma}} = 0
& \text{if} \quad \tau < Bn
\end{array}
\right.
\end{align*}
where $\boldsymbol{\tau}$ and $\boldsymbol{\dot{\gamma}}$ are the deviatoric stress and rate of strain tensors, respectively, and 
\begin{align*}
	\tau = \sqrt{\displaystyle \frac{1}{2}\  \boldsymbol{\tau}:\boldsymbol{\tau}} 
	\quad \& \quad 
	\dot{\gamma} = \sqrt{\displaystyle \frac{1}{2}\  \boldsymbol{\dot{\gamma}}:\boldsymbol{\dot{\gamma}}} 
\end{align*}
represent second invariants of these tensors. The Bingham number, $Bn$, is defined by:
\begin{align}
Bn = \frac{\hat{\tau}_{y}}{\hat{\mu}\hat{\Omega}} 
\label{eq:Bn}
\end{align}
where $\tau_{y}$ and $\hat{\mu}$ are  the fluid's yield stress and plastic viscosity.

The flow is governed by the Cauchy's equations of motion and continuity while the dye concentration  is described by an advection-diffusion equation,
\begin{equation}
\label{eq:Cauchy}
\begin{aligned}
	&\frac{\partial \boldsymbol{u}}{\partial {t}} 
	+ 
	 \boldsymbol{u} \cdot \boldsymbol{\nabla} \boldsymbol{u} 
	 +
	  \boldsymbol{\nabla} P 
	  = 
	  \frac{1}{c\ Re} \boldsymbol{\nabla} \cdot \boldsymbol{\tau} 
	  \\
	&\boldsymbol{\nabla} \cdot \boldsymbol{u} 
	= 
	0 \\
	&\frac{\partial \alpha}{\partial t} 
	+ 
	\boldsymbol{\nabla} \cdot (\boldsymbol{u} \alpha) 
	=
	\frac{1}{c\ Pe} \nabla^2 \alpha
\end{aligned}
\end{equation}

Here, $\boldsymbol{u}$ and $P$ are the dimensionless velocity and pressure, respectively. The characteristic scales for length, time, velocity, shear stress, and pressure are $\hat{r}_o$, $\hat{\Omega}$, $\hat{r}_o \hat{\Omega}$, $\hat{\mu} \hat{\Omega}$, and $\hat{\rho} \hat{r}_o^2 \hat{\Omega}^2$. Results, however, are presented in terms of the stirrer's period, $\hat{T}_{stirrer}= 2\pi/\hat{\Omega}$,
\begin{align}
T = \frac{\hat{t}}{\hat{T}_{stirrer}} = \frac{t}{2\pi}
\end{align}

The Reynolds ($Re$) and Peclet numbers ($Pe$) are defined as,
\begin{align}
& Re = \frac{\hat{\rho} \hat{\Omega} \hat{r}_o \hat{d}_s}{\hat{\mu}} \\
& Pe = \frac{\hat{\Omega} \hat{r}_o \hat{d}_s}{\hat{D}_m} 
\label{eq:Peh}
\end{align}
where $\hat{D}_m$ is the diffusion coefficient of dye. In this study, Peclet number is held constant at $Pe = 500$. 

Definitions and ranges of the relevant dimensionless groups are summarized in Table \ref{tab:dimLess}.

\begin{table}
	\begin{center}
		\def~{\hphantom{0}}
		\begin{tabular}{lccc}
			Dimensionless group & 
			\quad Definition  & 
			Value/Range
			\\
			Reynolds ($Re$) &
			$\displaystyle\frac{\hat{\rho} \hat{\Omega} \hat{r}_{o}\hat{d}_s}{\hat{\mu}}$ &
			$25-150$
			\\[1cm]
			Peclet ($Pe$) & 
			$\displaystyle\frac{\hat{\Omega} \hat{r}_{o} \hat{d}_{s} }{ \hat{D}}$ &
			$500$
			\\[1cm]
			Bingham ($Bn$) & 
			$\displaystyle\frac{\hat{\tau}_{y}}{\hat{\mu}\hat{\Omega}}$ & 
			$0-5$
			\\[.5cm]
		\end{tabular}
		\caption{Dimensionless groups governing the model problem.}
		\label{tab:dimLess}
	\end{center}
\end{table}

%%%%%%%%%%%%%%%%%%%%%%%%%%
%
%\subsection{Quantifying mixing and kinetic energy}

To characterize the rate of mixing, a normalized variance of the dye, $\sigma^{2}_{R_{sd}}$, is defined over a circular subdomain $A_{R_{sd}}$ of radius $R_{sd}$ that is concentric with the stirrer's path,
\begin{equation}
	\label{eq:variance}
	\sigma^{2}_{R_{sd}} = \frac{1}{A_{R_{sd}}} \int_{A_{R_{sd}}} \left(1 - \frac{\alpha}{\bar{\alpha}}\right)^{2} dA
\end{equation}

Here, $\bar{\alpha}$  is the average dye concentration over the subdomain, for all $R_{sd}$. For given values of the governing dimensionless parameters, $R_{sd}$ is chosen to ensure the subdomain captures the region where dye concentration is affected by the stirring (within the timeframe of interest). 

To quantify the kinetic energy, $\ke$ is defined as,
\begin{equation}
	\label{eq:velNorm}
	\ke 
%	= {\color{red}||\boldsymbol{u}||} 
	= \sqrt{ 
	\int_{A} |\boldsymbol{u}|^{2} dA}
\end{equation}
where $\ke$ is the kinetic energy, $|\boldsymbol{u}|$ is the speed, and $A$ represents the flow domain.

\subsection{Numerical Method}

\begin{figure}
	\centering
	\subfloat[\label{fig:meshVariance}]{
		\includegraphics[trim=0cm 0cm 0cm 0cm, clip=true,height=.25\textwidth]{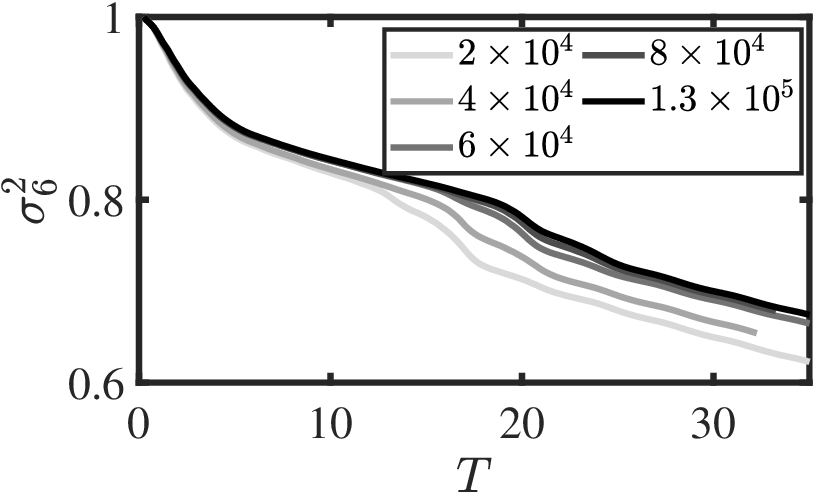}
	}~
	\subfloat[\label{fig:meshErrVariance}]{
		\includegraphics[trim=0cm 0cm 0cm 0cm, clip=true,height=.25\textwidth]{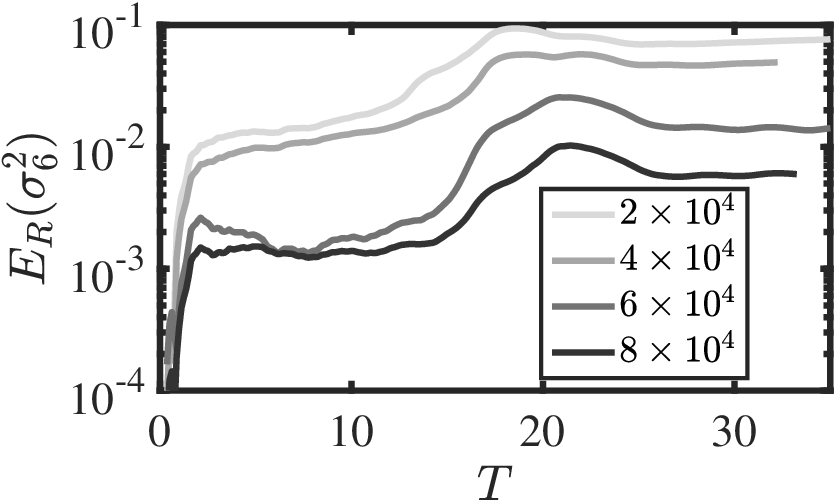}
	}\\
	\subfloat[\label{fig:meshNorm}]{
		\includegraphics[trim=0cm 0cm 0cm 0cm, clip=true,height=.25\textwidth]{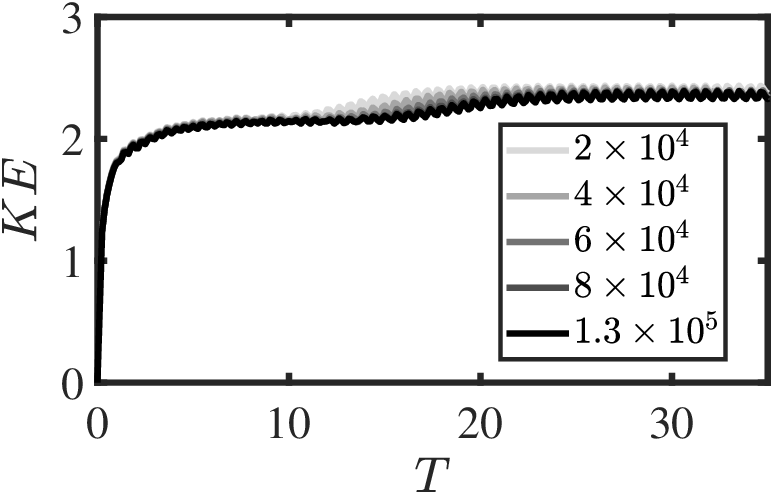}
	}~
	\subfloat[\label{fig:meshErrNorm}]{
		\includegraphics[trim=0cm 0cm 0cm 0cm, clip=true,height=.25\textwidth]{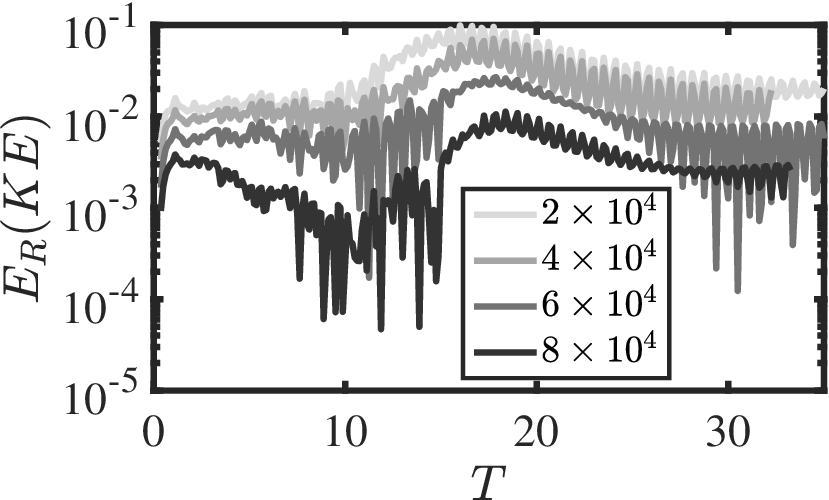}
	}
	
	\caption{Time evolution of (a) normalized variance of dye concentration for different mesh sizes, (b) relative error in dye concentration variance, $\displaystyle E_R(\sigma^2_{6}) = \frac{\sigma^2_{6} - \sigma^2_{6,\ 1.3\times10^5}}{\sigma^2_{6,\ 1.3\times10^5}}$, (c) kinetic energy for different mesh sizes, and (d) relative error in kinetic energy, $\displaystyle E_R(\ke) = \frac{\ke - \ke_{1.3\times10^5}}{\ke_{1.3\times10^5}}$, for $Re = 150$ and $Bn = 2$.}
	\label{fig:girdTest}
\end{figure}

\cite{OpenFOAM} version 6 was used for numerical simulations. The \textit{twoLiquidMixingFoam} solver, which employs the PIMPLE or PISO-SIMPLE algorithm to decouple pressure and velocity in the governing equations, was utilized. For temporal discretization, we applied the second-order Crank-Nicolson scheme, while spatial discretization was second-order as well. Adaptive time stepping was implemented based on a constant Courant-Friedrichs-Lewy (CFL) number, set to 0.05.

In this study, we adopted a modified version of the bi-viscosity model originally developed by \cite{tanner1983numerical}:
%Bi-viscosity regularization \cite{tanner1983numerical} has been used to regularize the Bingham model at $\dot{\gamma}=0$,
\begin{equation}
	\boldsymbol{\tau}(\dot{\gamma}) = \begin{cases}	
	\displaystyle(1 + \frac{Bn}{\dot{\gamma}_{c}})  \boldsymbol{\dot{\gamma}}
	& \text{if} \quad \dot{\gamma} \leq \dot{\gamma}_{c} \\
	\\[3mm]
	\displaystyle(1 + \frac{Bn}{\dot{\gamma}}) \boldsymbol{\dot{\gamma}}
	& \text{if} \quad \dot{\gamma} > \dot{\gamma}_{c}
	\end{cases}
\end{equation}
	
Here, $\dot{\gamma}_{c}$ is chosen to be significantly smaller than the characteristic strain rate of the problem, with $\dot{\gamma}_{c} = \hat{\dot{\gamma}}_c / \hat{\Omega} = 10^{-4}$.

To verify grid independence, five different mesh sizes were tested. Figure \ref{fig:girdTest} shows the evolution of normalized variance and velocity norm, along with the corresponding relative errors for different mesh sizes. The finest mesh ($1.3\times 10^5$) was used to estimate the relative error. For the remainder of the simulations, we used a mesh size of $8\times10^4$, which resulted in a relative error of less than 1\% for both the normalized variance and velocity norm. Further details on the benchmarking and validation of the numerical solver can be found in \cite{garmroodi2024mixing}.

%-------------------------------------------------------------------------------------------------------------------------	
\section{Results and discussion}
\label{sec:results}
%\begin{enumerate}
%	\item Newtonian $\Rightarrow$ explain one way coupling, A Re that implies shedding, stages of mixing development, sigma 2, mixing area expanding to infinity, stages of flow development, strouhal?, why two escaping vortices, connection w flow pasta cylinder? mixing mechanisms, 
%	\item mixing in VPF $\Rightarrow$ low to high Bn, mixing slowing down, mixing mechanims, flow regimes, interaction of the stirrer and  its tail, shedding, flow dynamics, decay of free vortices away from the center
%	
%	\item $r_y$, $r_m$, flow regimes and critical conditions?
%	
%\end{enumerate}

%-------------------------------------
	
\subsection{The Newtonian limit}
\begin{figure}%{r}{0.4\textwidth}
	%	\vspace{-15mm}
	\centering
	\subfloat[$R_{sd}=5$\label{fig:illustrative_Newtonian_concentrationA}]{
		\includegraphics[trim=2cm 0.5cm 5.5cm 1.25cm, clip=true,height=\subfigthree\textwidth]{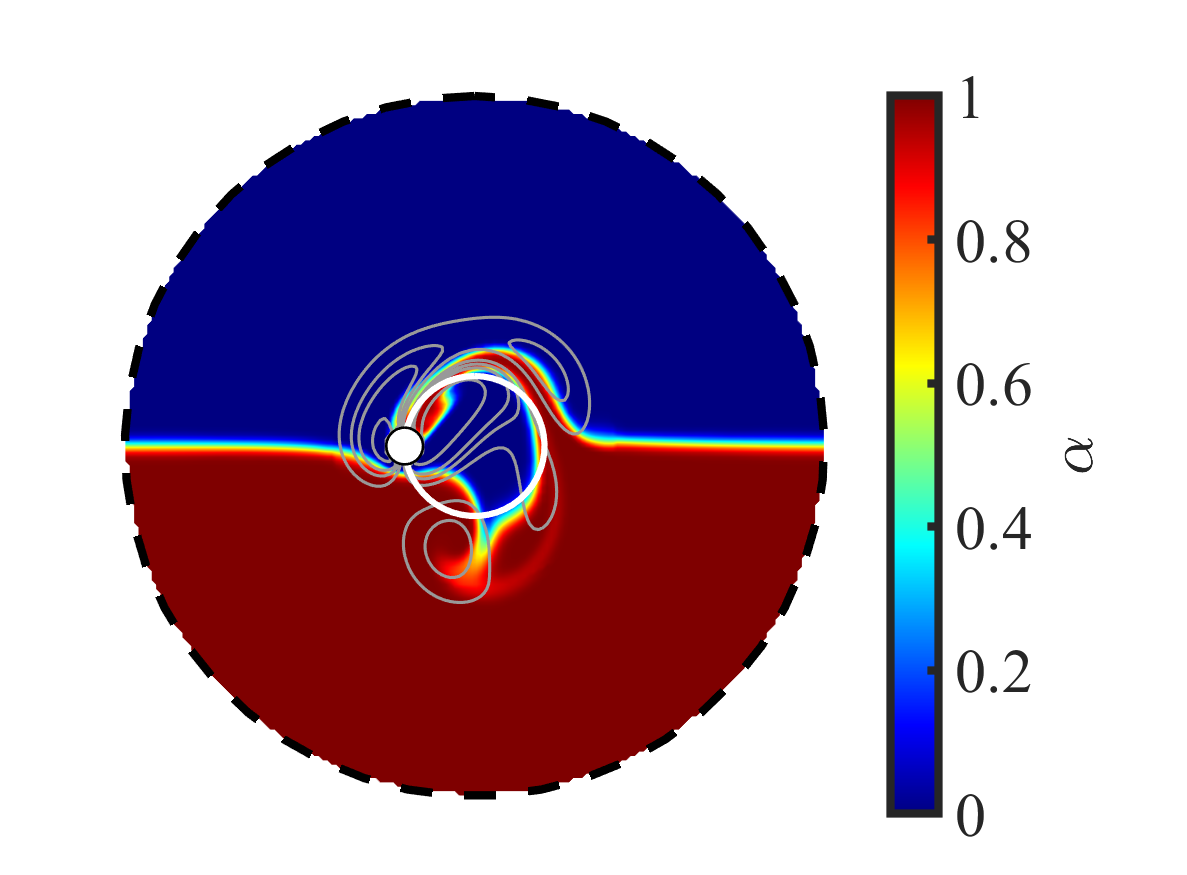}
	}~
	\hspace{0.2cm}
	\subfloat[$R_{sd}=5$\label{fig:illustrative_Newtonian_concentrationB}]{
		\includegraphics[trim=2cm 0.5cm 5.5cm 1.25cm, clip=true,height=\subfigthree\textwidth]{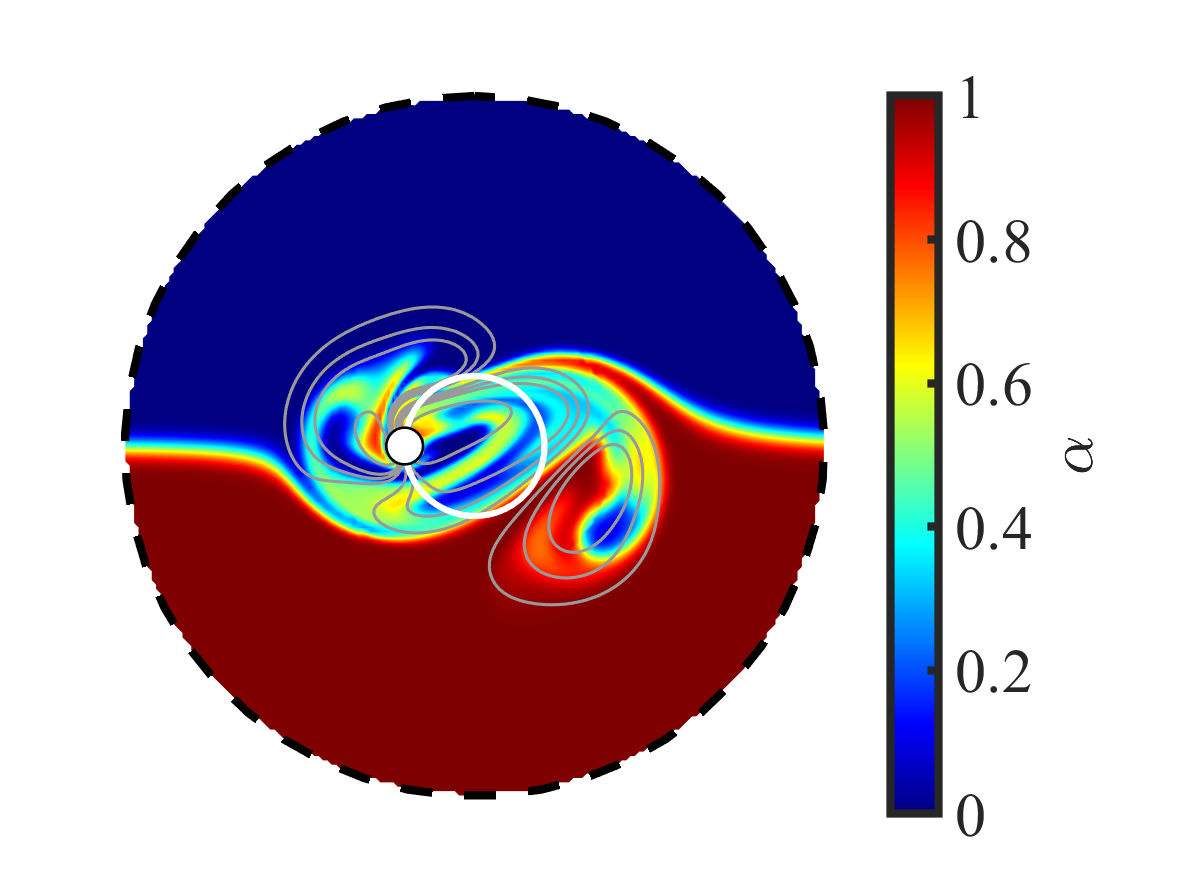}
	}~
	\hspace{0.2cm}
	\subfloat[$R_{sd}=5$\label{fig:illustrative_Newtonian_concentrationC}]{
		\includegraphics[trim=2cm 0.5cm 1.5cm 1.25cm, clip=true,height=\subfigthree\textwidth]{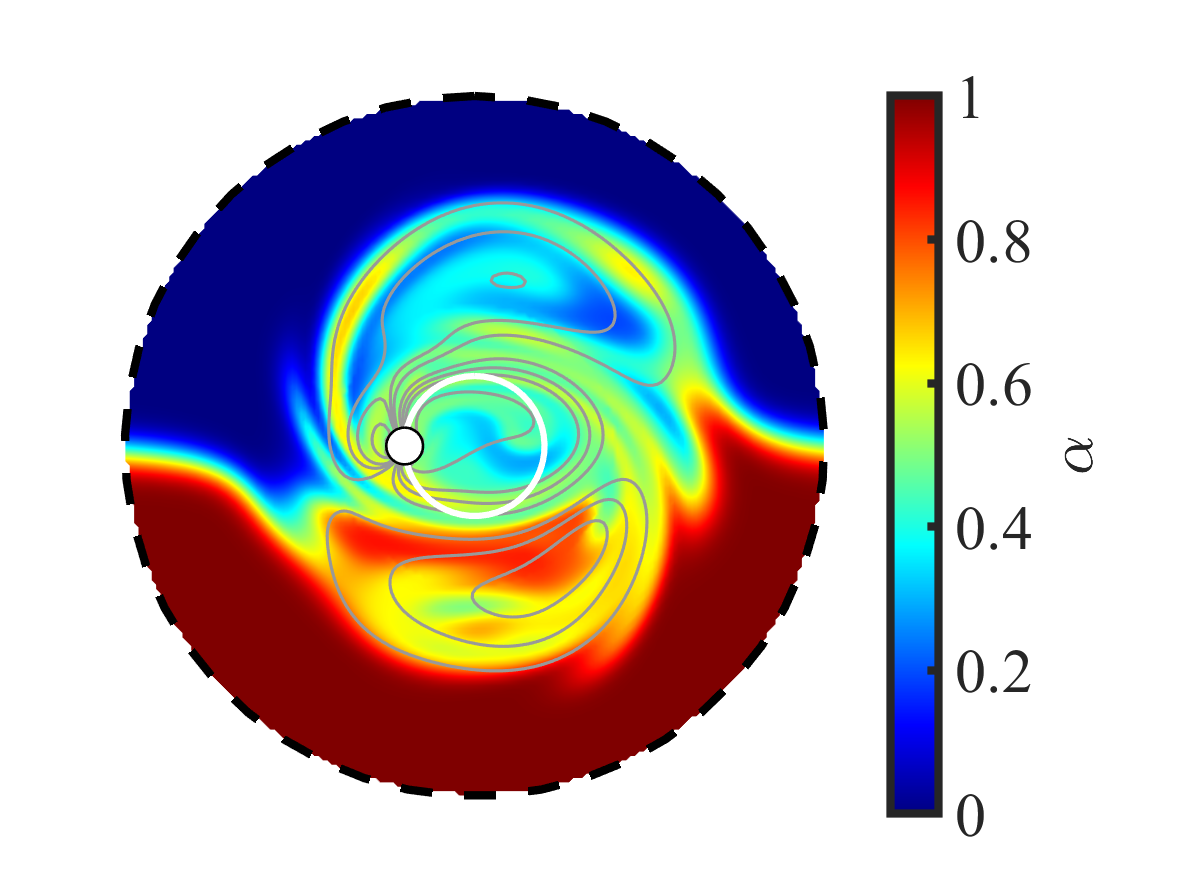}
	}\\
	\vspace{-0.3cm}
	\subfloat[$R_{sd}=5$\label{fig:illustrative_Newtonian_concentrationD}]{
		\includegraphics[trim=2cm 0.5cm 5.5cm 1.25cm, clip=true,height=\subfigthree\textwidth]{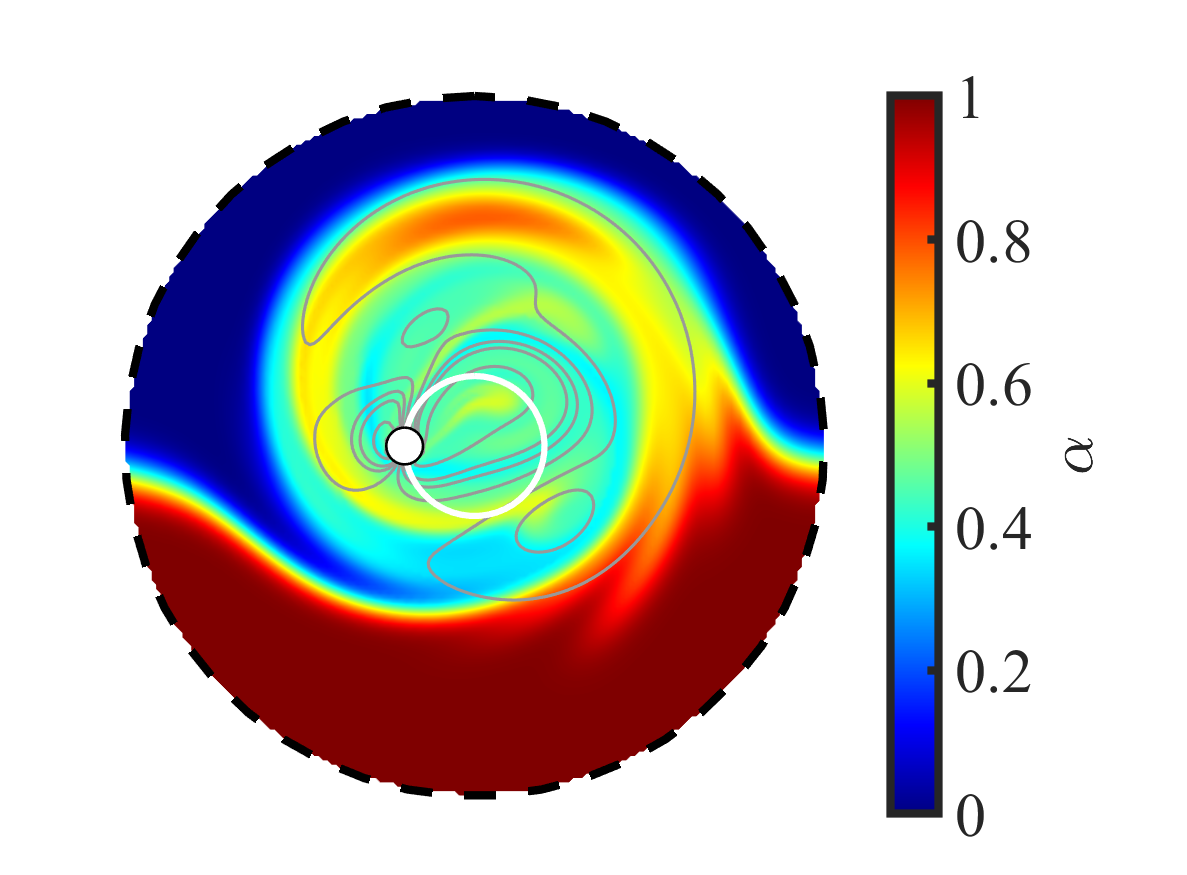}
	}~
	\hspace{0.2cm}
	\subfloat[$R_{sd}=8$\label{fig:illustrative_Newtonian_concentrationE}]{
		\includegraphics[trim=2cm 0.5cm 5.5cm 1.25cm, clip=true,height=\subfigthree\textwidth]{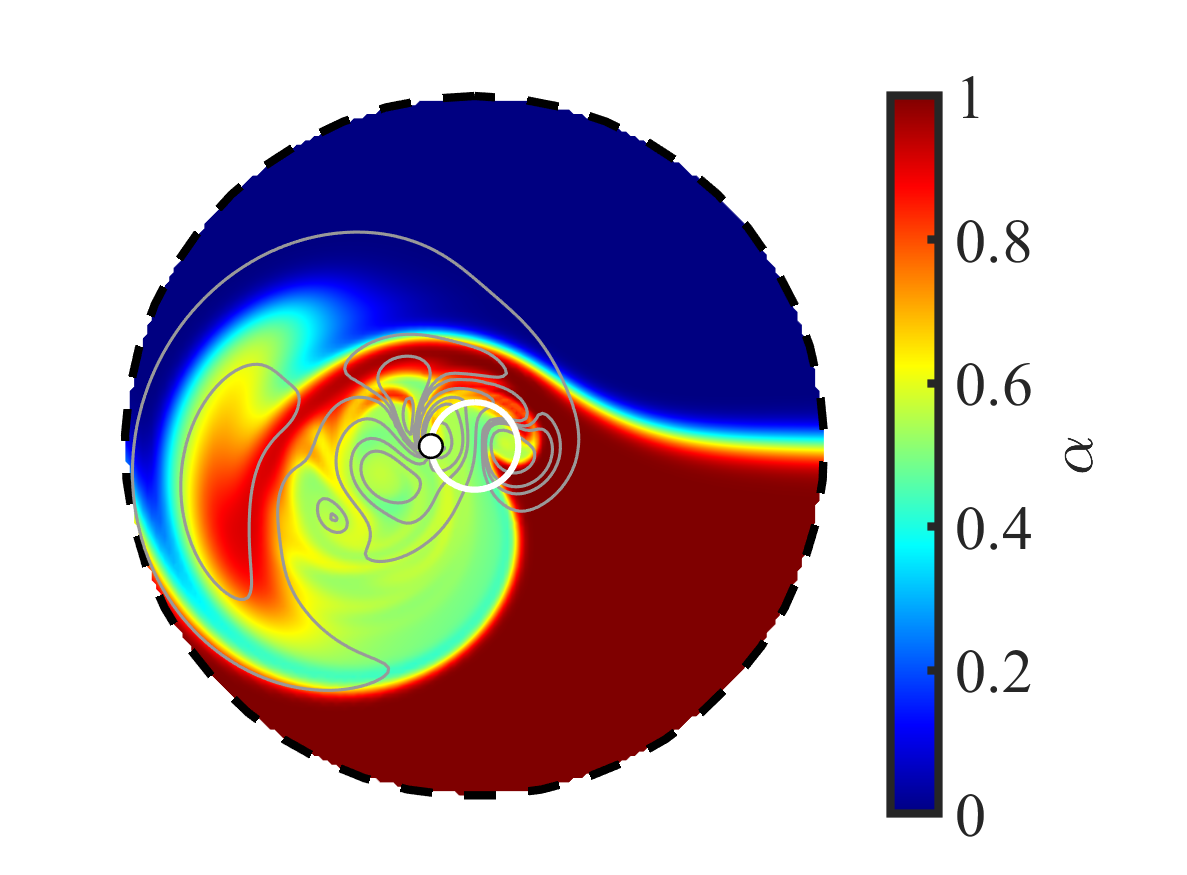}
	}~
	\hspace{0.2cm}
	\subfloat[$R_{sd}=15$\label{fig:illustrative_Newtonian_concentrationF}]{
		\includegraphics[trim=2cm 0.5cm 1.5cm 1.25cm, clip=true,height=\subfigthree\textwidth]{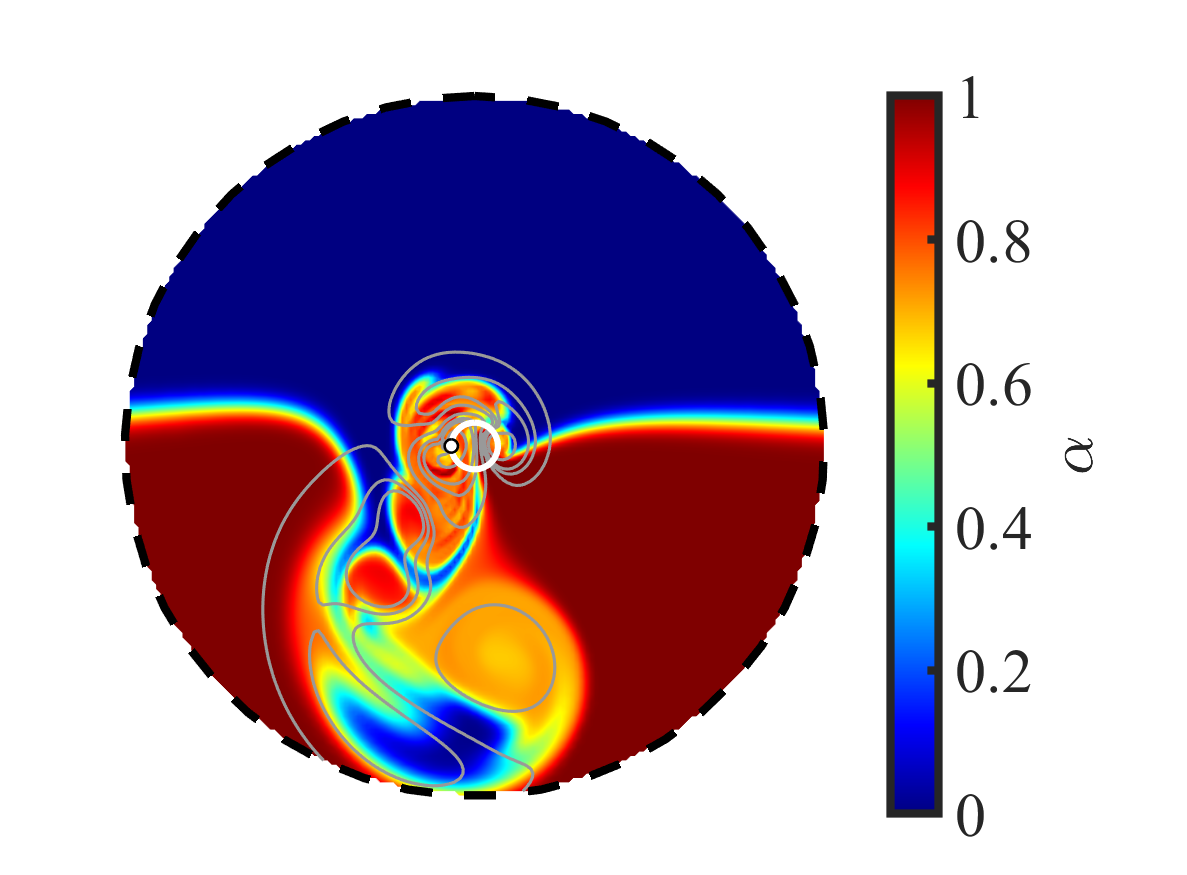}
	}\\
%	\vspace{-0.3cm}
	\subfloat[\label{fig:concentration_Newtonian_variance}]{
		\includegraphics[trim=0cm 0cm 0cm 0cm, clip=true,height=.3\textwidth]{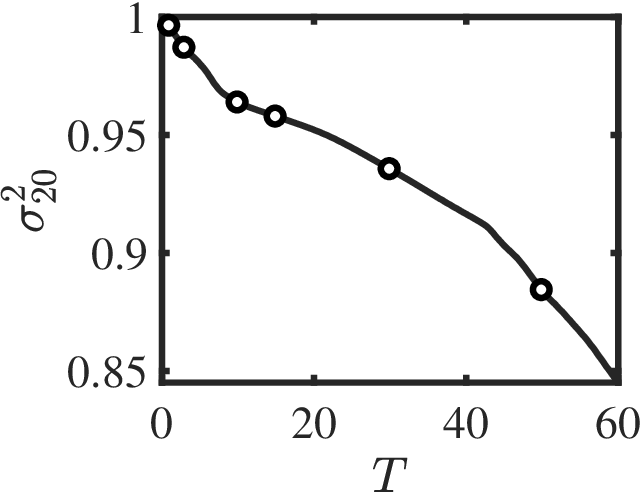}
	}	
	\caption{(a–f) Snapshots of the dye concentration field in a Newtonian fluid, $Bn = 0$, $Re = 50$, at times $\newTime = 1, 3, 10, 15, 30, \text{and}~50$. The white circle marks the stirrer's path, while grey lines represent the streamlines. The radius of the field of view is provided in the caption. (g) Time evolution of the normalized variance, with circular markers indicating the time instances of the snapshots in (a–f).
	}
	\label{fig:illustrative_Newtonian_concentration}
\end{figure}

%{\color{red}
%\begin{itemize}
%	\item {\bf The Newtonian limit}  
%	\item describe different stages of  mixing in Newtonian fluid  (adv, diff, adv...) 
%	\item discuss different mixing mechanism and draw connections w the vorticity field
%	\item Strouhal number and shedding frequency
%	\item Discuss interaction of the shed vortices w the stirrer 
%	\item discuss the possibility of the stirrer cutting off its own tail (Williamson 1996 "recirculation region" or "wake", Kundu: "wake" or "pair of attached vortices") 
%\end{itemize}
%}	
Figure \ref{fig:illustrative_Newtonian_concentration} shows snapshots of dye concentration in a Newtonian fluid ($Bn=0$) at $Re=50$. To facilitate the illustration of concentration development, the snapshots show subdomains of different radii (indicated as $R_{sd}$ in the captions). The white and grey lines show the stirrer's path and the streamlines, respectively. A normalized variance of the dye concentration, $\sigma_{20}^2$, is shown in Figure \ref{fig:concentration_Newtonian_variance}. The markers in Figure \ref{fig:concentration_Newtonian_variance} indicate the time instances of the snapshots.

When stirring begins, the stirrer crosses the dye interface periodically, stretching and folding it (see, e.g., Figures \ref{fig:illustrative_Newtonian_concentrationA} and \ref{fig:illustrative_Newtonian_concentrationB}). This action creates a striated pattern near the stirrer's path, from hereon referred to as the `central region'. The stretched interface and thin striations promote local diffusion. As a result, mixing proceeds rapidly during this stage, characterized by a sharp decline in $\sigma^2_{20}$  (see Figure \ref{fig:concentration_Newtonian_variance}). Mixing slows down once the dye concentration becomes nearly uniform along the stirrer's path ($T\approx 10$; see Figure \ref{fig:illustrative_Newtonian_concentrationC}). 

For $10 \lesssim T \lesssim 20$, the dye concentration appears approximately axisymmetric within the central region, and mixing is temporarily dominated by radial diffusion across the streamlines within this region and across the boundary of this region (Figure \ref{fig:illustrative_Newtonian_concentrationD}). A well-mixed subdomain, where the dye concentration is $\alpha \approx 0.5$, thus develops in the central region. However, this region is not quiescent, as the flow is not axisymmetric. Figure \ref{fig:illustrative_Newtonian_concentrationE} shows that the well-mixed region is slowly advected by the flow, following an approximately helical trajectory. Consequently, the dye interface is brought back into the stirrer's path, where it is once again stretched and folded (Figure \ref{fig:illustrative_Newtonian_concentrationF}). The acceleration of mixing during this stage is congruent with the downward concavity of $\sigma_{20}^2$ for $T \gtrsim 20$ (Figure \ref{fig:concentration_Newtonian_variance}). 

Another mechanism contributing to mixing during this phase ($T \gtrsim 20$) is the extensive stretching of the interface beyond the central region (see Figure  \ref{fig:illustrative_Newtonian_concentrationF}). As the well-mixed region drifts away from the center, interface stretching extends beyond the stirrer's direct influence. This contributes to the sharp decay rates observed in $\sigma^2_{20}$ during this period ($T \gtrsim 20$).

\begin{figure}%{r}{0.4\textwidth}
	%	\vspace{-15mm}
	\centering
	\subfloat[$R_{sd}=5$ \label{fig:illustrative_Newtonian_vorticityA}]{
		\includegraphics[trim=2cm 0.5cm 5.5cm 1.25cm, clip=true,height=\subfigthree\textwidth]{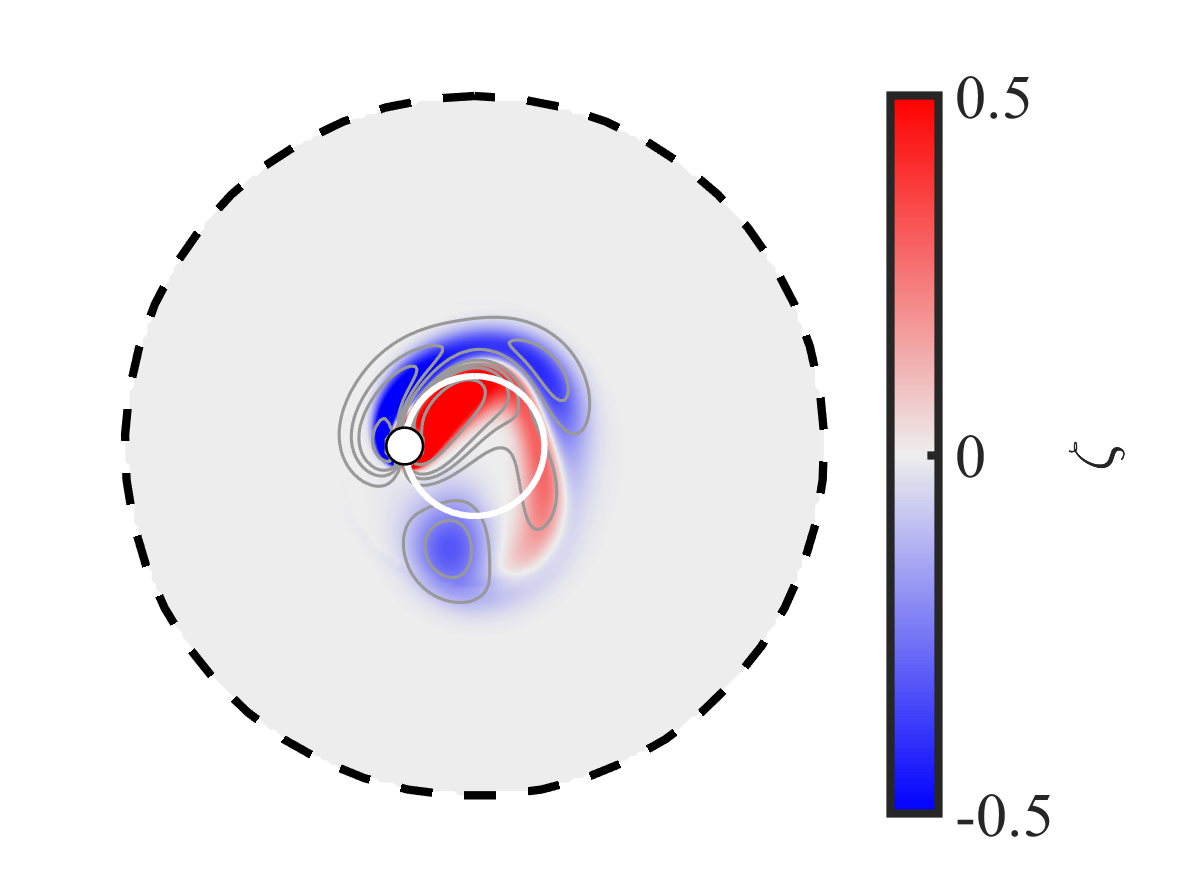}
	}~
	\hspace{0.2cm}
	\subfloat[$R_{sd}=5$\label{fig:illustrative_Newtonian_vorticityB}]{
		\includegraphics[trim=2cm 0.5cm 5.5cm 1.25cm, clip=true,height=\subfigthree\textwidth]{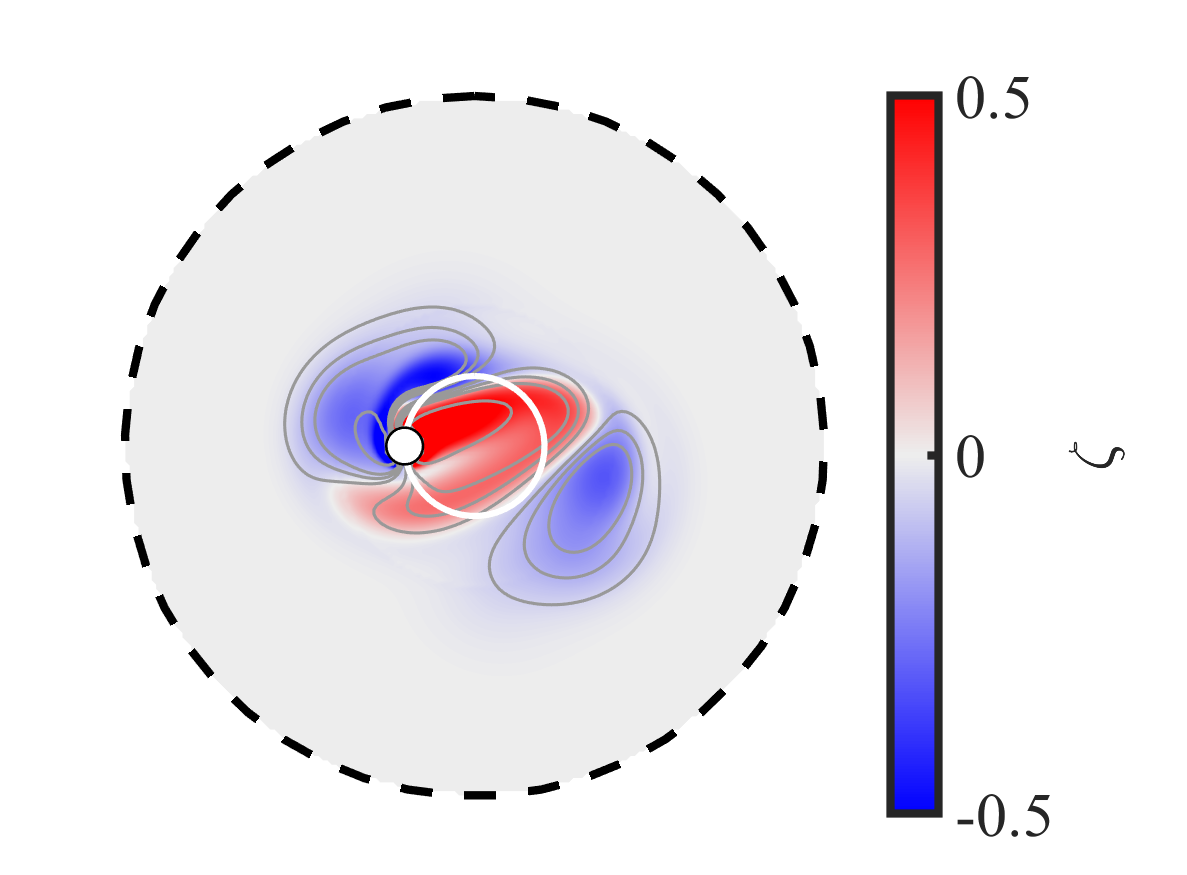}
	}~
	\hspace{0.2cm}
	\subfloat[$R_{sd}=5$\label{fig:illustrative_Newtonian_vorticityC}]{
		\includegraphics[trim=2cm 0.5cm 1cm 1.25cm, clip=true,height=\subfigthree\textwidth]{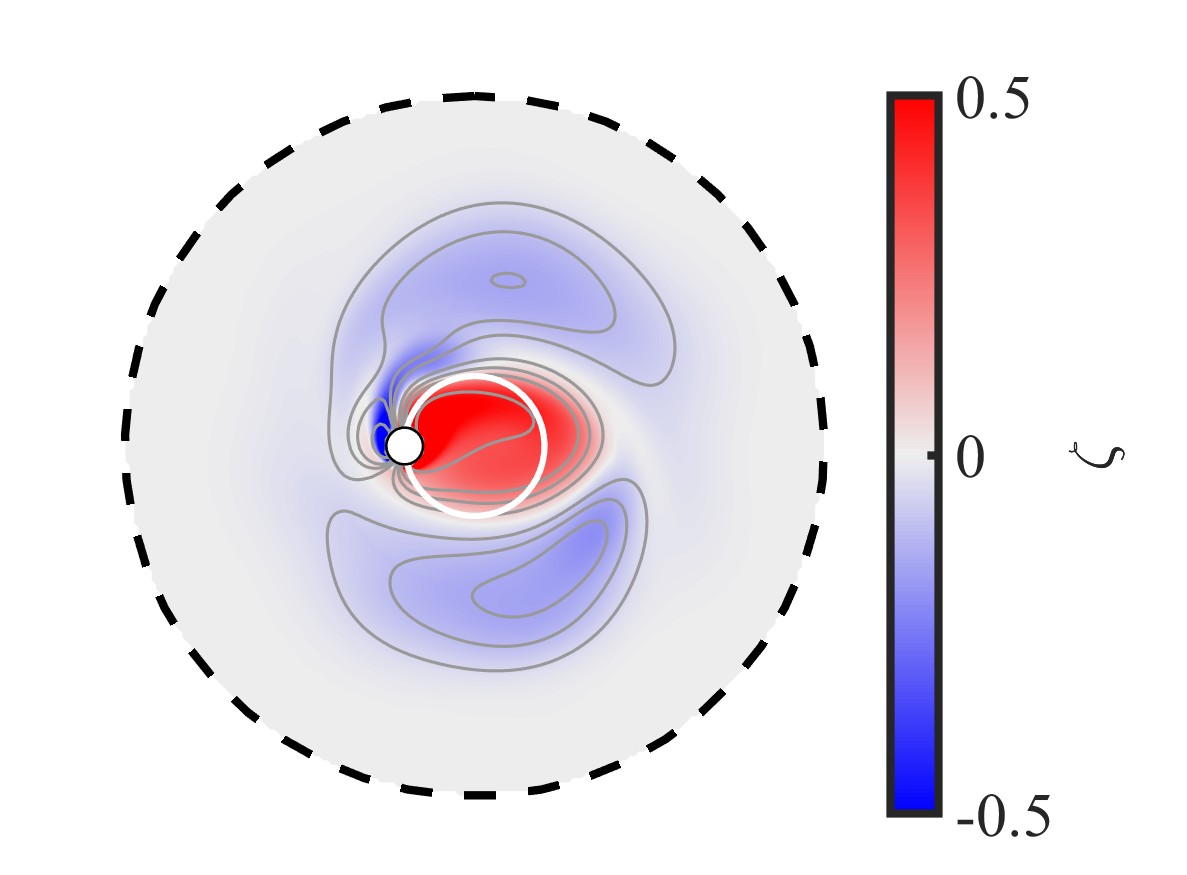}
	}\\
	\vspace{-0.3cm}
	\subfloat[$R_{sd}=5$ \label{fig:illustrative_Newtonian_vorticityD}]{
		\includegraphics[trim=2cm 0.5cm 5.5cm 1.25cm, clip=true,height=\subfigthree\textwidth]{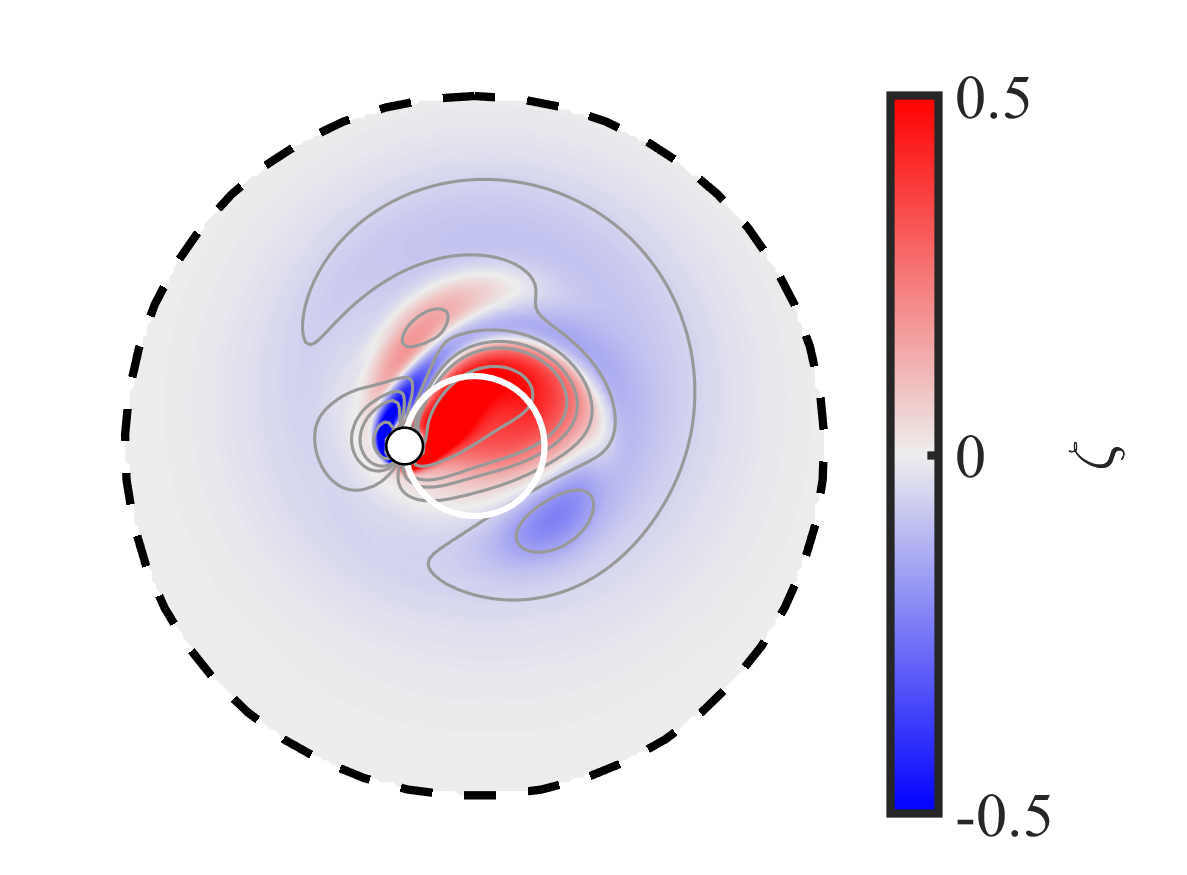}
	}~
	\hspace{0.2cm}
	\subfloat[$R_{sd}=8$\label{fig:illustrative_Newtonian_vorticityE}]{
		\includegraphics[trim=2cm 0.5cm 5.5cm 1.25cm, clip=true,height=\subfigthree\textwidth]{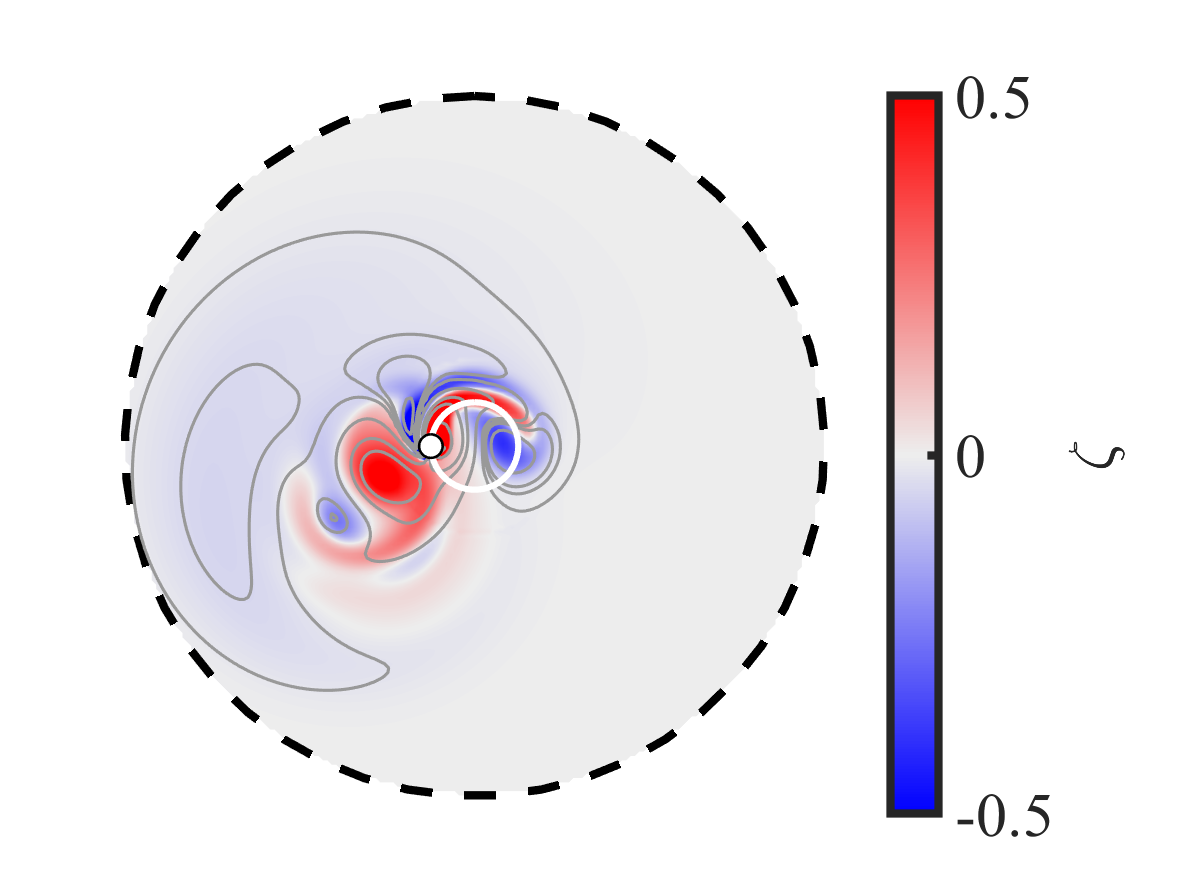}
	}%~
	\hspace{0.2cm}
	\subfloat[$R_{sd}=15$\label{fig:illustrative_Newtonian_vorticityF}]{
		\includegraphics[trim=2cm 0.5cm 1cm 1.25cm, clip=true,height=\subfigthree\textwidth]{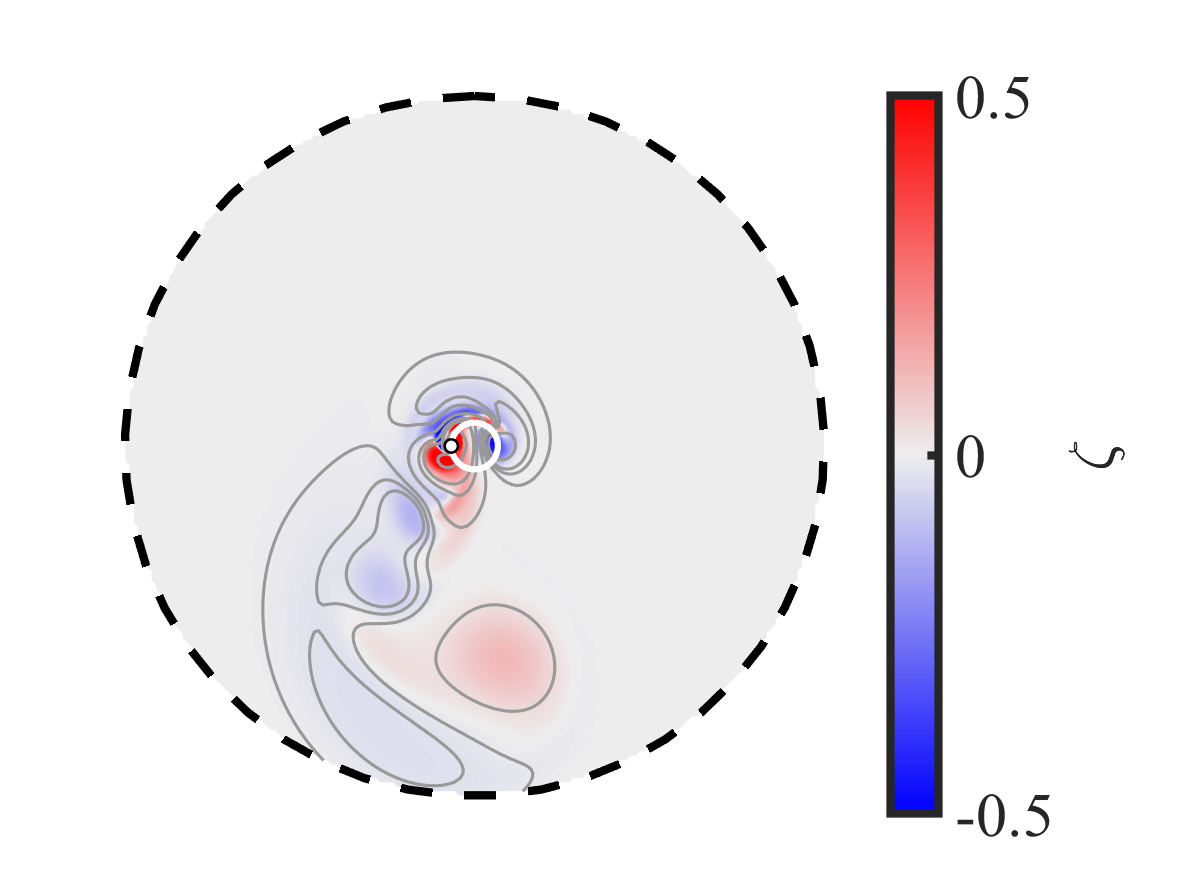}
	}

	\caption{(a–f) Snapshots of the vorticity field in a Newtonian fluid, $Bn = 0$, $Re = 50$,  at times $\newTime = 1, 3, 10, 15, 30, \text{and}~50$.  The white circle marks the stirrer's path, while grey lines represent the streamlines. The radius of the field of view is provided in the caption.}
	\label{fig:illustrative_Newtonian_vorticity}
\end{figure}

The development of the dye concentration is governed by the momentum balance in the flow field; i.e., the coupling between the dye concentration and momentum transfer is one-way because the rheology and density are independent of the dye concentration. Figure \ref{fig:illustrative_Newtonian_vorticity} illustrates the development of the vorticity field in Newtonian fluids at $Re=50$, at the same time instances as Figure \ref{fig:illustrative_Newtonian_concentration}. Similar to Figure \ref{fig:illustrative_Newtonian_concentration}, the white circular line is the stirrer's path and the grey lines are streamlines. To facilitate the illustration of the vorticity field, a scaled vorticity magnitude ($\zeta$) is defined as follows
\begin{align*}
\zeta = \left\{
\begin{array}{lr}
\displaystyle \log(\omega_z + 1) & \text{if} \quad \omega_z \ge 0 \\
-\log(|\omega_z| + 1) & \text{if} \quad \omega_z < 0
\end{array}
\right.
\end{align*}

Note that $\zeta$ retains the sign of $\omega_z$; i.e., positive and negative $\zeta$ indicate counter clockwise (CCW) and clockwise (CW) rotation, respectively.

When stirring starts, two small attached eddies appear behind the stirrer. More vortices are shed as the stirrer moves along its path (see Figure \ref{fig:illustrative_Newtonian_vorticityA}). The rotation of these vortices is consistent with the stirrer's movement with CCW and CW vortices shed inside and outside the stirrer's path, respectively. This is reminiscent of the archetypal problem of the flow around a cylinder that moves at the constant velocity of $\hat{U}_o = \hat{r}_o\hat{\Omega}$. In the archetypal problem, laminar shedding is expected at $40 \lesssim Re \lesssim 150$ \citep{roshko1954development}. 

Figure \ref{fig:illustrative_Newtonian_vorticityB} represents the early dynamics following the onset of stirring. The periodic passage of the stirrer disrupts the formation and advection of CCW vortices, while CW vortices remain outside the stirrer’s path, drifting slowly away (compare Figures \ref{fig:illustrative_Newtonian_vorticityB} \& \ref{fig:illustrative_Newtonian_vorticityC}). Meanwhile, as the attached eddies develop, the CCW eddy expands across the stirrer’s path. When the stirrer moves through this eddy, it generates additional CCW vortices (see Figure \ref{fig:illustrative_Newtonian_vorticityD}). These CCW vortices also dissipate quickly before reaching beyond the central region. 

Figure \ref{fig:illustrative_Newtonian_vorticityA} also shows that the stirrer completes a period before the previously shed CW vortices advect away from the central region; e.g., during the second period ($1\le T \le 2$), the stirrer interacts with the CW vortices shed during the first period. This mechanism interferes with vortex shedding: temporarily ($2 \le T < 20$), the stirrer's path is surrounded by two CW vortices that interact with and follow the stirrer slowly moving away from the domain center. The first two stages of mixing (stretching and folding of the interface followed by diffusion) take place concurrently with the vorticity dynamics discussed above.

At $T \approx 20$, the attached CCW eddy expands sufficiently to escape the stirrer's path (\ref{fig:illustrative_Newtonian_vorticityE}). As this eddy gradually moves away from the central region, the approximate symmetry of the vorticity field, along with the corresponding symmetry in dye concentration in the central region (Figure \ref{fig:illustrative_Newtonian_concentrationE}), is disrupted. From this point onward, vortices are shed away from the central region, traveling far across the flow domain (Figure \ref{fig:illustrative_Newtonian_vorticityF}).

%{\color{red}When a cylinder travels on a straight line, the dimensionless shedding frequency is  \citep{roshko1954development}
%\begin{align*}
%	& 0.1\lesssim St=\frac{\hat{n} \hat{d}_s}{\hat{r}_o \hat{\Omega}} \lesssim 0.2
%\end{align*}
%where $\hat{n}$ and $St$ are the shedding frequency and the Strouhal number, respectively. It follows that,
%\begin{align*}
%	0.1 c \lesssim \frac{\hat{n}}{\hat{\Omega}} \lesssim 0.2 c
%\end{align*}
%
%This is comparable to the shedding frequency observed here. It is expected that as the radius of the path approaches infinity the characteristic $St$, corresponding to a cylinder moving on a straight line, is recovered.
%}

\begin{figure}
	\centering
	\subfloat[$R_{sd}=3$\label{fig:locationVortexNewtonianA}]{
		\includegraphics[trim=0cm 0cm 0cm 0cm, clip=true,width=.33\textwidth]{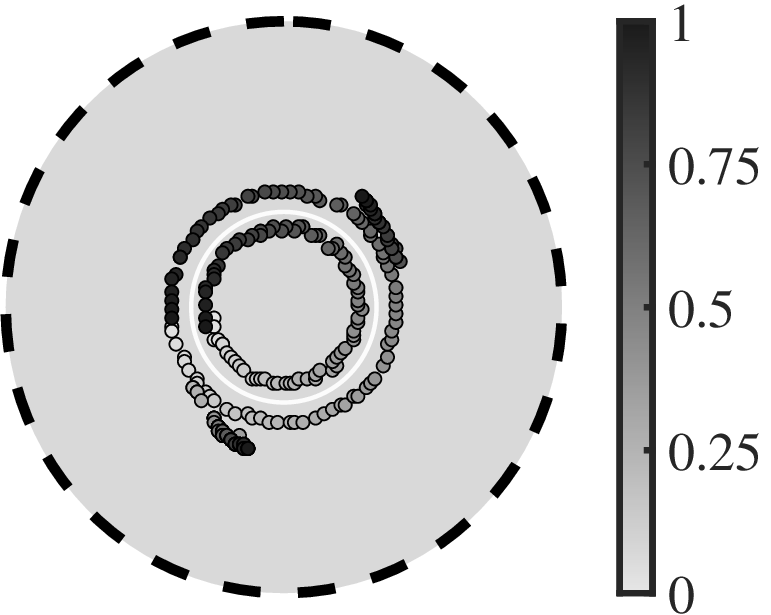}
	}~
%	\hspace{0.2cm}
	\subfloat[ $R_{sd}=3$\label{fig:locationVortexNewtonianB}]{
		\includegraphics[trim=0cm 0cm 0cm 0cm, clip=true,width=.315\textwidth]{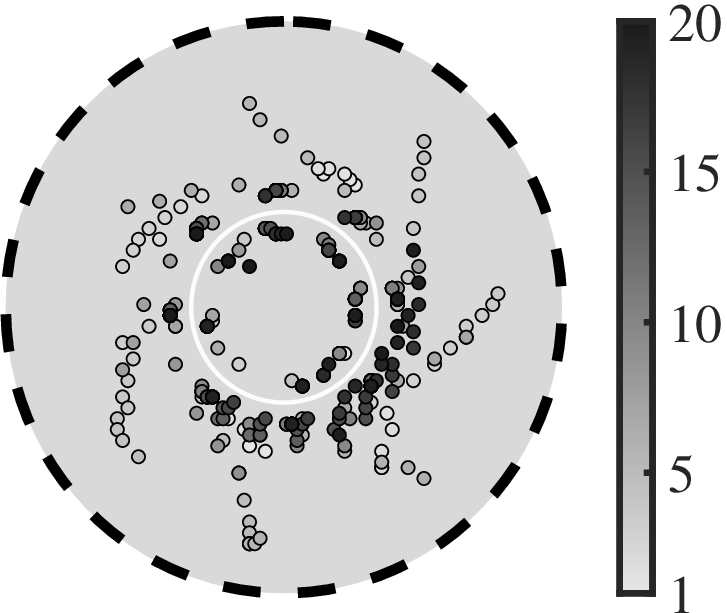}
	}~
%	\hspace{0.2cm}
	\subfloat[$R_{sd}=10$\label{fig:locationVortexNewtonianC}]{
		\includegraphics[trim=0cm 0cm 0cm 0cm, clip=true,width=.315\textwidth]{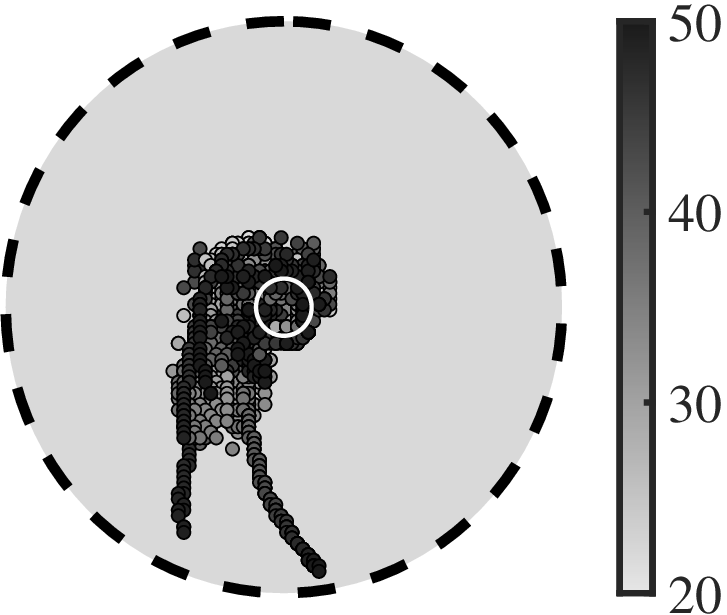}
	} 
%	\hspace{0.35cm}
%	\subfloat[$R_{sd}=15$\label{fig:locationVortexNewtonianD}]{
%		\includegraphics[trim=0cm 0cm 0cm 0cm, clip=true,width=.28\textwidth]{/Illustrative/locationVortexNewtonianD.eps}
%	} 
	\caption{Time evolution of the vortex centers in a Newtonian fluid, $Bn = 0$, $Re = 50$,  over different time intervals, as indicated by the colorbar. Lighter shades correspond to earlier times within each subfigure. The white solid line represents the stirrer's path, while the black dashed lines denote the subdomain boundary.}
	\label{fig:vorticesCenters_Newtonian}
\end{figure}

To better illustrate the stages of vortex development, Figure \ref{fig:vorticesCenters_Newtonian} shows the approximate paths of vortex centers. When stirring begins, a pair of attached eddies forms and travels with the stirrer (Figure \ref{fig:locationVortexNewtonianA}). Simultaneously, two clockwise (CW) vortices are shed outside the stirrer's path within the first period ($T \le 1$). As shown in Figure \ref{fig:locationVortexNewtonianB} , these shed vortices initially remain close to the central region, drifting away slowly. This quasi-periodic flow pattern leads to an approximately axisymmetric concentration field and a diffusion-dominated mixing phase. Finally, the vortices escape the central region, causing the well-mixed region to advect outward (Figure \ref{fig:locationVortexNewtonianC}) - a stage marked by interface stretching and folding beyond the central region. The kinetic energy of escaping vortices decays exponentially due to viscous dissipation, allowing them to advect indefinitely. Consequently, mixing continues unbounded as the vortices transport dye farther into the flow domain.

Comparing the development of mixing and vorticity field, we identify three key mechanisms promoting mixing during different stages of the flow: (i) local stretching and folding of the dye interface with the movement of the stirrer, (ii) diffusion dominated mixing when flow structure is approximately steady, and (iii) the advection of dye and stretching of the interface with vortices that escape the central region.  %To illustrate this, Figure \ref{fig:vorticesCenters_Newtonian} display the approximate time evolution of vortex centers during the different stages of mixing. Minima and maxima of the vorticity field are used to estimate the locations of the vortex centers. In every subfigure, darker shades of grey represent later times within the given time period. The white solid line shows the stirrer's path. Note that different subfigures show subdomains of different sizes.

%arare localized in the central region and, with both radial and tangential motion, facilitate the diffusion of striated dye. During the temporary diffusive time period shown in Figure \ref{fig:locationVortexNewtonianB}, the vortices moved at a nearly constant radial distance. The gradual rotation of vortex centers is presented in Figure \ref{fig:locationVortexNewtonianC}. Traveling vortices and newly formed vortices from the central region due to vortex shedding are presented in Figure \ref{fig:locationVortexNewtonianD}, which resulted in the advection of blobs of dye and the stretching and folding of the interface.

%%--------------------------------		
\subsection{Blending a viscoplastic fluid}
\label{sec:vpf}
%{\color{red}
%\begin{itemize}
%%\item {\bf Stirring a viscoplastic fluid} 
%	\item illustrated the change in mixing when the fluid has a yield stress: 
%	\item when $Bn\ll 1$ similar to Newtonian except "free vortices" decay within a finite distance
%	\item at higher Bn, vortices remain trapped near the stirrer
%	\item at $Bn \sim 1$ only two eddies attached to the stirrer
%	\item drawn analogies with the classic problem of the flow past a cylinder to provide a fludi mech perspective
%\end{itemize}
%}		

\begin{figure}%{r}{0.4\textwidth}
	%	\vspace{-15mm}
	\centering
	\subfloat[$\newTime=1$ \label{fig:Bn025_c_A}]{
		\includegraphics[trim=2cm 0.5cm 5.5cm 1.25cm, clip=true,height=\subfigthree\textwidth]{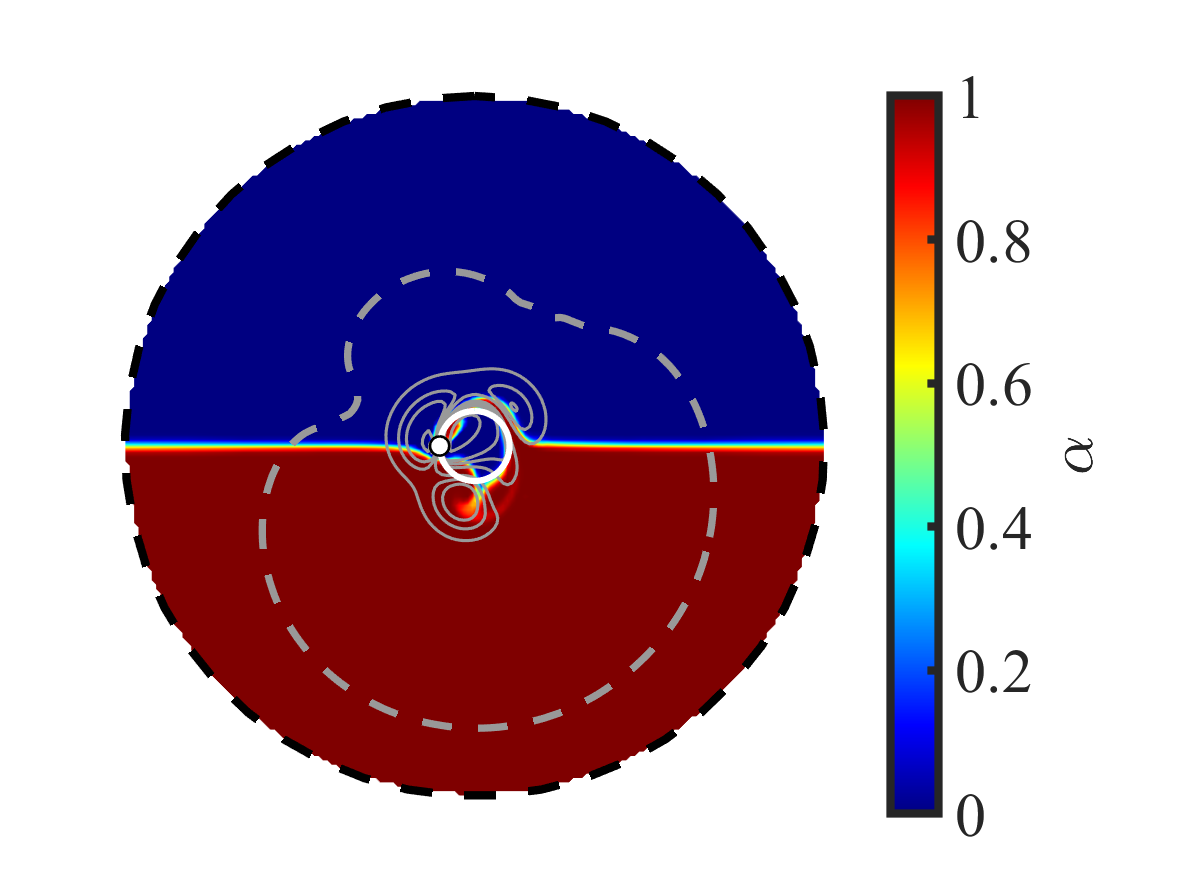}
	}~
	\hspace{0.2cm}
	\subfloat[$\newTime=3$  \label{fig:Bn025_c_B}]{
		\includegraphics[trim=2cm 0.5cm 5.5cm 1.25cm, clip=true,height=\subfigthree\textwidth]{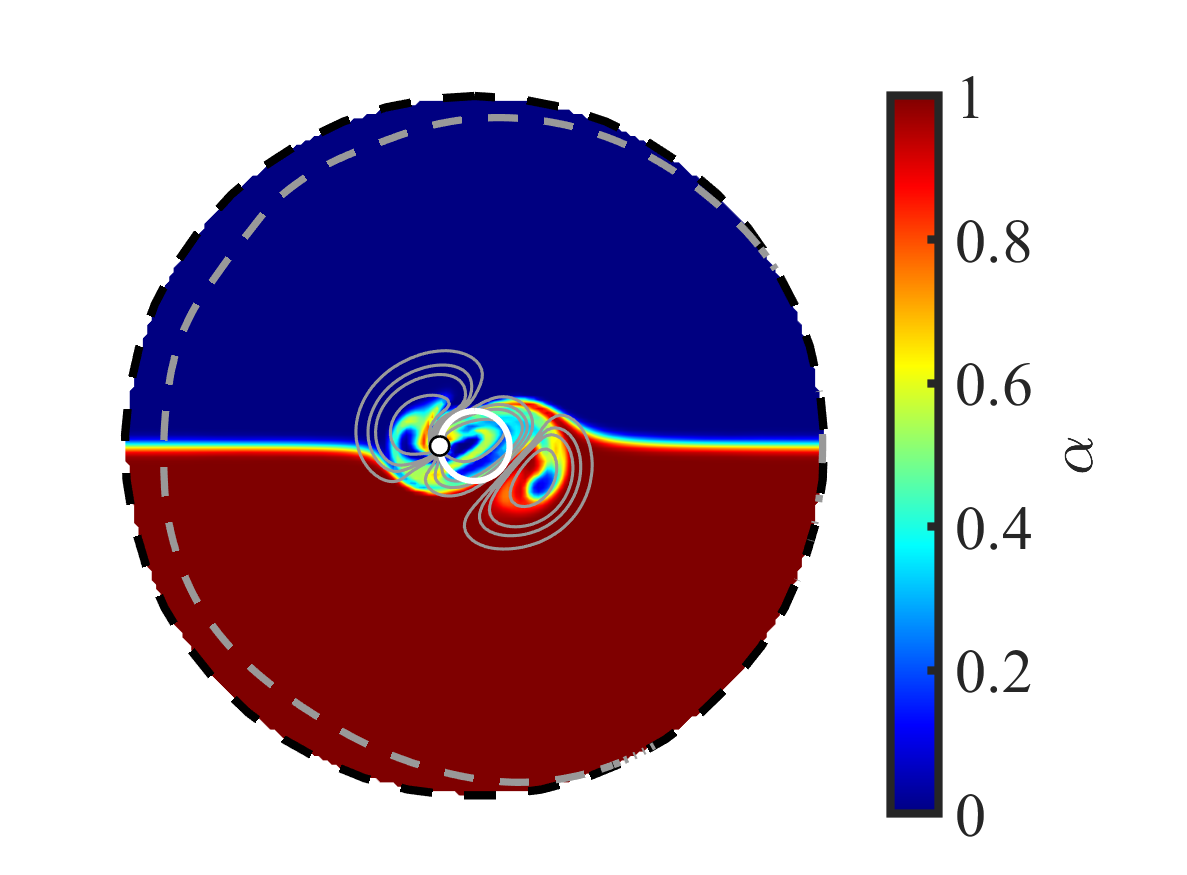}
	}~
	\hspace{0.2cm}
	\subfloat[$\newTime=10$\label{fig:Bn025_c_C}]{
		\includegraphics[trim=2cm 0.5cm 1.5cm 1.25cm, clip=true,height=\subfigthree\textwidth]{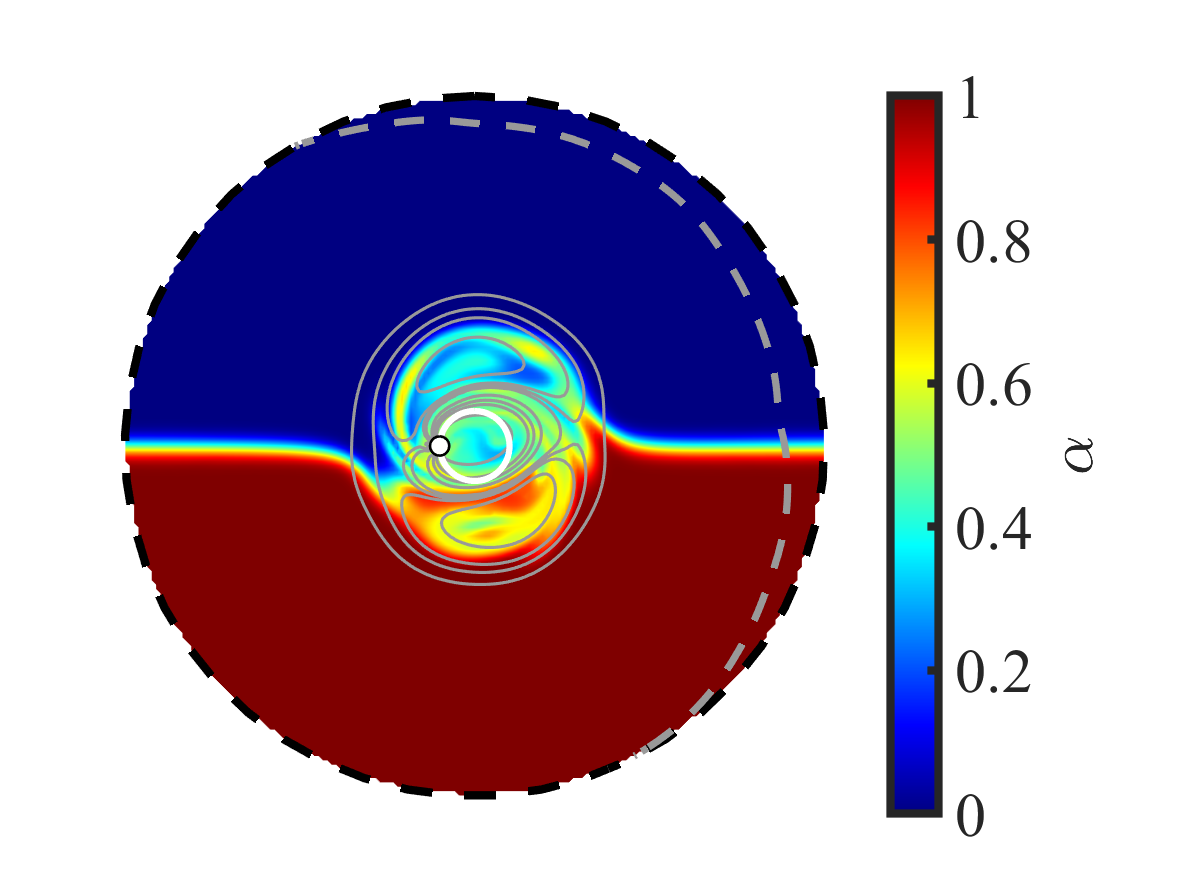}
	}\\
	\vspace{-0.3cm}
	\subfloat[$\newTime=15$\label{fig:Bn025_c_D}]{
		\includegraphics[trim=2cm 0.5cm 5.5cm 1.25cm, clip=true,height=\subfigthree\textwidth]{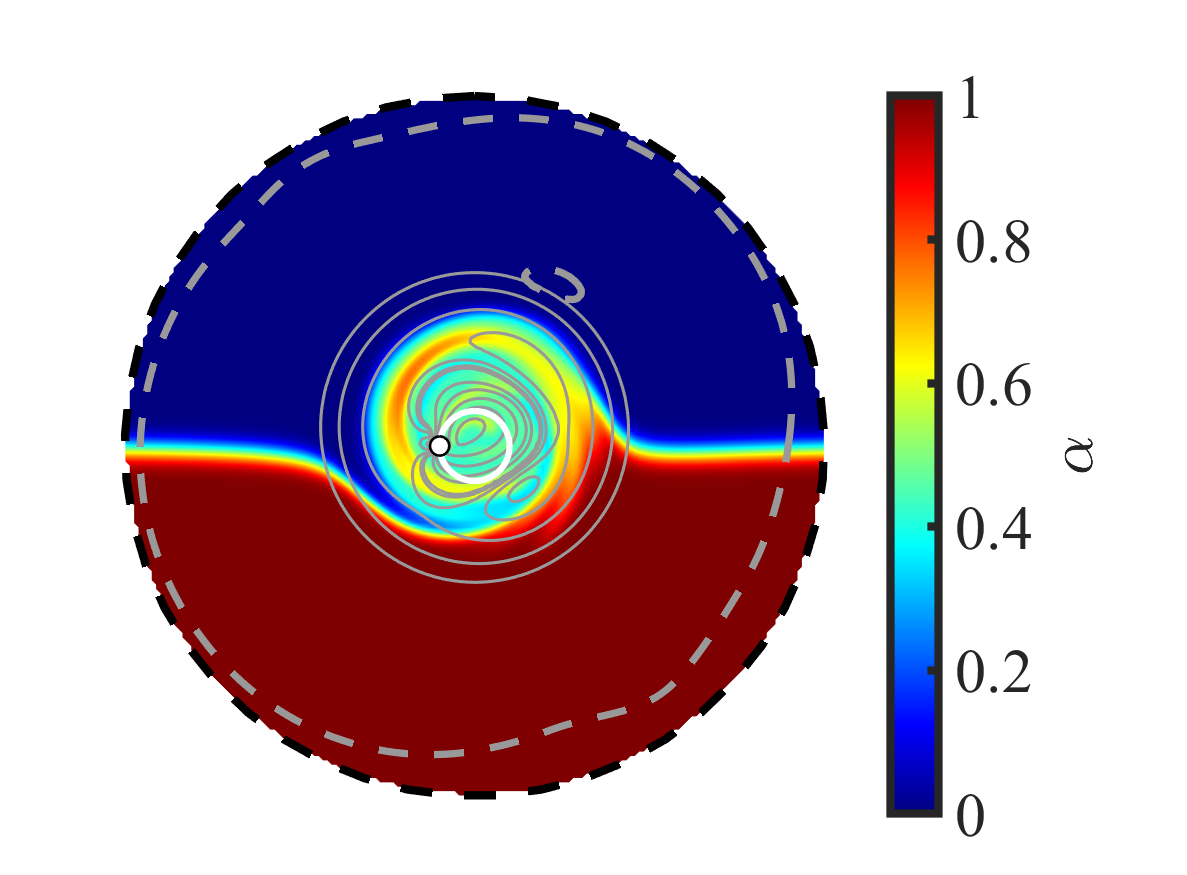}
	}~
	\hspace{0.2cm}
	\subfloat[$\newTime=30$\label{fig:Bn025_c_E}]{
		\includegraphics[trim=2cm 0.5cm 5.5cm 1.25cm, clip=true,height=\subfigthree\textwidth]{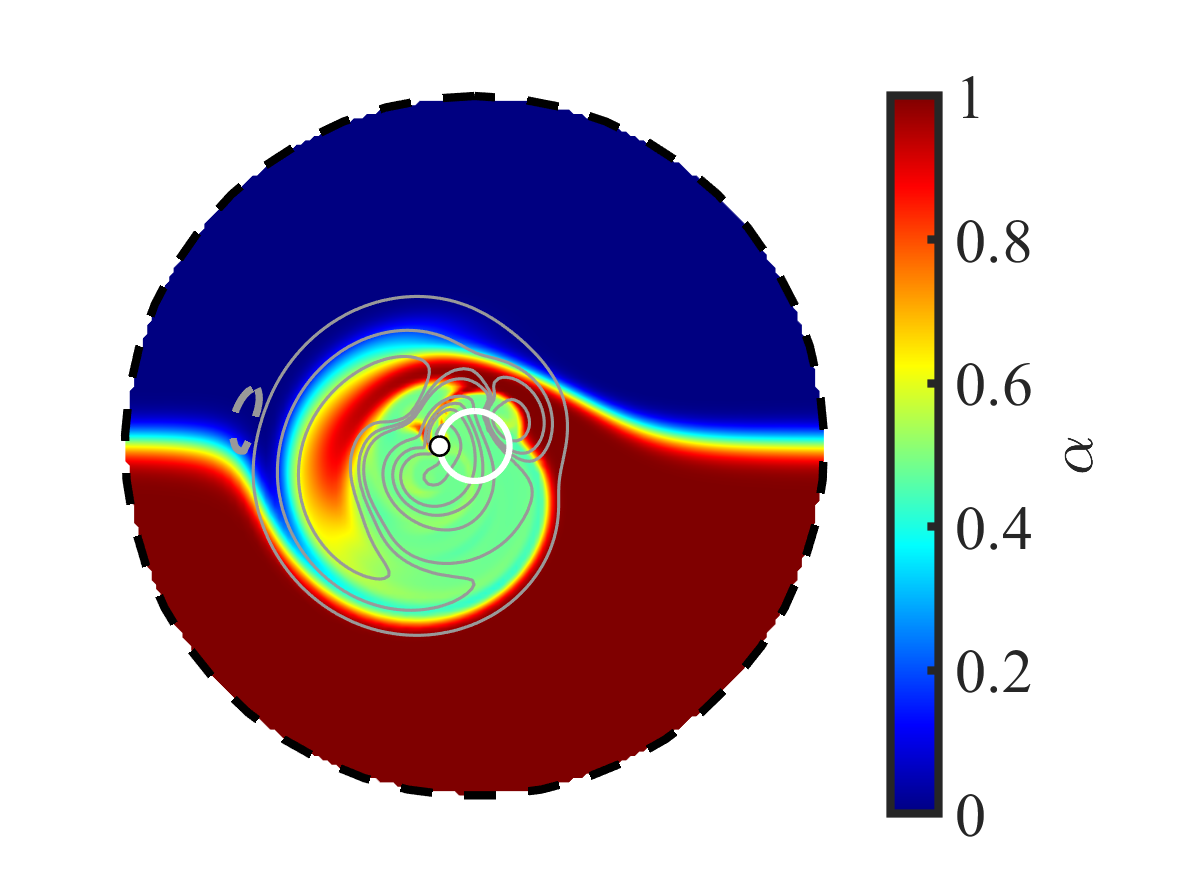}
	}~
	\hspace{0.2cm}
	\subfloat[$\newTime=50$\label{fig:Bn025_c_F}]{
		\includegraphics[trim=2cm 0.5cm 1.5cm 1.25cm, clip=true,height=\subfigthree\textwidth]{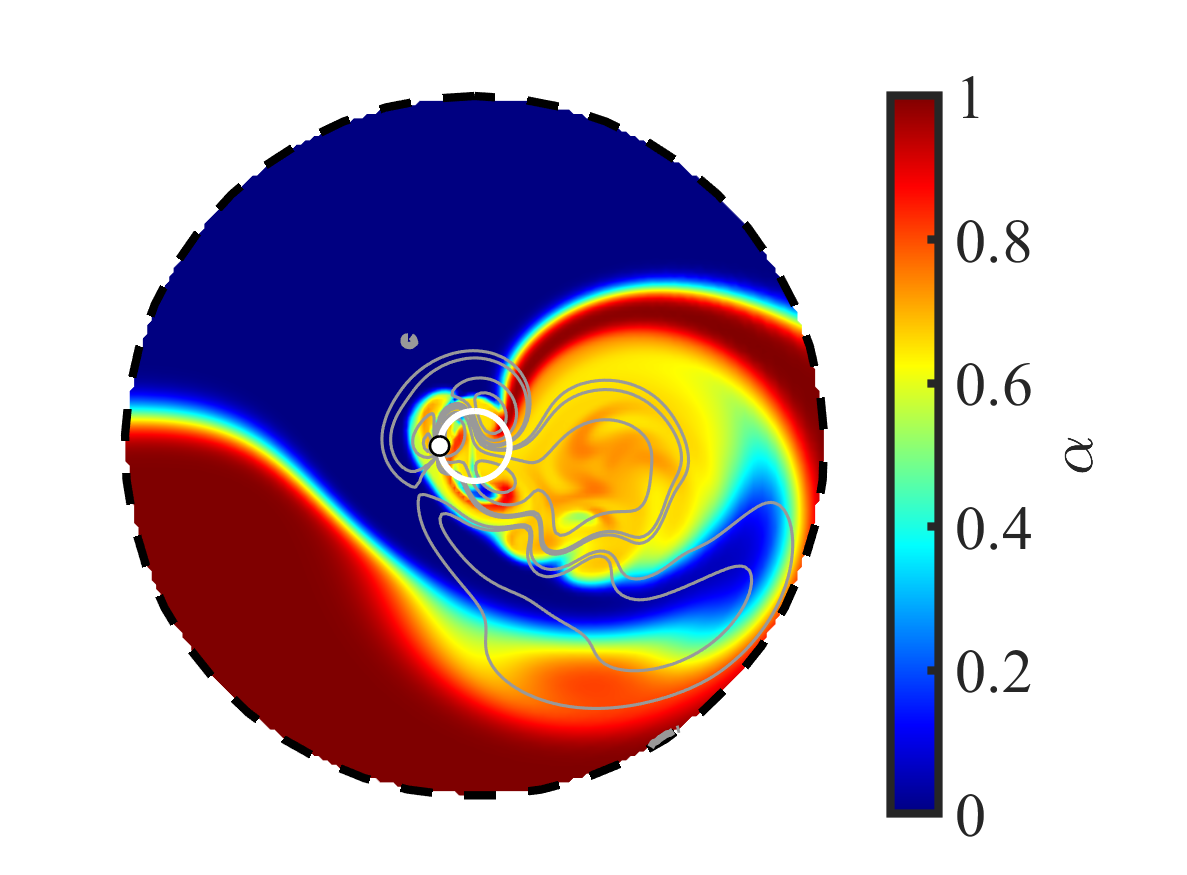}
	}
	
	\caption{Snapshots of the dye concentration field in a VPF with a low yield stress, $Bn = 0.025$, $Re = 50$, and $R_{sd} = 10$. The white circle indicates the stirrer’s path, grey lines represent streamlines, and dashed grey lines show contours of $\tau = 1.05Bn$.}
	\label{fig:Bn025_c}
\end{figure}

Figure \ref{fig:Bn025_c} shows the development of the dye concentration when the fluid has a small yield stress, $Re=50$ and $Bn=0.025$. The dashed grey lines display the contours of $\tau=1.05Bn$, a conservative estimate of the boundary of the yielded region. As before, the white and grey solid lines represent the stirrer's path and the streamlines, respectively. 

The primary stages of mixing resemble those observed in the Newtonian case, beginning with the stretching and folding of the dye interface (Figure \ref{fig:Bn025_c_A}\&\ref{fig:Bn025_c_B}). This is followed by the formation of an approximately axisymmetric concentration profile and the emergence of a well-mixed region within the central area (Figures \ref{fig:Bn025_c_C}\&\ref{fig:Bn025_c_D}, respectively). As the well-mixed region drifts outward from the center, the interface undergoes extensive stretching, accelerating the mixing process (Figure \ref{fig:Bn025_c_E}\&\ref{fig:Bn025_c_F}). A comparison between Figures \ref{fig:illustrative_Newtonian_concentration} and \ref{fig:Bn025_c} further shows that the yield stress has little influence on the concentration field until the well-mixed region moves beyond the central area.

\begin{figure}%{r}{0.4\textwidth}
	%	\vspace{-15mm}
	\centering
	\subfloat[$\newTime=1$\label{fig:Bn025_v_A}]{
		\includegraphics[trim=2cm 0.5cm 5.5cm 1.25cm, clip=true,height=\subfigthree\textwidth]{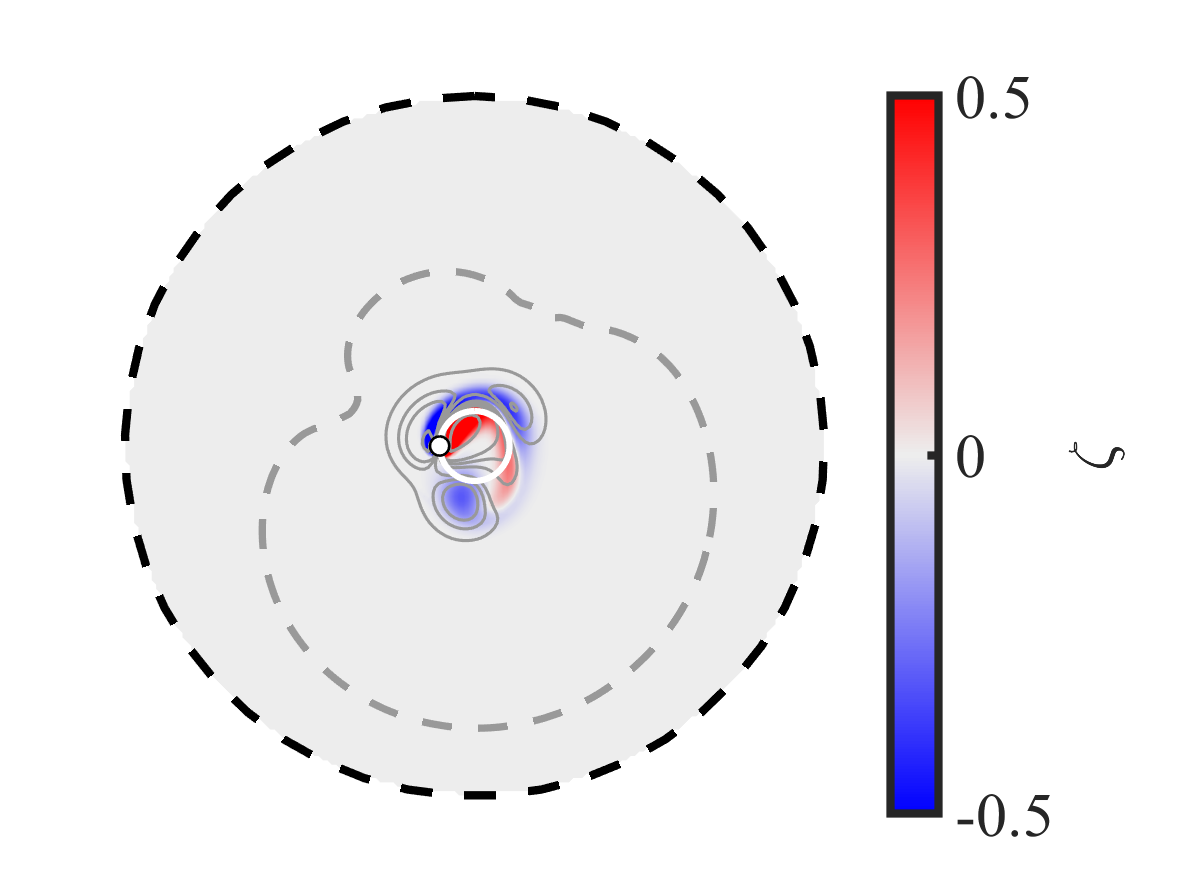}
	}~
	\hspace{0.2cm}
	\subfloat[$\newTime=3$\label{fig:Bn025_v_B}]{
		\includegraphics[trim=2cm 0.5cm 5.5cm 1.25cm, clip=true,height=\subfigthree\textwidth]{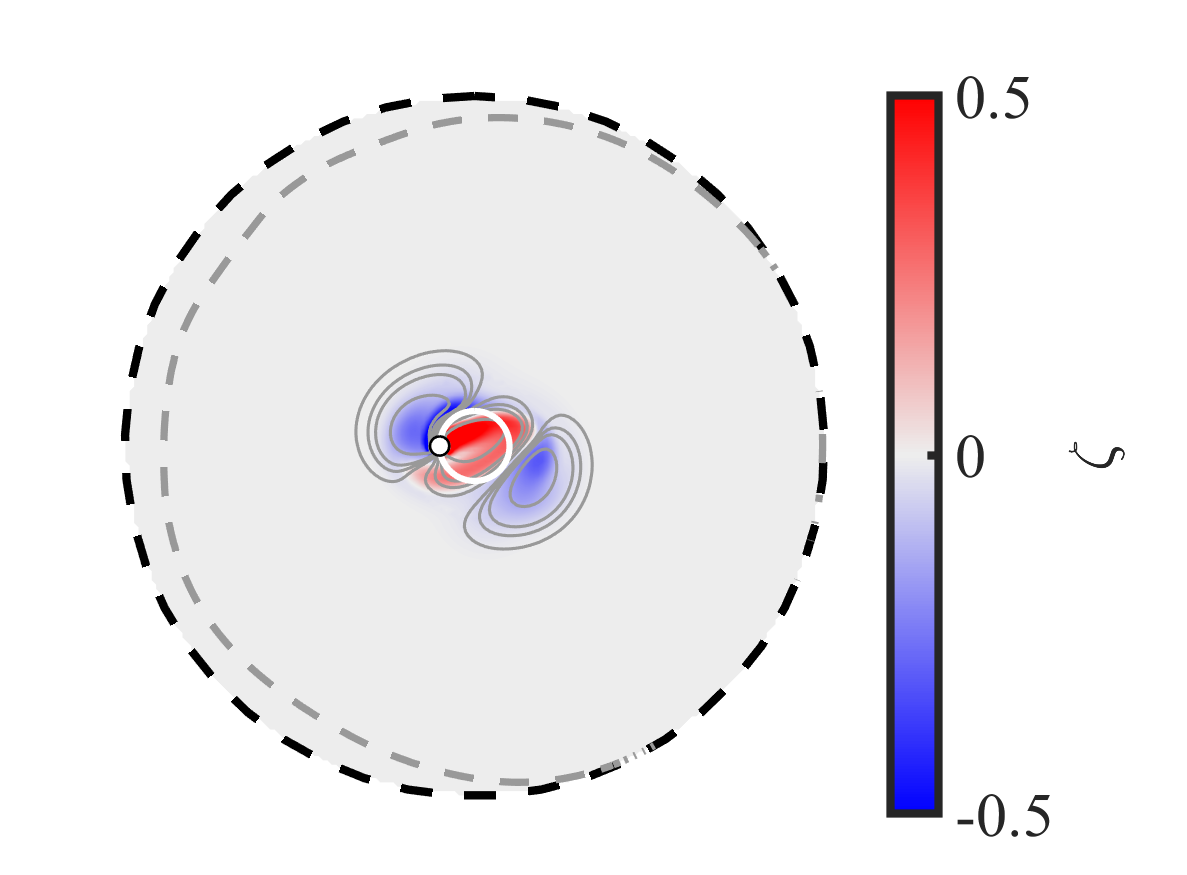}
	}~
	\hspace{0.2cm}
	\subfloat[$\newTime=10$\label{fig:Bn025_v_C}]{
		\includegraphics[trim=2cm 0.5cm 1cm 1.25cm, clip=true,height=\subfigthree\textwidth]{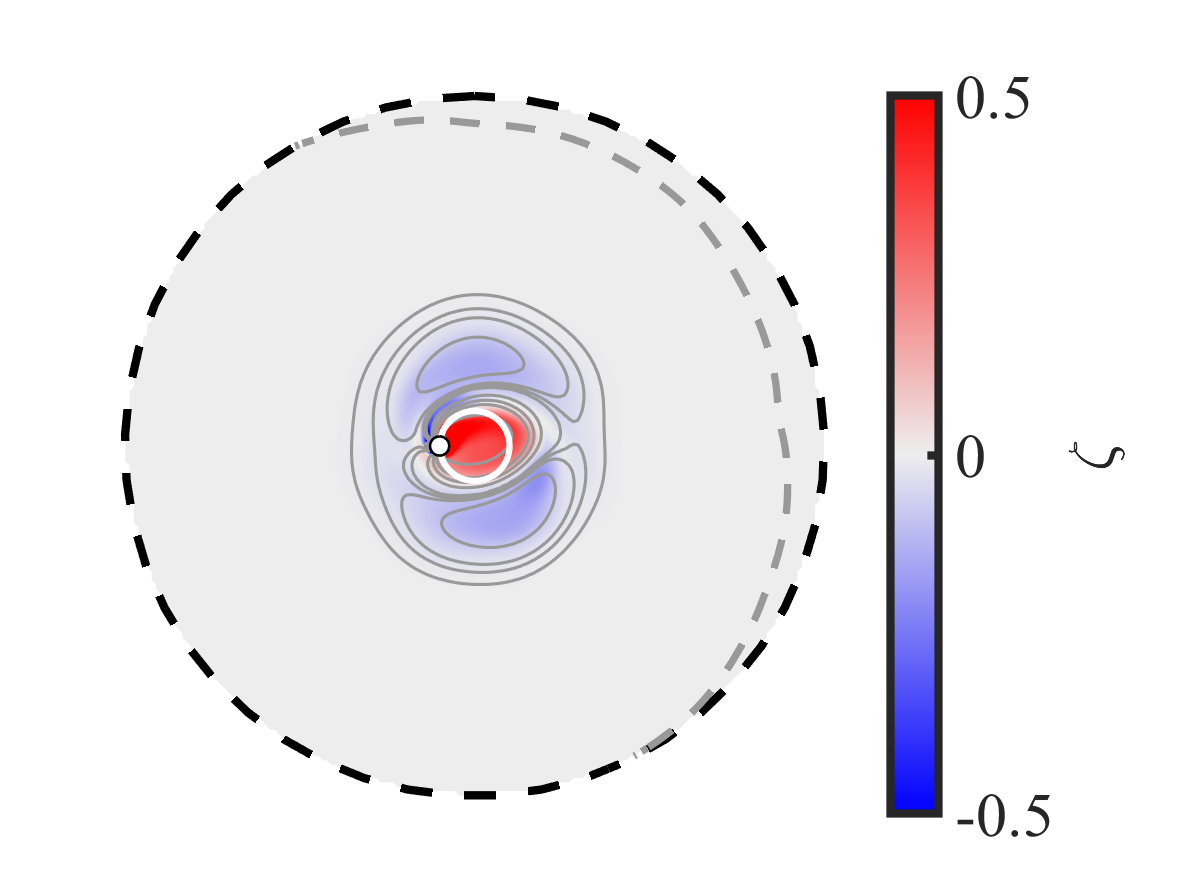}
	}\\
	\vspace{-0.3cm}
	\subfloat[$\newTime=15$ \label{fig:Bn025_v_D}]{
		\includegraphics[trim=2cm 0.5cm 5.5cm 1.25cm, clip=true,height=\subfigthree\textwidth]{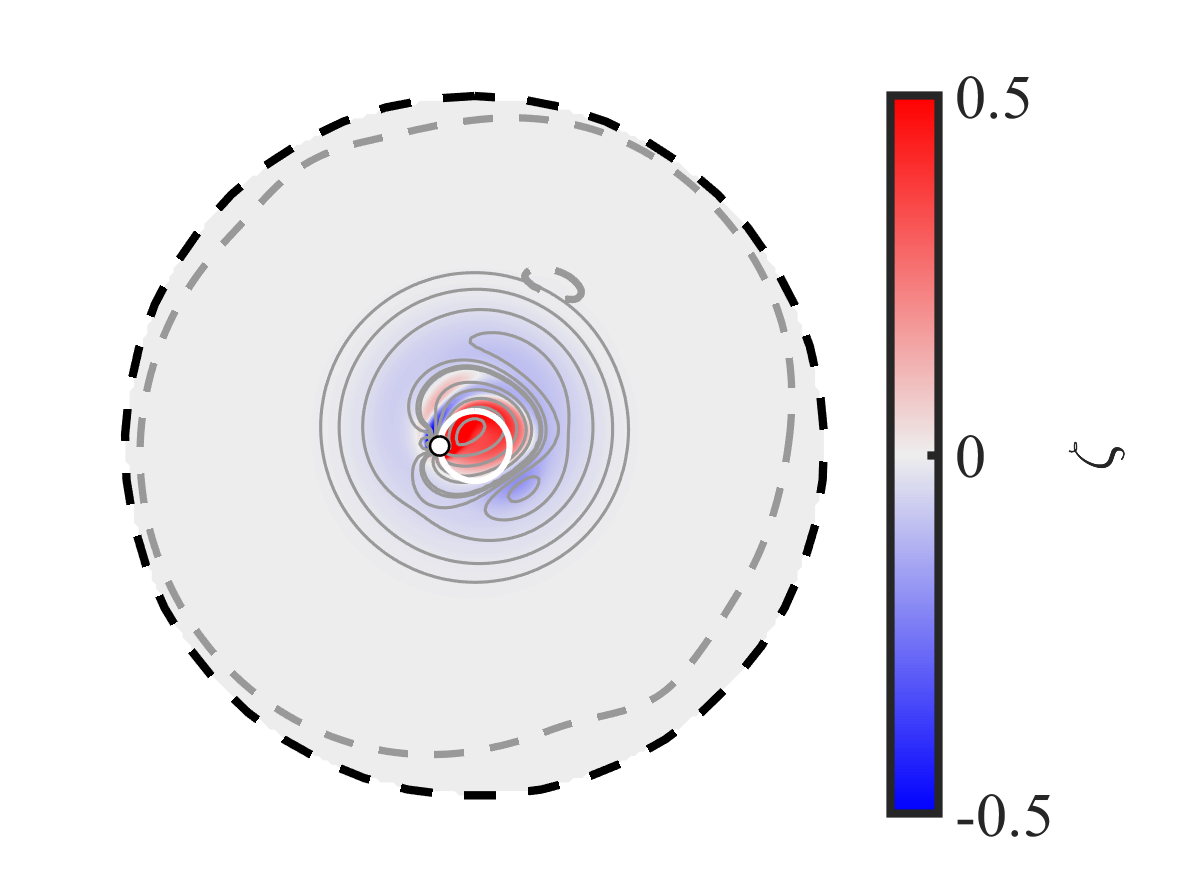}
	}~
	\hspace{0.2cm}
	\subfloat[$\newTime=30$\label{fig:Bn025_v_E}]{
		\includegraphics[trim=2cm 0.5cm 5.5cm 1.25cm, clip=true,height=\subfigthree\textwidth]{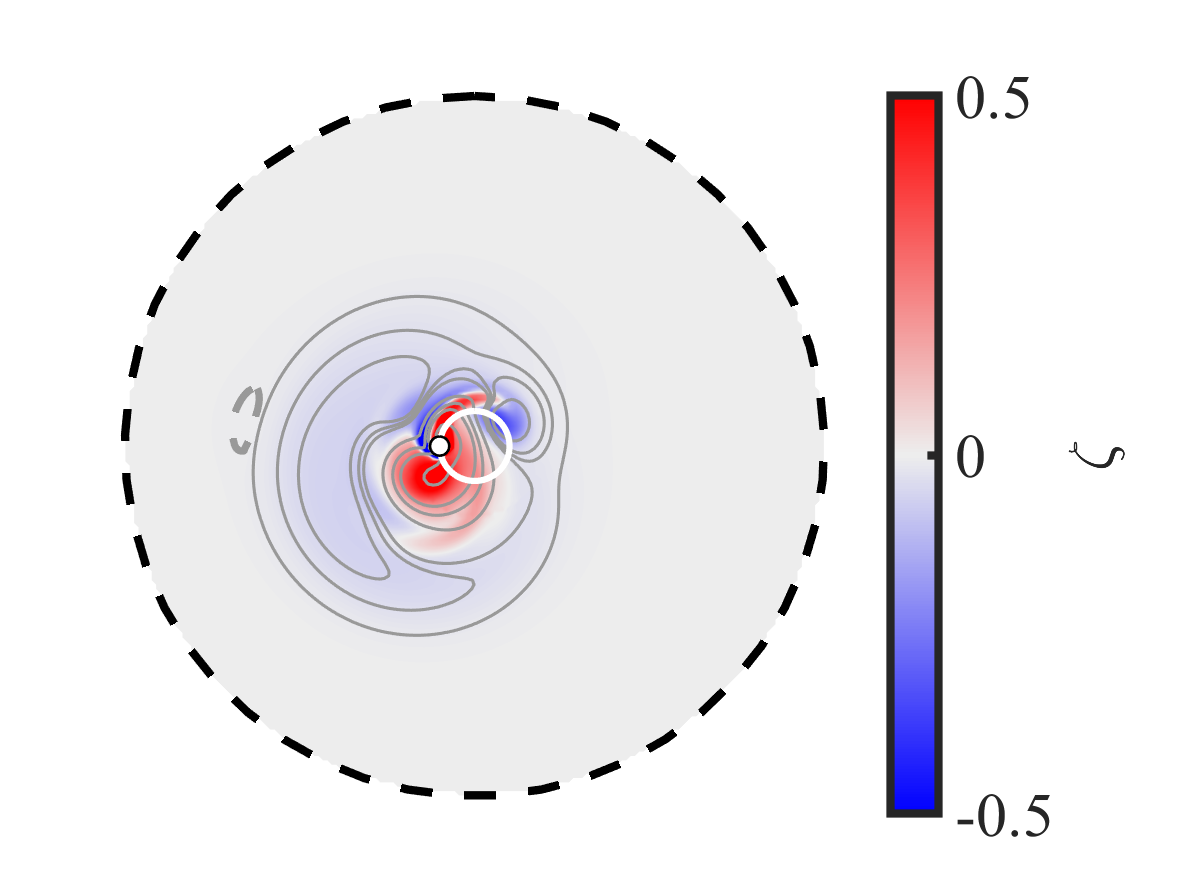}
	}%~
	\hspace{0.2cm}
	\subfloat[$\newTime=50$\label{fig:Bn025_v_F}]{
		\includegraphics[trim=2cm 0.5cm 1cm 1.25cm, clip=true,height=\subfigthree\textwidth]{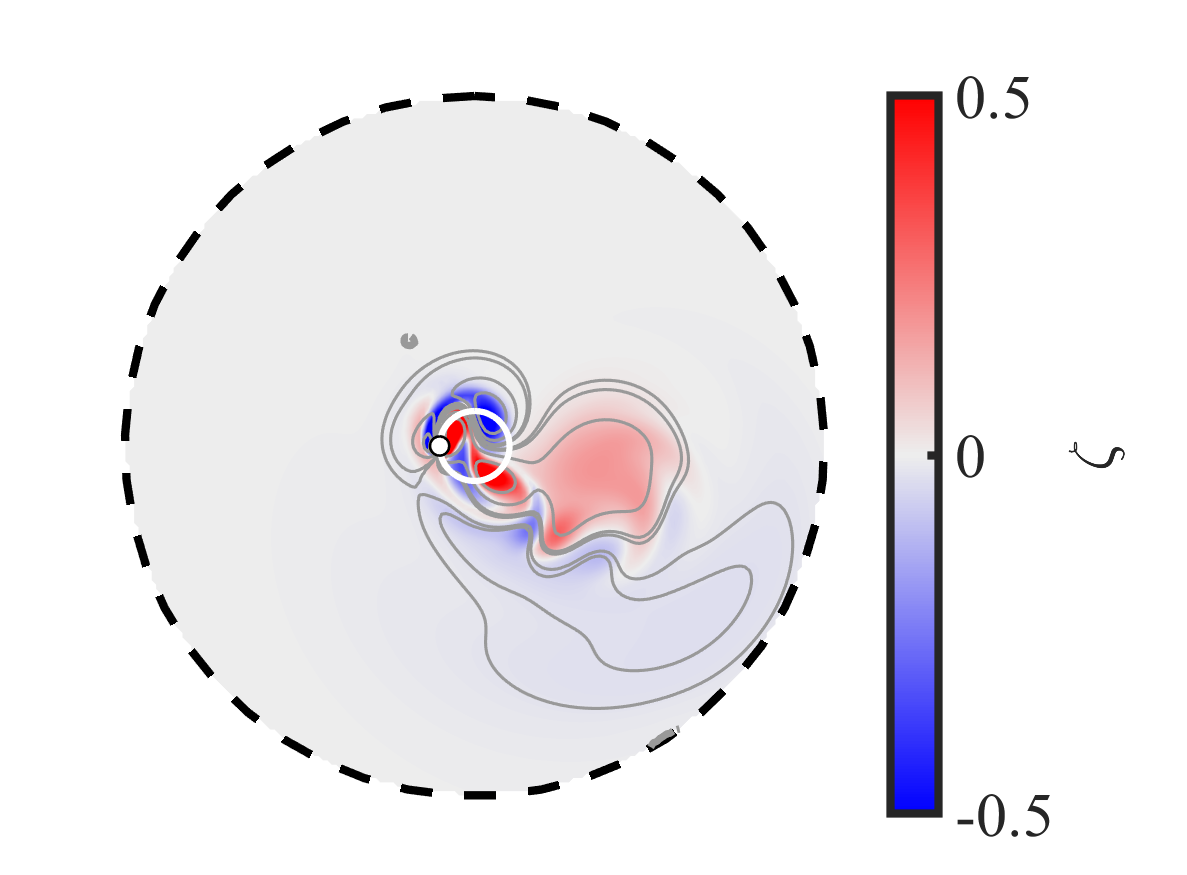}
	}	
	
	\caption{Snapshots of the vorticity field in a Newtonian fluid, a VPF with a low yield stress, $Bn = 0.025$, $Re = 50$, and $R_{sd} = 10$.  The white circle indicates the stirrer’s path, grey lines represent streamlines, and dashed grey lines show contours of $\tau = 1.05Bn$.}
	\label{fig:Bn025_v}
\end{figure}

Figure \ref{fig:Bn025_v} presents the evolution of the vorticity field at the same time instances as Figure \ref{fig:Bn025_c}. Comparison with Figure \ref{fig:illustrative_Newtonian_vorticity} illustrates that the vorticity fields are quite similar until vortices escape the central region. This indicates that within the central region, the influence of the yield stress is negligible compared to purely viscous effects. However, outside this region, energy dissipation due to yield stress becomes increasingly significant with distance from the domain center. Far enough from the center, the fluid is expected to remain unyielded.

%As the CW eddy drifts away from the central region, the influence of the yield stress becomes increasingly pronounced. A comparison of the vorticity fields for the Newtonian and $Bn = 0.025$ cases at $t = 50$, for example, reveals that 
In the viscoplastic case, the escaped vortices predominantly drift in the azimuthal direction (see Figures \ref{fig:Bn025_v_E}\&\ref{fig:Bn025_v_F}). In contrast, in the Newtonian case, the radial displacement of the vortices is more pronounced. This highlights the first mechanism of mixing localization in viscoplastic fluids: when the fluid has a yield stress, escaped vortices are confined within a finite distance from the stirrer, effectively localizing mixing. This behavior contrasts with that of Newtonian fluids, where vortices can theoretically advect without bounds.

\begin{figure}
	\centering
	\subfloat[$R_{sd}=3$\label{fig:Bn025_vc_A}]{
		\includegraphics[trim=0cm 0cm 0cm 0cm, clip=true,width=.33\textwidth]{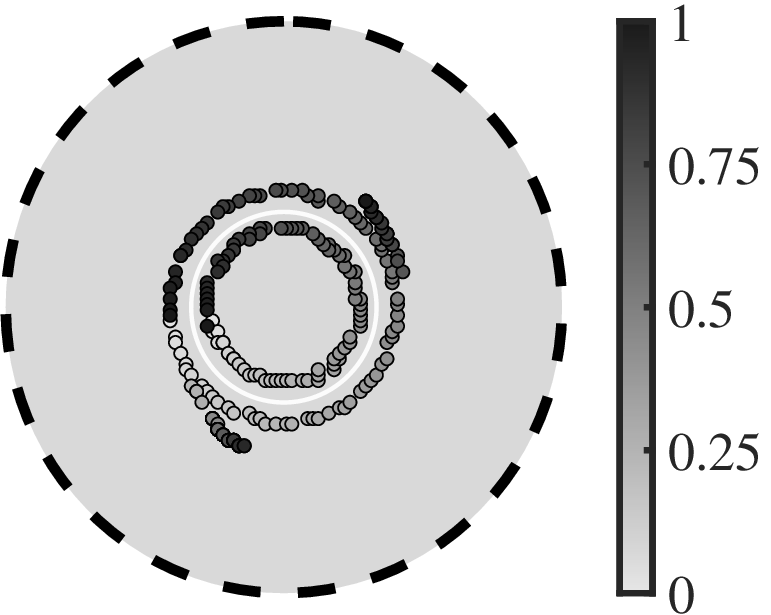}
	}~
%	\hspace{0.2cm}
	\subfloat[ $R_{sd}=3$\label{fig:Bn025_vc_B}]{
		\includegraphics[trim=0cm 0cm 0cm 0cm, clip=true,width=.315\textwidth]{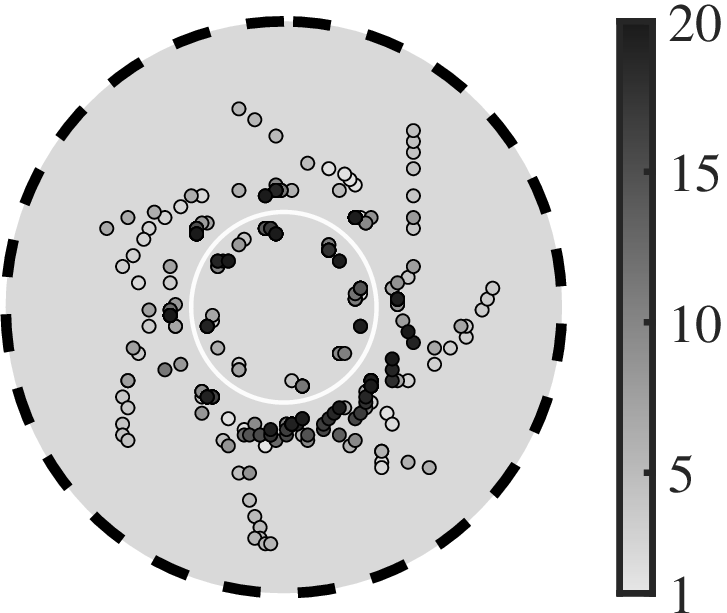}
	}~
%	\hspace{0.2cm}
	\subfloat[$R_{sd}=10$\label{fig:Bn025_vc_C}]{
		\includegraphics[trim=0cm 0cm 0cm 0cm, clip=true,width=.315\textwidth]{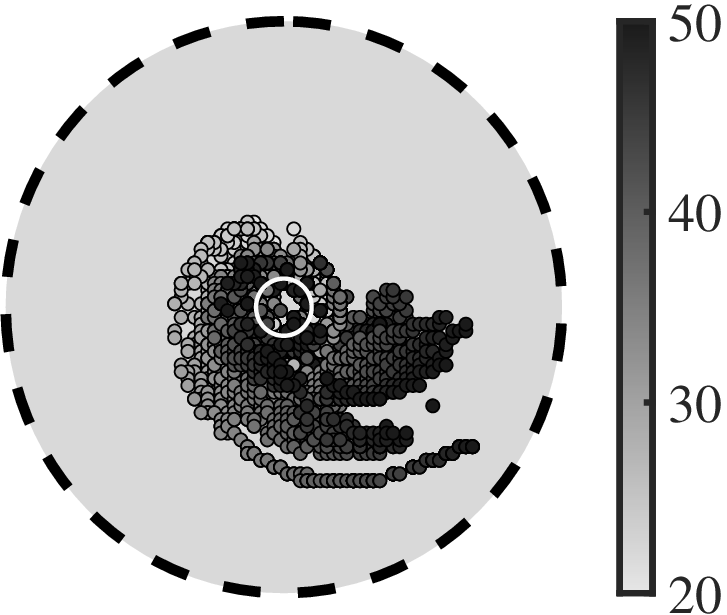}
	} 
	
	\caption{Time evolution of the vortex centers in a VPF with a low yield stress, $Bn = 0.025$, $Re = 50$, over different time intervals, as indicated by the colorbar. Lighter shades correspond to earlier times within each subfigure. The white solid line represents the stirrer's path, while the black dashed lines denote the subdomain boundary. The radius of the field of view is provided in the caption.}
	\label{fig:vorticesCenters_Bn=0.025}
\end{figure}

The effect of yield stress on vortex dynamics is further demonstrated in Figure \ref{fig:vorticesCenters_Bn=0.025} which presents the time evolution of the approximate vortex centers over the same intervals as shown in Figure \ref{fig:vorticesCenters_Newtonian}. The similar development of vortices in the central region is evident in Figures \ref{fig:Bn025_vc_A}\&\ref{fig:Bn025_vc_B}. However, a comparison of Figures \ref{fig:locationVortexNewtonianC}\&\ref{fig:Bn025_vc_C}, reveals that the azimuthal movement of the vortex centers is more pronounced when  $Bn>0$.

\begin{figure}%{r}{0.4\textwidth}
			%	\vspace{-15mm}
			\centering
			\subfloat[$\newTime=1$ \label{fig:Bn4_c_A}]{
				\includegraphics[trim=2cm 0.5cm 5.5cm 1.25cm, clip=true,height=\subfigthree\textwidth]{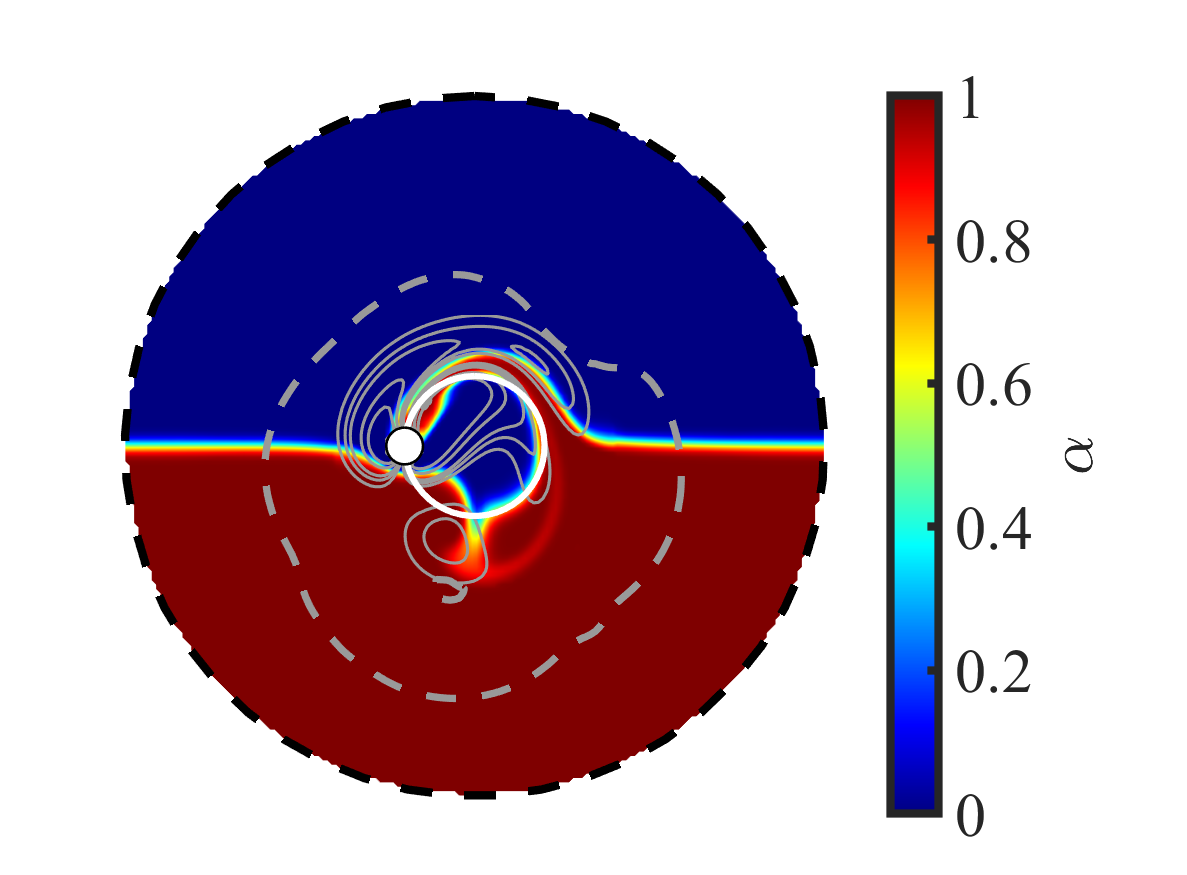}
			}~
			\hspace{0.2cm}
			\subfloat[$\newTime=3$\label{fig:Bn4_c_B}]{
				\includegraphics[trim=2cm 0.5cm 5.5cm 1.25cm, clip=true,height=\subfigthree\textwidth]{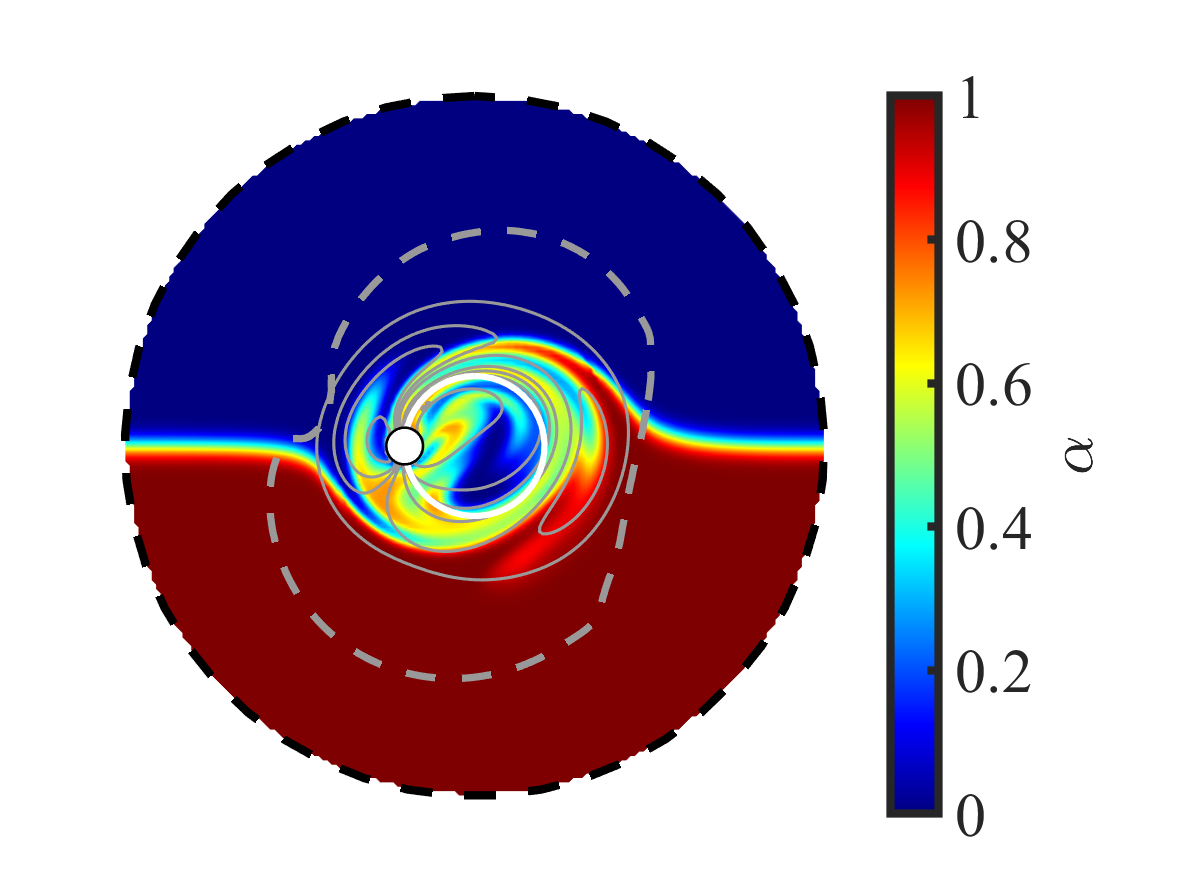}
			}~
			\hspace{0.2cm}
			\subfloat[$\newTime=10$\label{fig:Bn4_c_C}]{
				\includegraphics[trim=2cm 0.5cm 1.5cm 1.25cm, clip=true,height=\subfigthree\textwidth]{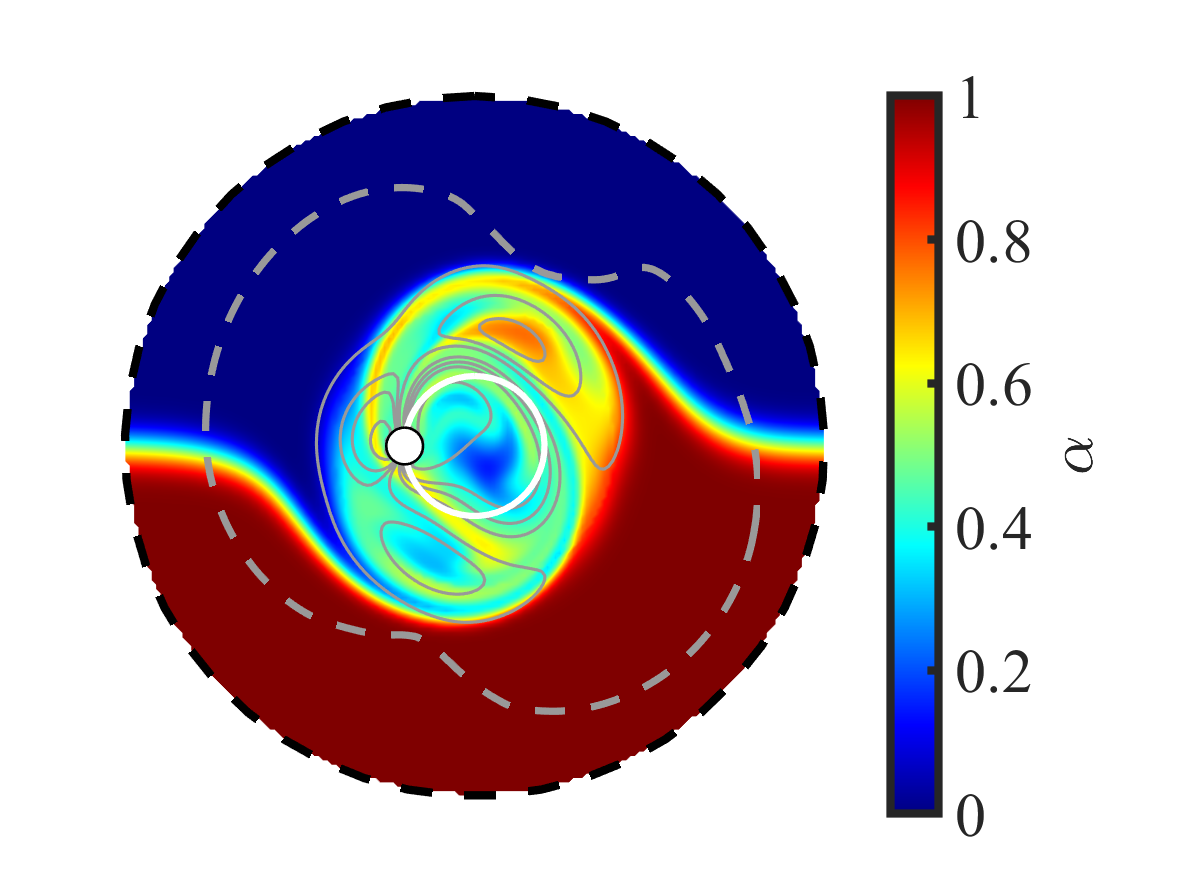}
			}\\
			\vspace{-0.2cm}
			\subfloat[$\newTime=20$ \label{fig:Bn4_c_D}]{
				\includegraphics[trim=2cm 0.5cm 5.5cm 1.25cm, clip=true,height=\subfigthree\textwidth]{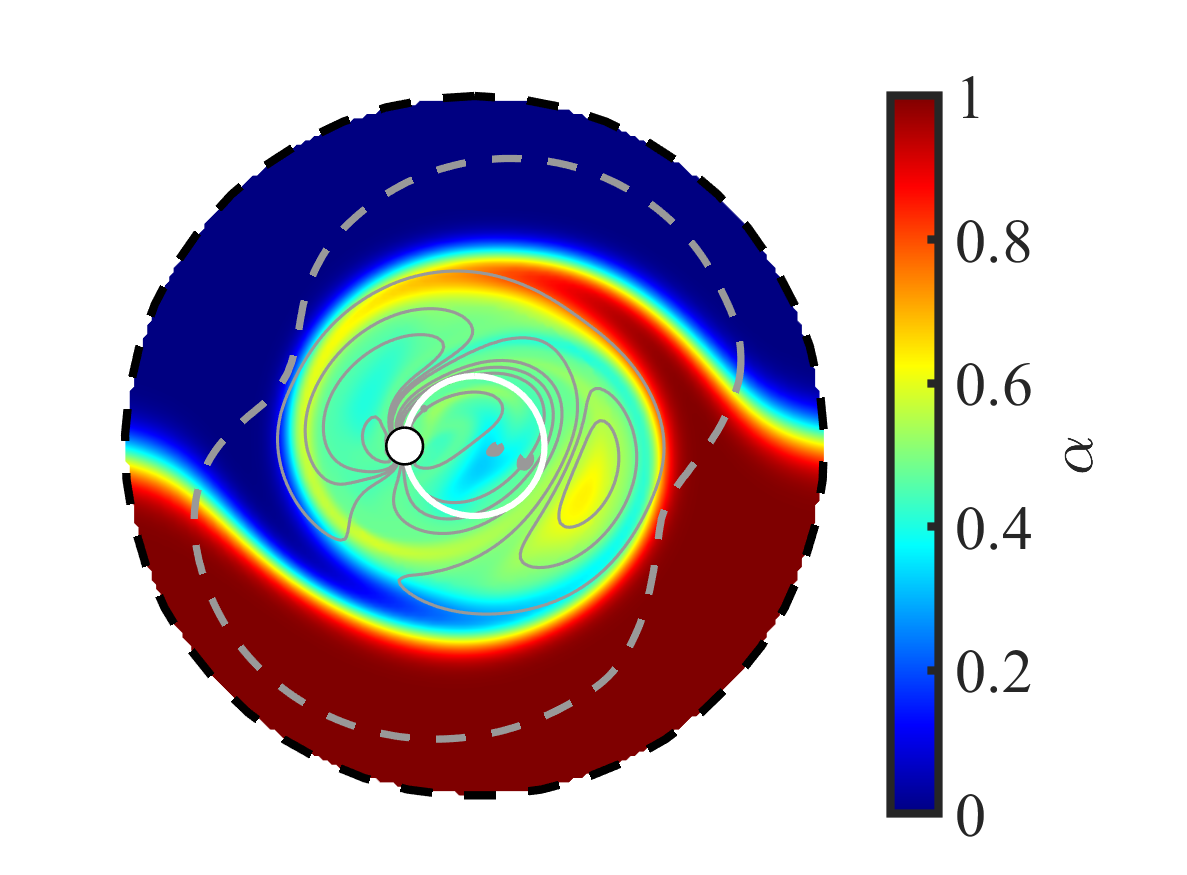}
			}~
			\hspace{0.2cm}
			\subfloat[$\newTime=30$\label{fig:Bn4_c_E}]{
				\includegraphics[trim=2cm 0.5cm 5.5cm 1.25cm, clip=true,height=\subfigthree\textwidth]{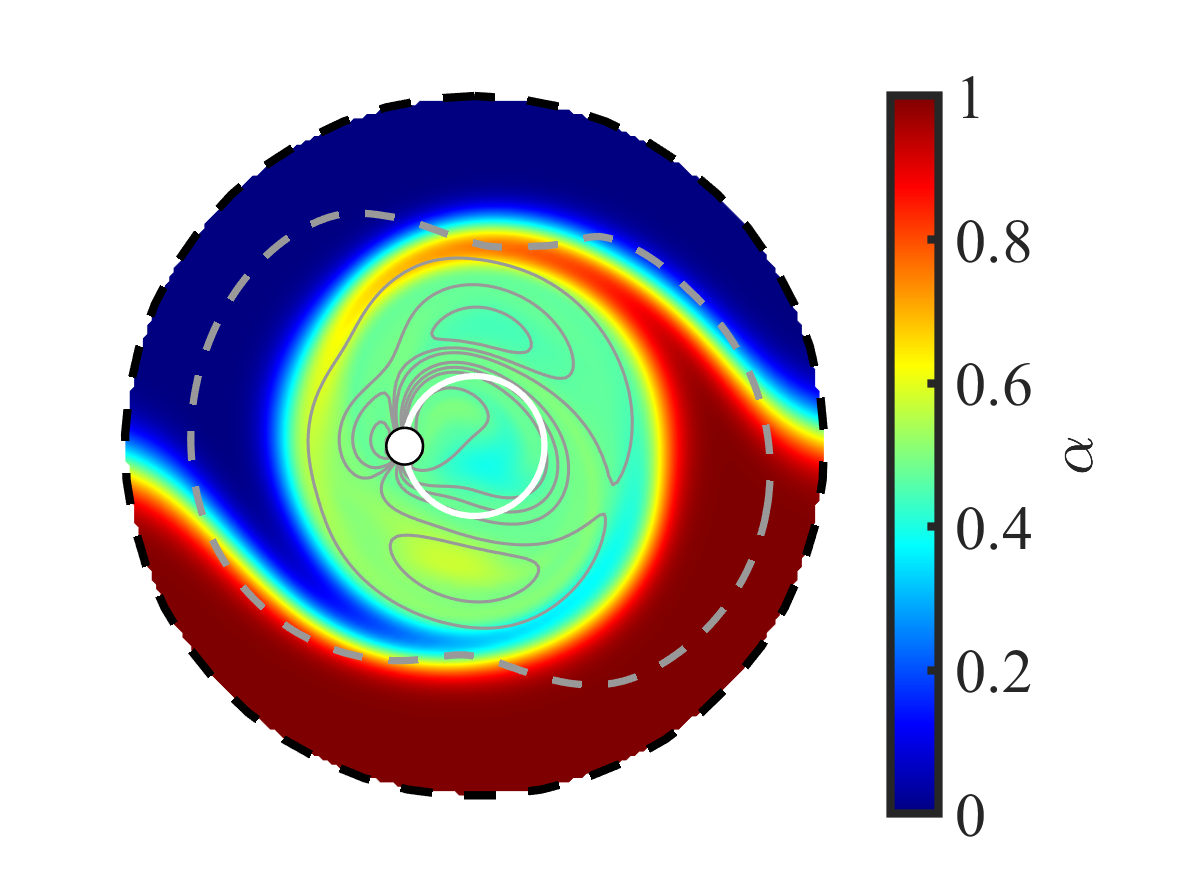}
			}~
			\hspace{0.2cm}
			\subfloat[$\newTime=50$\label{fig:Bn4_c_F}]{
				\includegraphics[trim=2cm 0.5cm 1.5cm 1.25cm, clip=true,height=\subfigthree\textwidth]{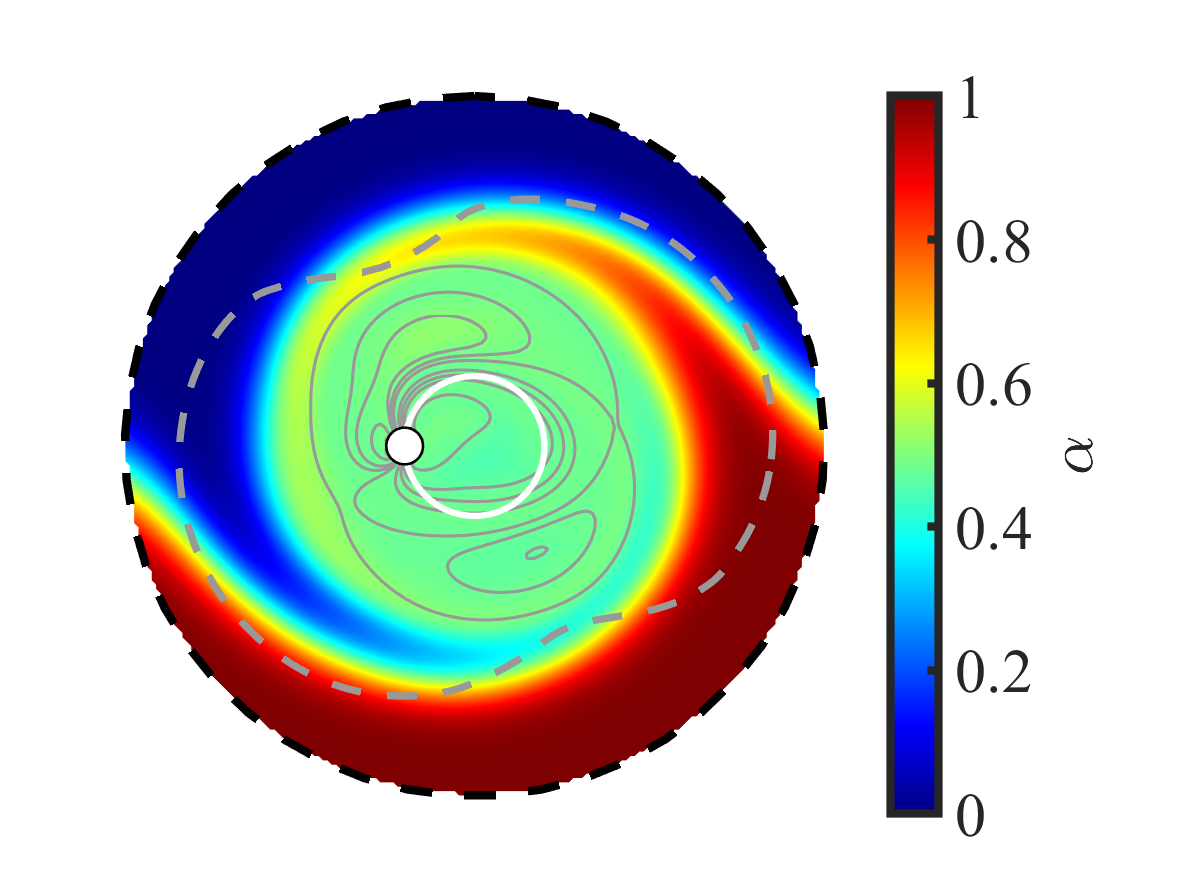}
			}			
			\caption{(a–f) Snapshots of the dye concentration field in a VPF with a moderate yield stress, $Bn = 0.4$, $Re = 50$, and $R_{sd} = 5$. The white circle indicates the stirrer’s path, grey lines represent streamlines, and dashed grey lines show contours of $\tau = 1.05Bn$.}
		\label{fig:Bn4_c}
		\end{figure}

Figure \ref{fig:Bn4_c} illustrates the evolution of dye concentration at a moderate yield stress, $Bn = 0.4$. The approximate size of the yielded region is significantly smaller compared to the case of $Bn = 0.025$ (see dashed grey lines), and mixing remains relatively confined to the central region. The initial two stages of mixing are qualitatively similar to the previous cases: stretching and folding of the interface are followed by diffusion-dominated mixing, leading to the formation of a well-mixed region. However, in this case, the well-mixed region remains largely quiescent. The deformation of the dye interface does not extend beyond the central region.

\begin{figure}
		\centering
		\subfloat[$\newTime=1$ \label{fig:Bn4_v_A}]{
			\includegraphics[trim=2cm 0.5cm 5.5cm 1.25cm, clip=true,height=\subfigthree\textwidth]{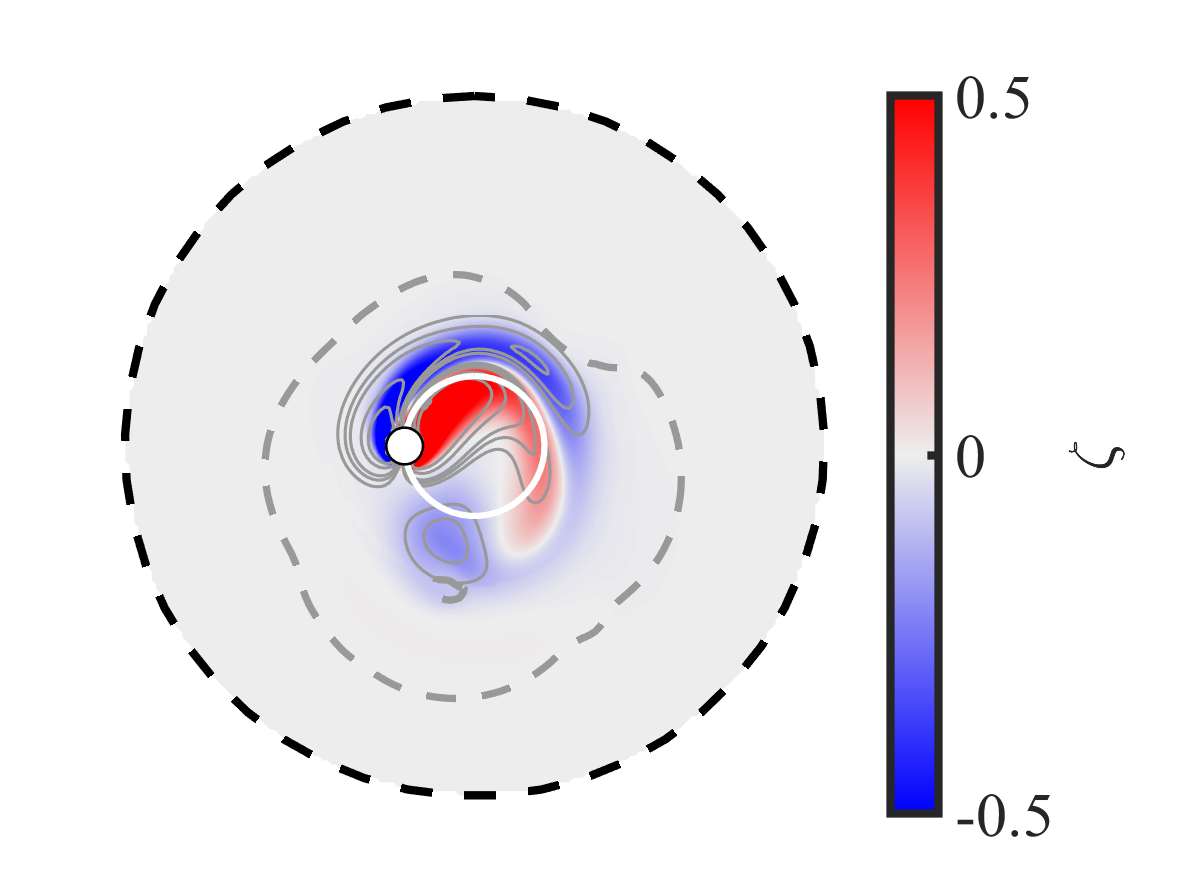}
		}~
		\hspace{0.2cm}
		\subfloat[$\newTime=3$\label{fig:Bn4_v_B}]{
			\includegraphics[trim=2cm 0.5cm 5.5cm 1.25cm, clip=true,height=\subfigthree\textwidth]{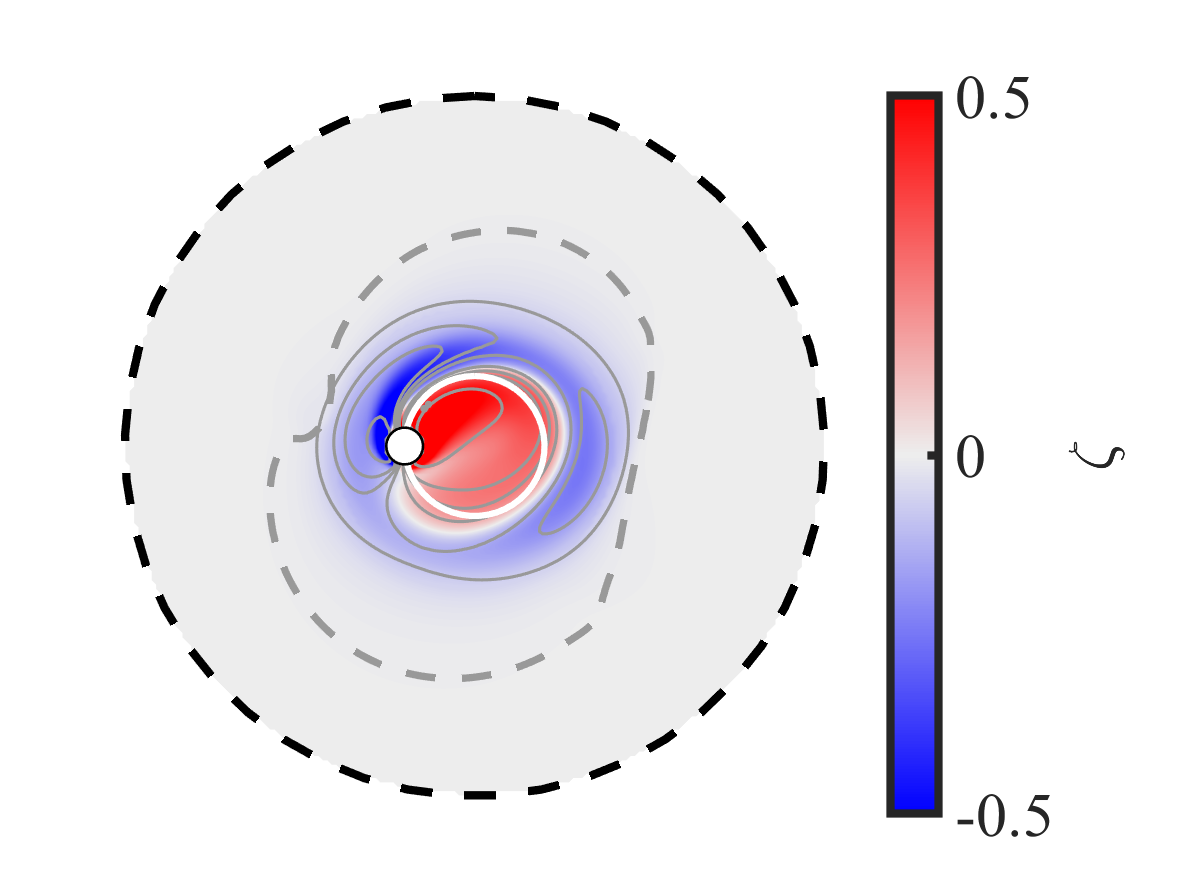}
		}~
		\hspace{0.2cm}
		\subfloat[$\newTime=10$\label{fig:Bn4_v_C}]{
			\includegraphics[trim=2cm 0.5cm 1cm 1.25cm, clip=true,height=\subfigthree\textwidth]{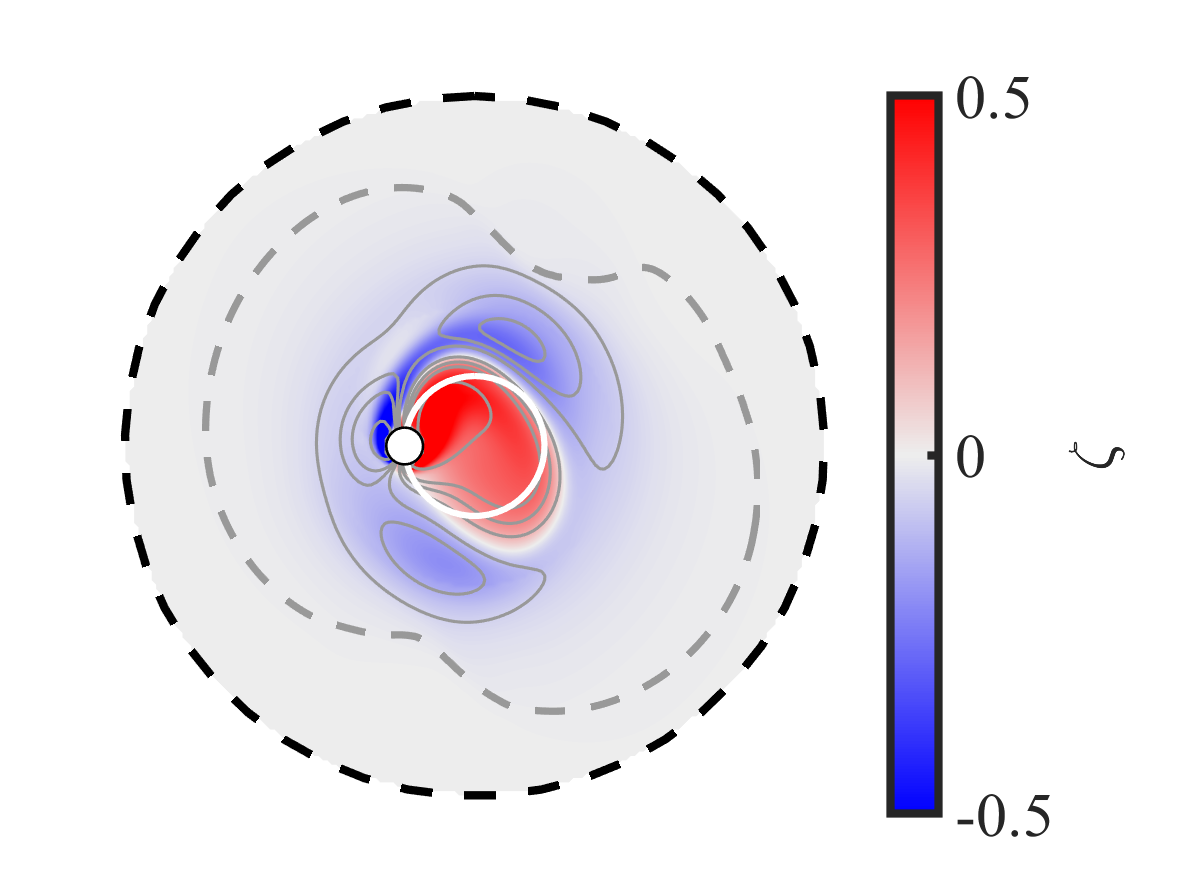}
		}\\
		\vspace{-0.2cm}
		\subfloat[$\newTime=20$ \label{fig:Bn4_v_D}]{
			\includegraphics[trim=2cm 0.5cm 5.5cm 1.25cm, clip=true,height=\subfigthree\textwidth]{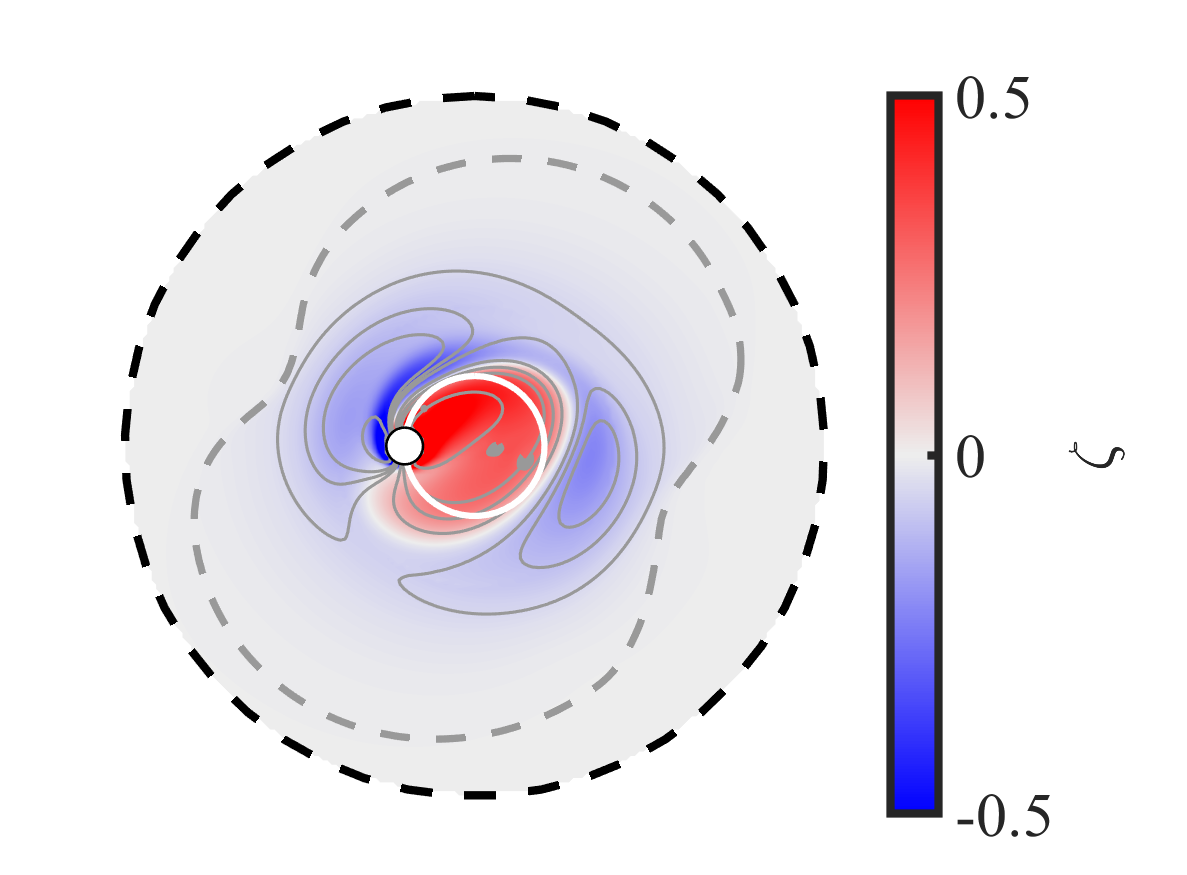}
		}~
		\hspace{0.2cm}
		\subfloat[$\newTime=30$\label{fig:Bn4_v_E}]{
			\includegraphics[trim=2cm 0.5cm 5.5cm 1.25cm, clip=true,height=\subfigthree\textwidth]{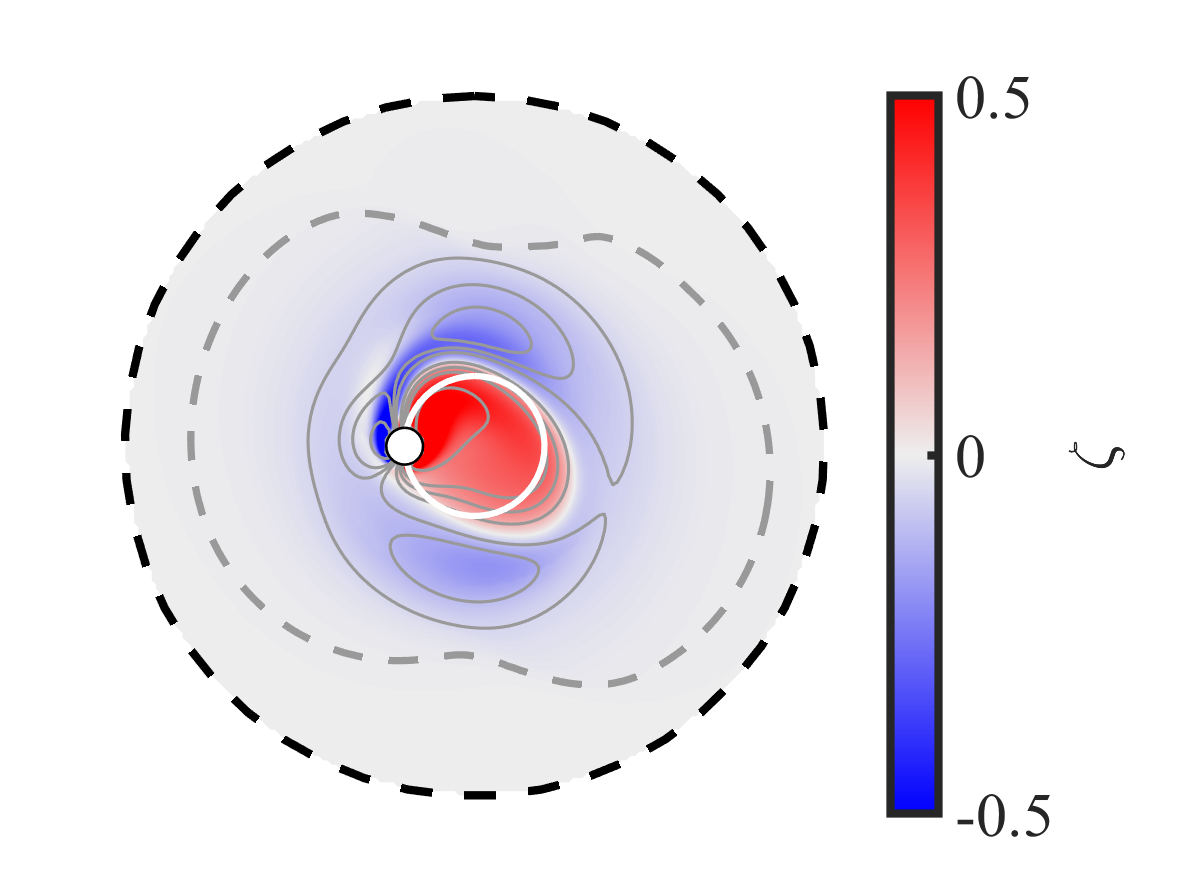}
		}~
		\hspace{0.2cm}
		\subfloat[$\newTime=50$\label{fig:Bn4_v_F}]{
			\includegraphics[trim=2cm 0.5cm 1cm 1.25cm, clip=true,height=\subfigthree\textwidth]{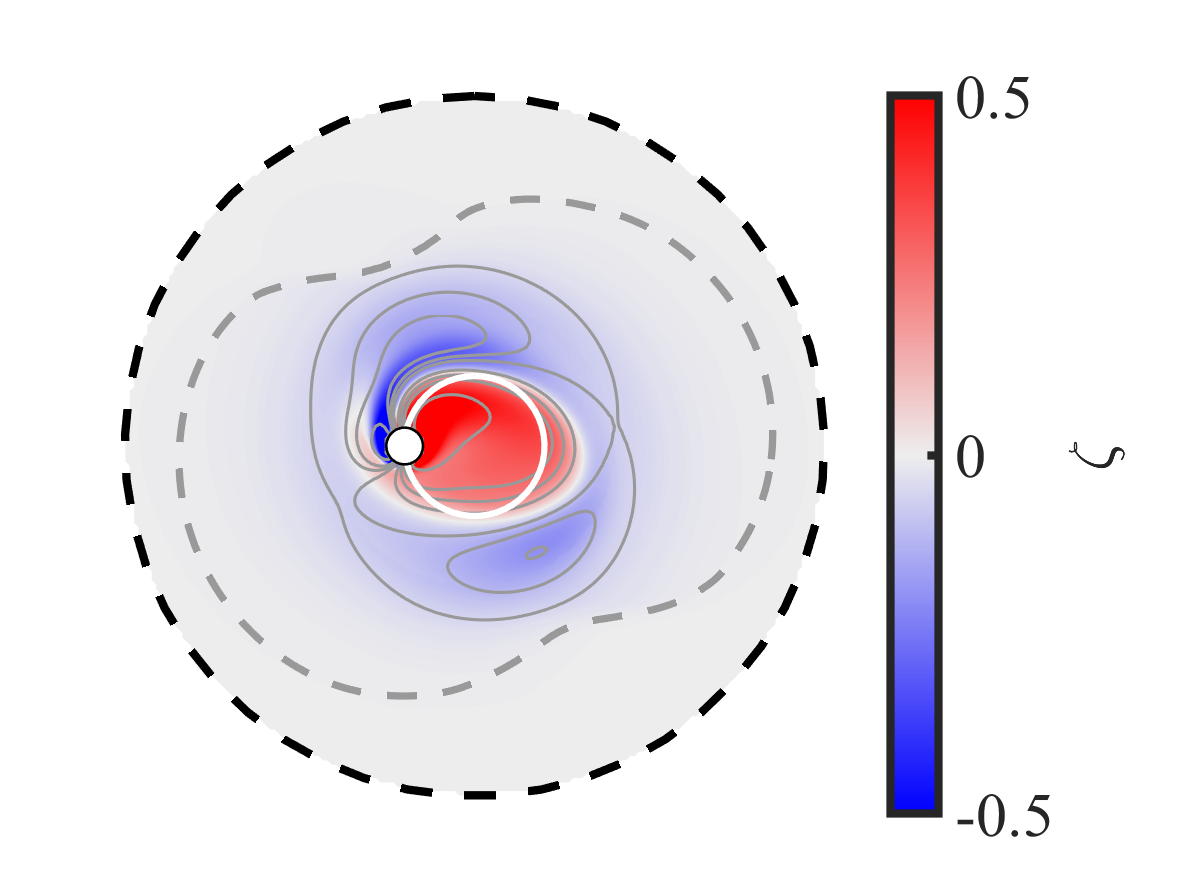}
		}	
		\caption{(a–f) Snapshots of the vorticity field in a VPF with moderate yield stress value. $Bn = 0.4$, $Re = 50$, $R_{sd}=5$. The white circle indicates the stirrer's path, while the grey lines depict the streamlines. The dashed grey lines display the contours of $\tau=1.05Bn$.}
		\label{fig:Bn4_v}
	\end{figure}

\begin{figure}
	\centering
	\subfloat[\label{fig:Bn4_vc_A}]{
		\includegraphics[trim=0cm 0cm 0cm 0cm, clip=true,width=.33\textwidth]{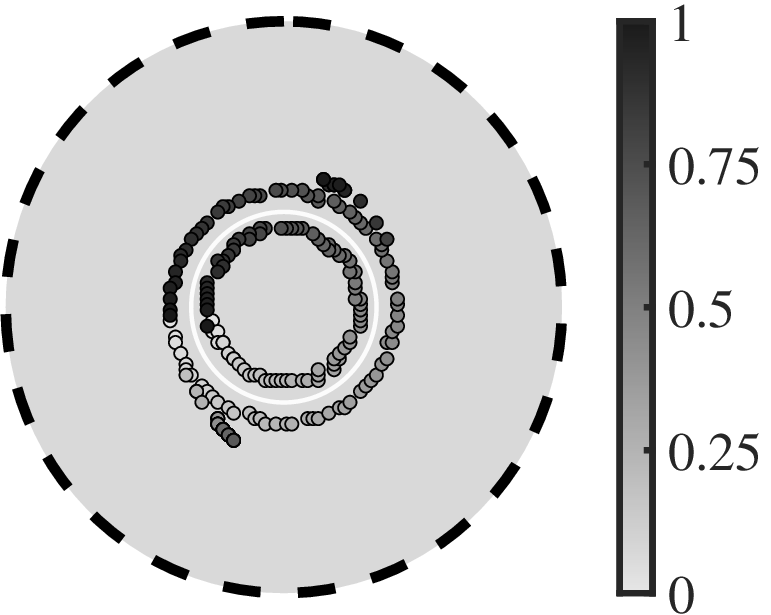}
	}~
%	\hspace{0.2cm}
	\subfloat[\label{fig:Bn4_vc_B}]{
		\includegraphics[trim=0cm 0cm 0cm 0cm, clip=true,width=.315\textwidth]{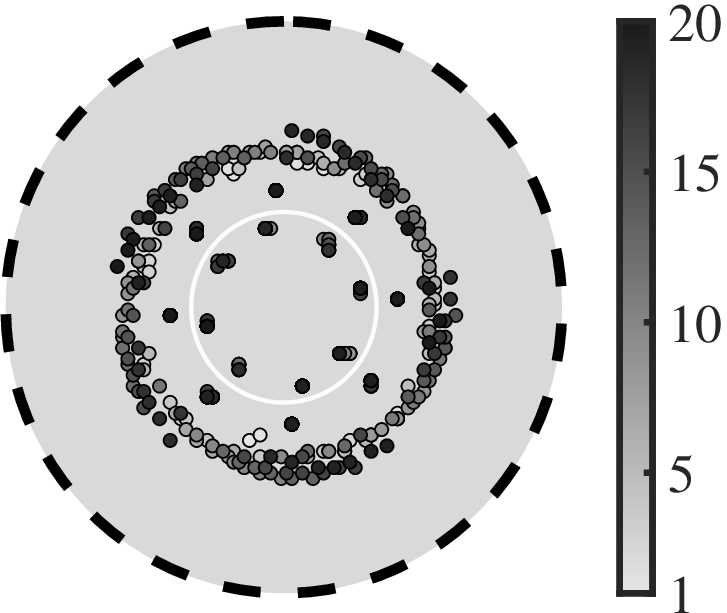}
	}~
%	\hspace{0.2cm}
	\subfloat[\label{fig:Bn4_vc_C}]{
		\includegraphics[trim=0cm 0cm 0cm 0cm, clip=true,width=.315\textwidth]{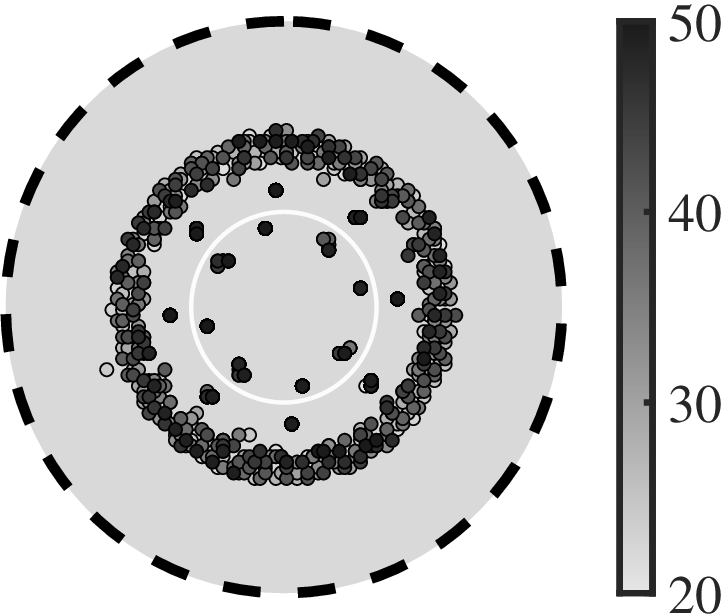}
	} 
	
	\caption{The time evolution of the vortex centers in the VPF with a moderate yield stress. $Bn = 0.4$, $Re = 50$ and $R_{sd}=3$, over different time intervals, as indicated by the colorbar. Lighter shades correspond to earlier times within each subfigure. The white solid line represents the stirrer's path, while the black dashed lines denote the subdomain boundary.}
	\label{fig:Bn4_vc}
\end{figure}

The evolution of the vorticity field at $Bn = 0.4$ is shown in Figure \ref{fig:Bn4_v}. The initial shedding pattern appears similar to the Newtonian case (Figure \ref{fig:Bn4_v_A}). However, the shed CW vortices stay closer to the stirrer and the CCW eddy remains mostly confined within the stirrer's path because the yield stress suppresses the advection of shed vortices and growth of attached eddies. Consequently, the size of the well-mixed region is reduced compared to the previous cases. This localization is closely connected to the development of vortices. Figure \ref{fig:Bn4_vc} illustrates the time evolution of the approximate vortex centers.  The formation of the attached eddies and vortex shedding during the first period is similar to the previous cases (Figure \ref{fig:Bn4_vc_A}). However, the shed vortices do not travel as far from the stirrer (Figure \ref{fig:Bn4_vc_B}). Instead, they remain trapped in the vicinity of the stirrer's path (Figure \ref{fig:Bn4_vc_C}). As a result, mixing remains confined to the central area throughout the process (Figure \ref{fig:Bn4_c_F}). This behavior represents the second mechanism of mixing localization due to yield stress: as shed vortices remain trapped near the stirrer, the stretching and folding of the interface are confined to this region, limiting the mixing rate.

\begin{figure}
	\centering
		\vspace{-0.4cm}
		\subfloat[$\newTime=1$\label{fig:concentration_Re=100_Bn=1A}]{
			\includegraphics[trim=2cm 0.5cm 5.15cm 1.25cm, clip=true,height=\subfigfour\textwidth]{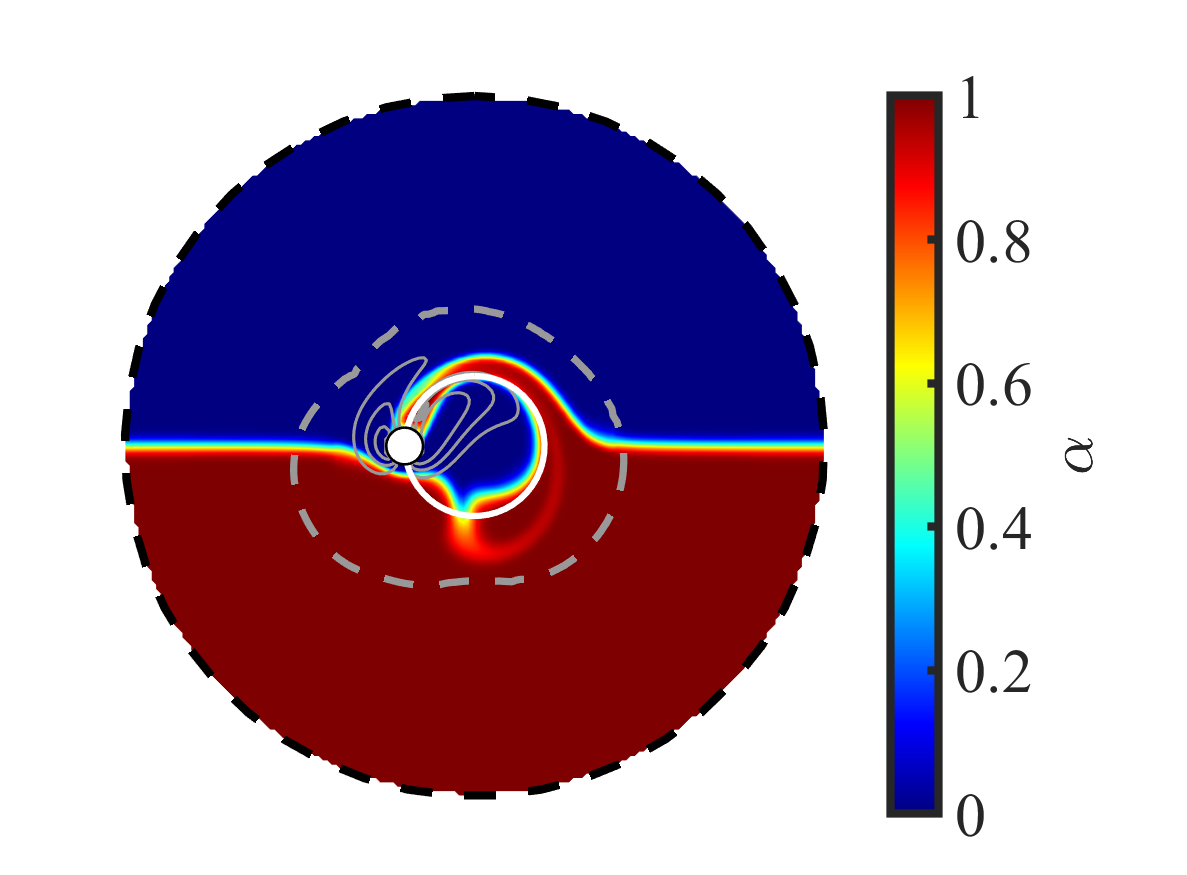}
		}
		\hspace{-0.25cm}
		\subfloat[$\newTime=15$ \label{fig:concentration_Re=100_Bn=1B}]{
			\includegraphics[trim=2cm 0.5cm 5.15cm 1.25cm, clip=true,height=\subfigfour\textwidth]{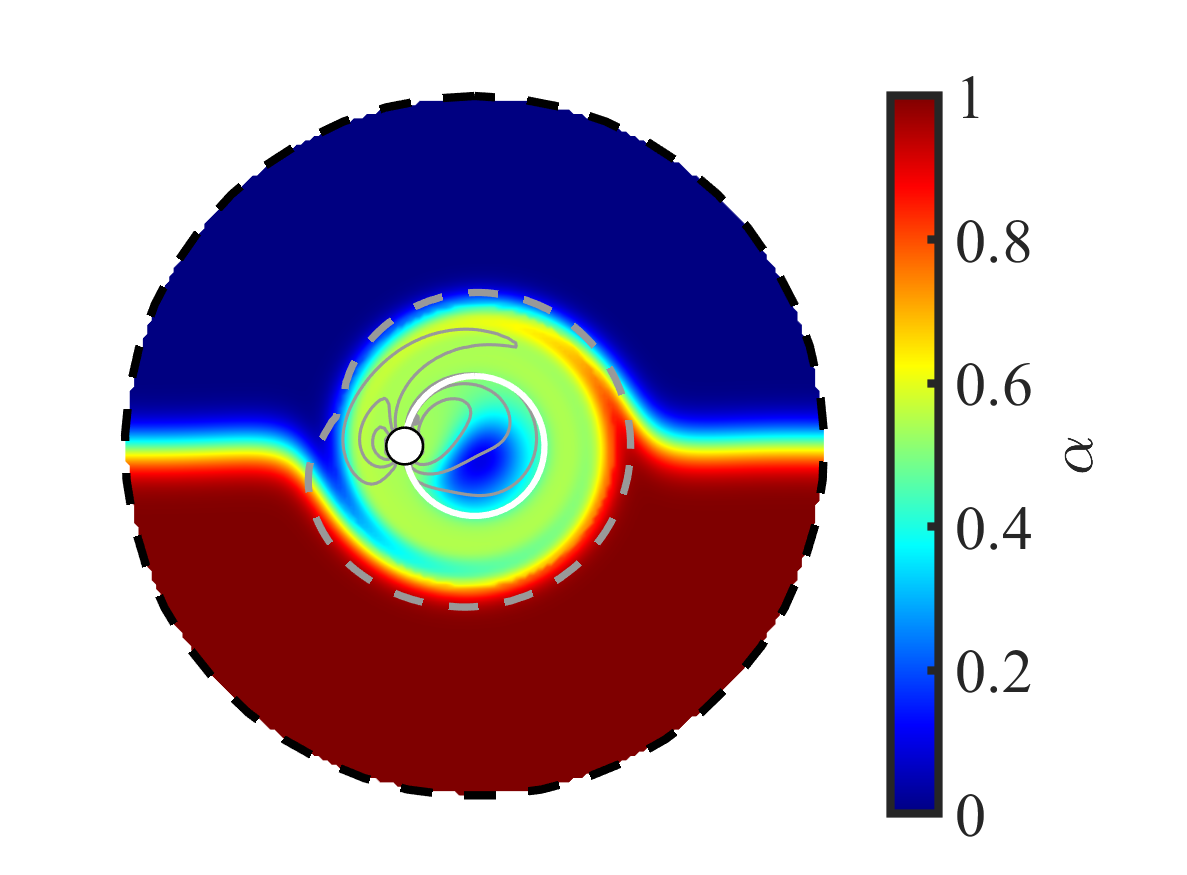}
		}
		\hspace{-0.25cm}
		\subfloat[$\newTime=30$\label{fig:concentration_Re=100_Bn=1C}]{
			\includegraphics[trim=2cm 0.5cm 5.15cm 1.25cm, clip=true,height=\subfigfour\textwidth]{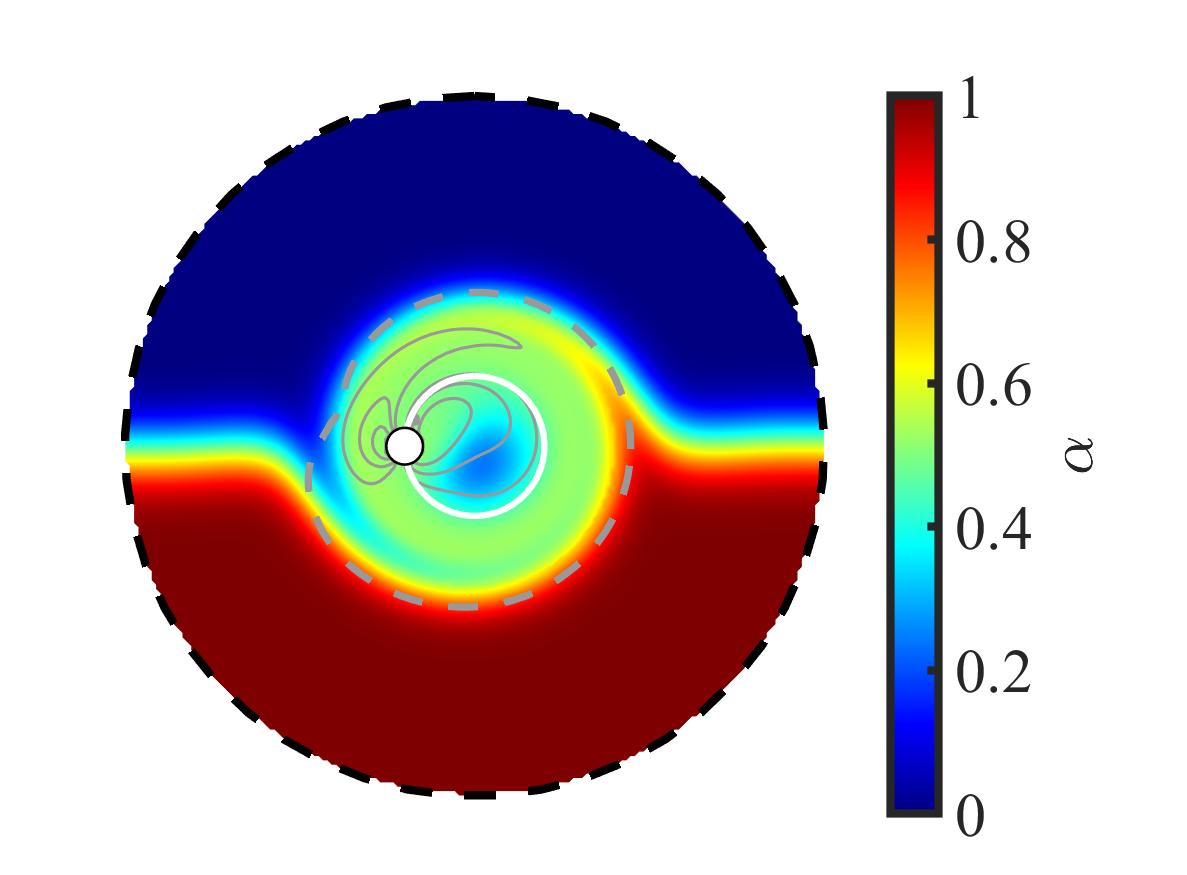}
		}
		\hspace{-0.25cm}
		\subfloat[$\newTime=50$\label{fig:concentration_Re=100_Bn=1D}]{
			\includegraphics[trim=2cm 0.5cm 1cm 1.25cm, clip=true,height=\subfigfour\textwidth]{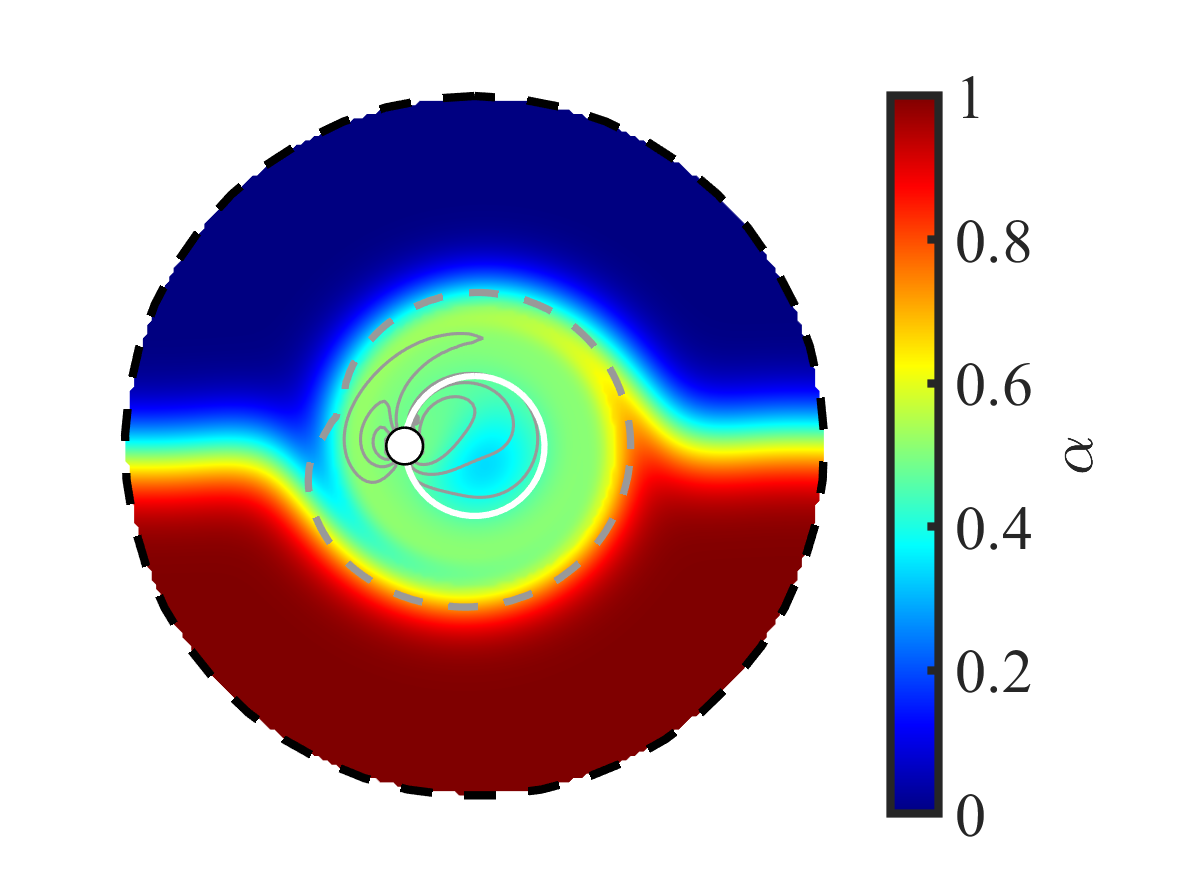}
		}\\
		\vspace{-0.4cm}
		\subfloat[$\newTime=1$ \label{fig:vorticity_Bn=1A}]{
			\includegraphics[trim=2cm 0.5cm 5.15cm 1.25cm, clip=true,height=\subfigfour\textwidth]{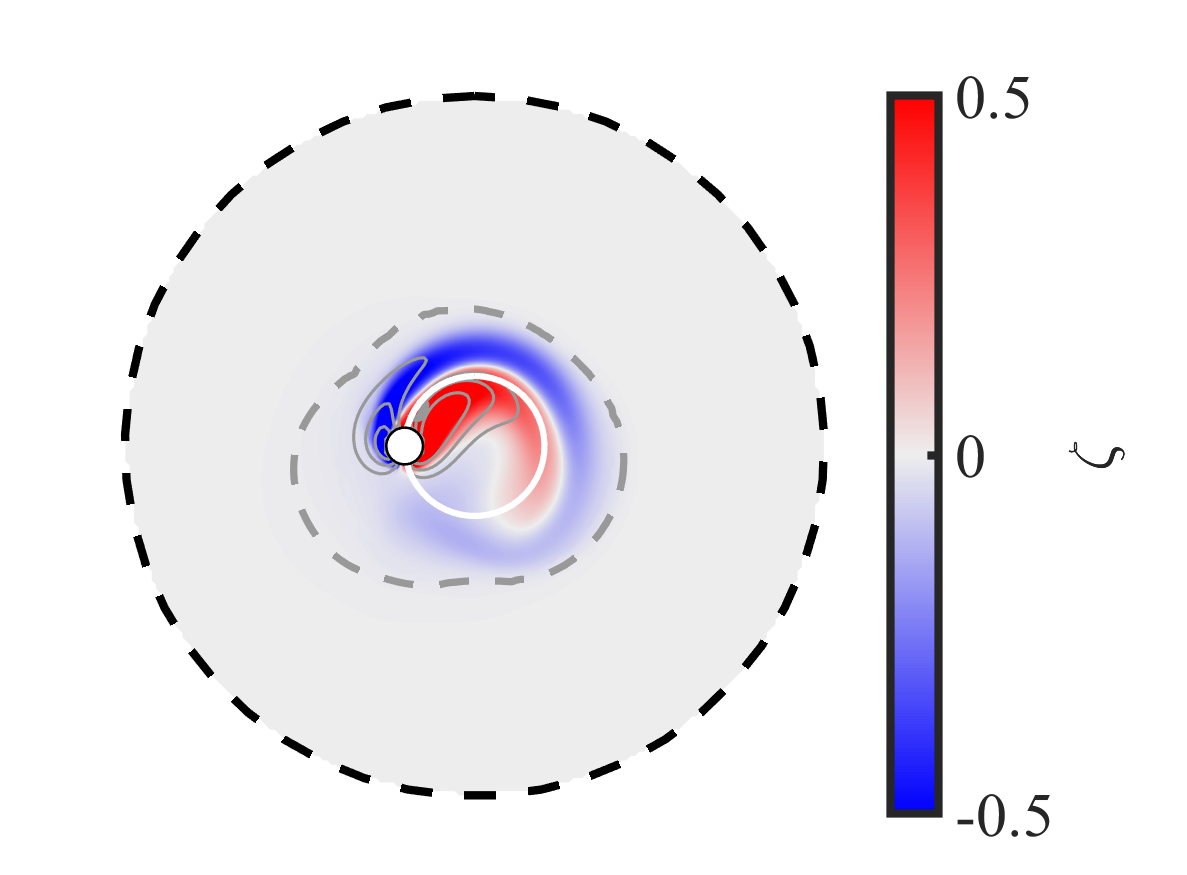}
		}~
		\hspace{-0.25cm}
		\subfloat[$\newTime=15$ \label{fig:vorticity_Bn=1B}]{
			\includegraphics[trim=2cm 0.5cm 5.15cm 1.25cm, clip=true,height=\subfigfour\textwidth]{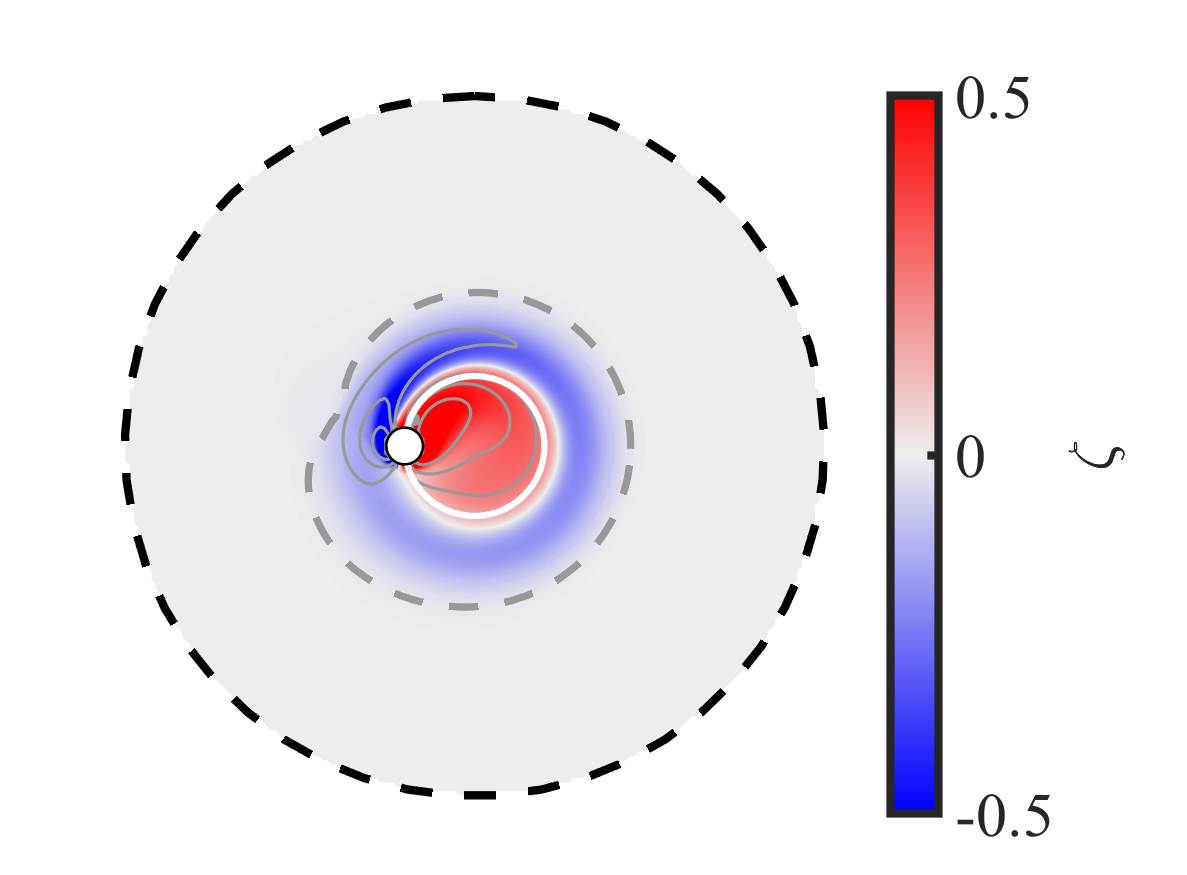}
		}~
		\hspace{-0.25cm}
		\subfloat[$\newTime=30$\label{fig:vorticity_Bn=1C}]{
			\includegraphics[trim=2cm 0.5cm 5.15cm 1.25cm, clip=true,height=\subfigfour\textwidth]{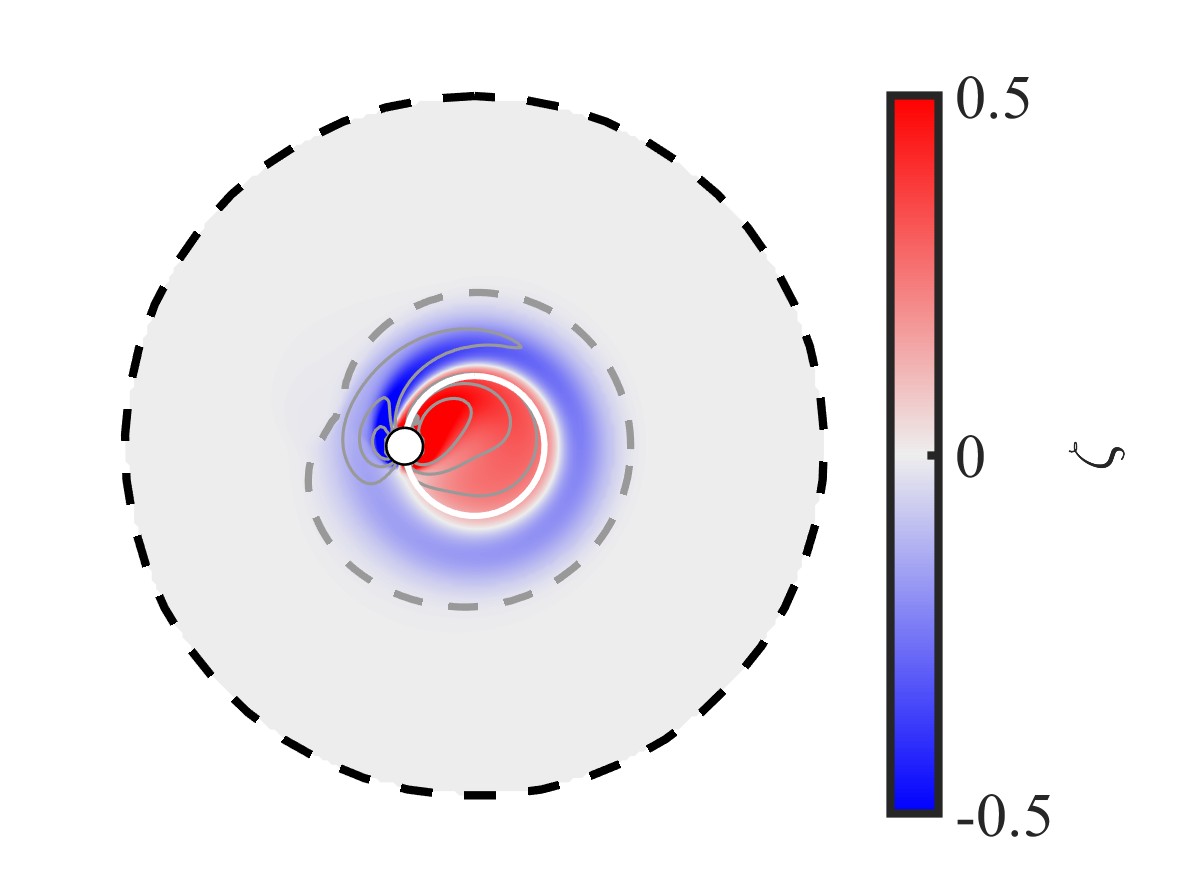}
		}~
		\hspace{-0.25cm}
		\subfloat[$\newTime=50$\label{fig:vorticity_Bn=1D}]{
			\includegraphics[trim=2cm 0.5cm 1cm 1.25cm, clip=true,height=\subfigfour\textwidth]{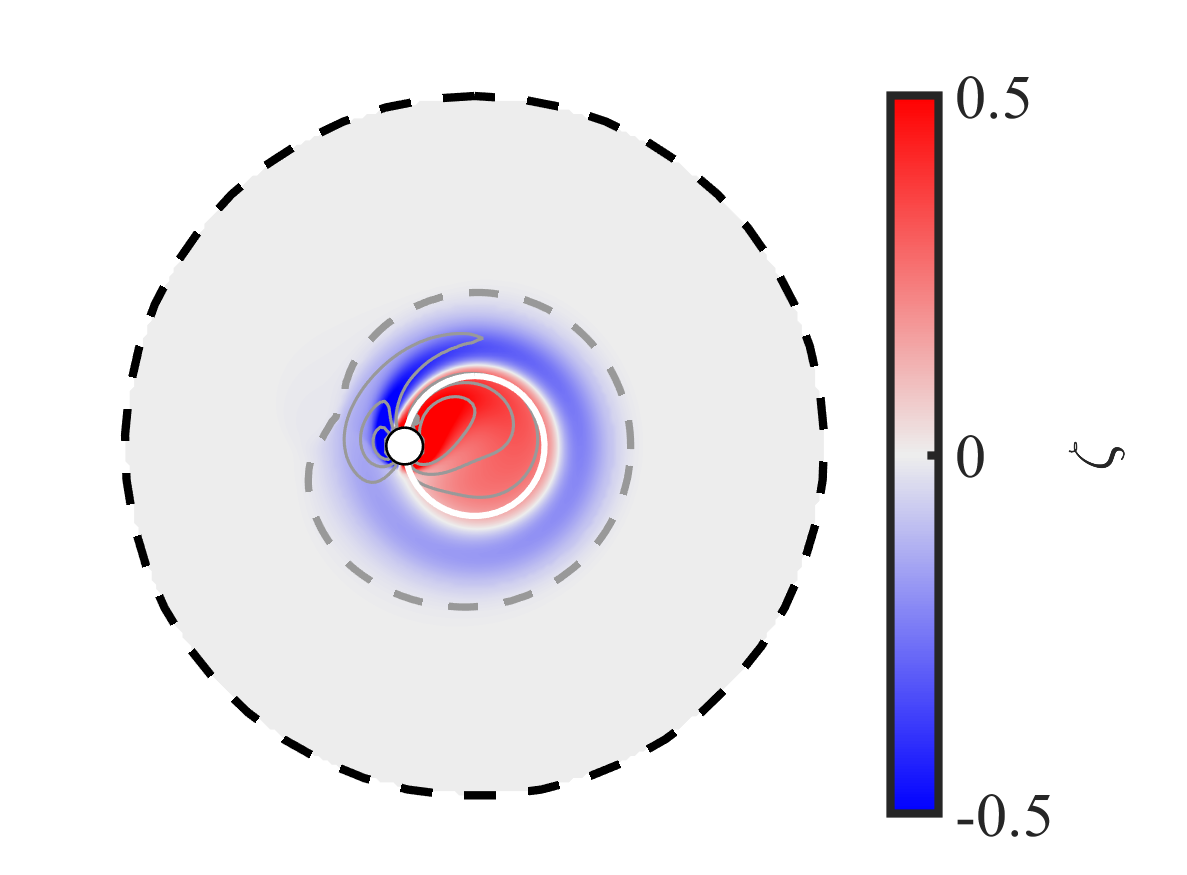}
		}
		\hspace{-0.25cm}		
	
		\caption{Snapshots of the dye concentration (a-d)  and vorticity (e-h)  fields field in a VPF with a high yield stress, $Bn=1$, $Re=50$ and $R_{sd}=5$. The white circle indicates the stirrer’s path, grey lines represent streamlines, and dashed grey lines show contours of $\tau = 1.05Bn$.}
		\label{fig:Bn10_cw}
	\end{figure}

\begin{figure}
	\centering
		\includegraphics[trim=0cm 0cm 0cm 0cm, clip=true,width=.33\textwidth]{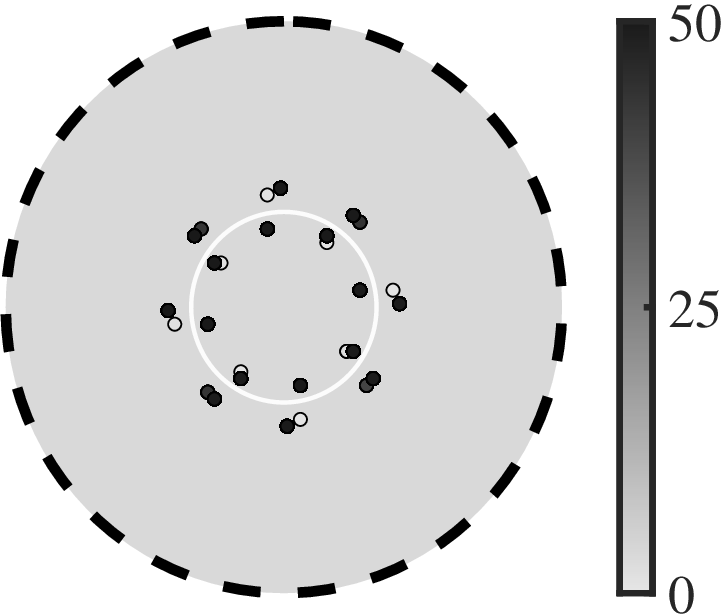}
	\caption{The time evolution of the vortex centers in the VPF with a high yield stress, $Bn=1$, $Re=50$, and $R_{sd}=3$. Lighter shades correspond to earlier times. The white solid line represents the stirrer's path, while the black dashed line denotes the subdomain boundary.}
	\label{fig:Bn10_vc}
\end{figure}

As expected, at higher yield stress values, mixing becomes increasingly confined to the central region. An illustrative case for $Bn=1$ is shown in Figure \ref{fig:Bn10_cw}, where the first and second rows display the evolution of dye concentration and vorticity fields, respectively. The dye concentration rapidly assumes an approximately axisymmetric distribution, after which mixing is primarily governed by radial diffusion. The two eddies remain attached to the stirrer, with no vortex shedding observed. This behavior is further illustrated in Figure \ref{fig:Bn10_vc}, showing the time evolution of the approximate vortex centers. This represents the third mode of localization: the complete suppression of vortex shedding, where interface deformation remains largely confined to the central region.  %{\color{red} explain somewhere we are not talking about unyielded regions and how they grow w yield stress}

%		\begin{figure}
%	\centering
%	\subfloat{
%		\includegraphics[trim=0cm 0cm 0cm 0cm, clip=true,width=.28\textwidth]{/Illustrative/locationVortexRe=50_Bn=1.eps}
%	}
%	\caption{\color{blue}The time evolution of the vortex centers in a viscoplastic fluid with a high yield stress. The lighter shades represent early times. The white circle and black dashed lines represent the stirrer's path and subdomain boundary, respectively. $Re=50, Bn=1$.}
%	\label{fig:vorticesCenters_Bn=1}
%\end{figure}

\begin{figure}
		\centering
%		\hspace{-0.25cm}
%		\subfloat[\label{fig:}]{
			\includegraphics[trim=0cm 0cm 0cm 0cm, clip=true,height=.32\textwidth]{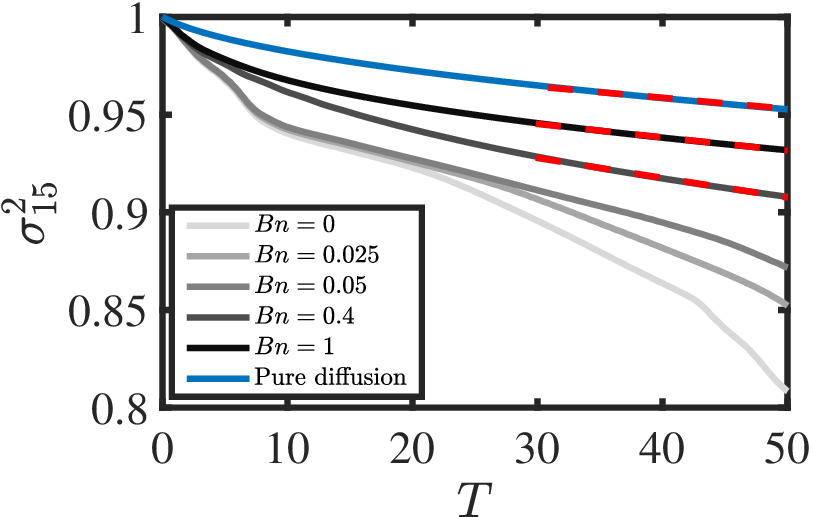}
%		}~
%		\hspace{-0.25cm}
%		\subfloat[\label{fig:}]{
%			\includegraphics[trim=0cm 0cm 0cm 0cm, clip=true,height=.3\textwidth]{/effectOfYieldStress/variance_Re=50_zoomed.eps}
%		}	
		\caption{Evolution of the normalized dye concentration variance for $0 \leq Bn \leq 1$. Darker shades of grey correspond to higher $Bn$ values. The solid blue line represents pure diffusion. The dashed lines depict the exponential fits for the diffusion-dominated stage. $Re = 50$.}
		\label{fig:variance_Re=50}
	\end{figure}

To compare mixing rates across different regimes, we present the normalized variance of dye concentration, $\sigma_{15}^2$, in Figure \ref{fig:variance_Re=50}. For reference, the solid blue line shows the purely diffusive case. As expected, mixing rates decrease with increasing $Bn$. At low $Bn$ (e.g., $Bn=0.05$), the initial mixing curves closely match the Newtonian case, indicating that at low yield stress, while the flow remains confined to the central region ($\newTime \lesssim 10$), yield stress has a negligible effect on local energy dissipation. However, the influence of yield stress becomes more pronounced in later stages, as the well-mixed region drifts outward (see $\newTime \approx 20$ and $\newTime \approx 30$ for $Bn = 0.025$ and $Bn = 0.05$, respectively). This shift occurs when the dye interface enters a region where energy dissipation and flow dynamics begin to be affected by the yield stress. From this stage onward, the $\sigma_{15}^2$ curves diverge, as the advection of escaped vortices is increasingly suppressed by the yield stress.

At moderate $Bn$, the divergence of $\sigma_{15}^2$ from the Newtonian case becomes apparent earlier. This is because, as $Bn$ increases, the dissipation caused by yield stress starts to influence smaller radii. The downward concavity in the variance is almost entirely suppressed, as vortices remain confined within the central region, preventing them from re-accelerating mixing. The mixing rate continues to decrease with increasing $Bn$, reflecting both the overall deceleration of the flow and the progressive suppression of vortices shed from the stirrer.

As described earlier, the final stage of mixing is diffusion-dominated for moderate and high Bingham numbers. In this range, the standard deviation of dye concentration approaches an exponential decay at long times,
\begin{align*}
	\sigma_{15}^2 \approx \sigma_0^2 \exp\left(-2\lambda \newTime\right)
\end{align*}
where $\lambda$ is the decay constant. The red dashed lines in Figure \ref{fig:variance_Re=50} depict the exponential fits. Following \cite{christov2009enhancement}, we define the enhancement factor $\eta_{\lambda} = \lambda/\lambda_p$, where $\lambda_p$ is the decay constant for the purely diffusive case. The intercept of the exponential fit, $\sigma_0$, can be viewed as an approximation of the standard deviation when the diffusion-dominated phase begins, providing a measure of mixing by the end of the advection-dominated stage. Here, a second enhancement factor is defined as $\eta_\sigma = \sigma_{op}/\sigma_0$, where $\sigma_{op}$ is the intercept of the exponential fit for the purely diffusive case. This enhancement factor, $\eta_\sigma$, enables comparison of the extent of mixing achieved during the initial phase dominated by interface stretching.

	\begin{figure}
		\centering
%		\hspace{-0.25cm}
		\subfloat[\label{fig:etal_Re50}]{
			\includegraphics[trim=0cm 0cm 0cm 0cm, clip=true,height=.27\textwidth]{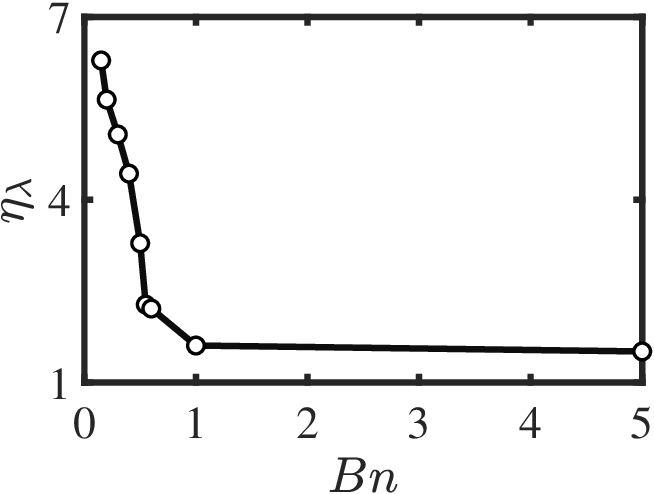}
		}\quad
		\hspace{-0.25cm}
		\subfloat[\label{fig:etas_Re50}]{
			\includegraphics[trim=0cm 0cm 0cm 0cm, clip=true,height=.27\textwidth]{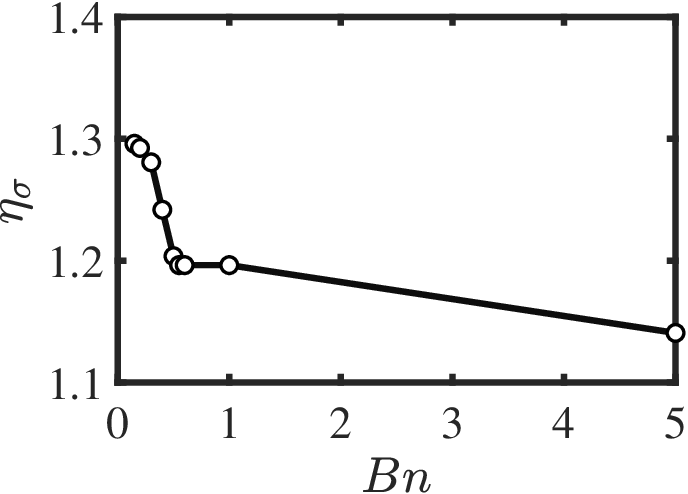}
		}	
		\caption{Variation of the enhancement factors (a) $\eta_\lambda$ and (b) $\eta_\sigma$ with $Bn$ at $Re=50$.}
		\label{fig:eta_Re50}
	\end{figure}

The variation of enhancement factors with $Bn$ is shown in Figure \ref{fig:eta_Re50}. Both enhancement factors decrease rapidly for small $Bn$, though Figure \ref{fig:etal_Re50} illustrates that changes in $\eta_\lambda$ are more pronounced than those in $\eta_\sigma$. This suggests that the long-term impact of yield stress on mixing can be more significant than its short-term effects: increasing the Bingham number by about two orders of magnitude reduces $\eta_\sigma$ by approximately $10\%$, attributed to the suppression of dye interface stretching during the initial mixing stage. Meanwhile, the same increase in $Bn$ results in more than an $80\%$ reduction in $\eta_\lambda$, as diffusion - the dominant mixing mechanism over longer timescales - becomes less effective when the interface area is reduced.
	
%%--------------------------------	
\subsection{Flow \& mixing regimes}
	\label{sec:regimeClassification}
%{\color{red}
%\begin{itemize} 
%	\item regimes SE (shedding, escaped), ST (shedding trapped), E (Eddies)
%	\item $||u||$ for diff $Bn$
%	\item describe the oscillations
%	\item fft 
%	\item evaluated $r_y$ as a measure of the localization of the flow
%	\item change of $r_y$ with Re and Bn in different regimes
%	\item presented approximate critical criteria for transition between these regimes.
%	\item introduced a definition of the effective Re based on the above and illustrated the similarity w Newotonian critical Re
%\end{itemize}
%}		

In the previous section, we identified three distinct mixing regimes:

\textbf{Regime \SE}: At sufficiently low $Bn$, mixing is initially dominated by the stretching and folding of the interface. This is followed by a brief period where mixing is driven primarily by radial diffusion. In this regime, vortices escape the central region, carrying dye and stretching the interface deep into the domain. Once the vortices escape, mixing re-accelerates, driven by advection and further interface stretching.

\textbf{Regime \ST}: At moderate $Bn$, mixing does not re-accelerate after the diffusion-dominated phase. In this regime, the shed vortices remain trapped in the central region, unable to escape. They are periodically influenced by the passage of the stirrer but remain confined, limiting further advection-driven mixing.

\textbf{Regime \E}: At sufficiently high $Bn$, mixing is quickly dominated by diffusion. The yield stress completely suppresses vortex shedding, confining interface stretching to the immediate vicinity of the stirrer.
	
 The mixing regimes described above are closely tied to the flow development and vortex dynamics. To facilitate comparison and distinction of the regimes, we examine the time evolution of kinetic energy (approximated by the velocity norm, $||u||$) and the average radius of the yielded region,  $\bar{r}_y$, 
\begin{equation}
	\label{eq:averagedRadius}
	\bar{r}_y = \frac{1}{2\pi} 
	\int_{0}^{2\pi} r_{y}\ d\theta
\end{equation}

Here, $r_y$ is a point on the estimated boundary of the quiescent yielded region.

\begin{figure}
		\centering
		\subfloat[\label{fig:yieldedTime_Re=50}]{
			\includegraphics[trim=0cm 0cm 0cm 0cm, clip=true,height=.27\textwidth]{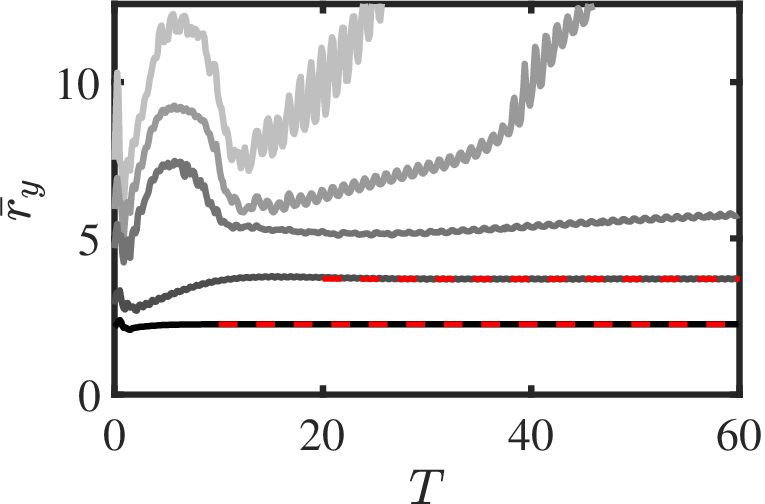}
		}
		\subfloat[\label{fig:unormRe50}]{
			\includegraphics[trim=0cm 0cm 0cm 0cm, clip=true,height=.27\textwidth]{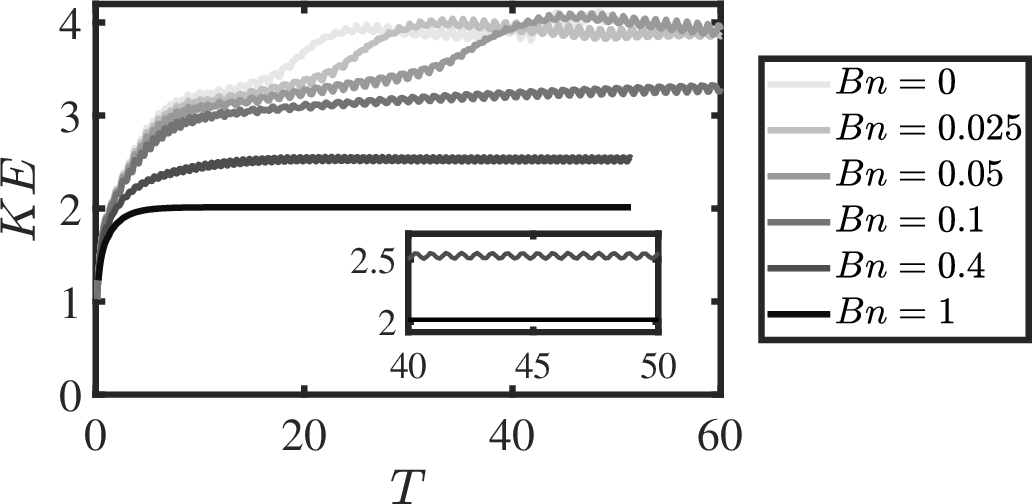}
		}
	\caption{(a) Time evolution of averaged radius of the yielded region,$\overline{r}_y$, at different yield stress values. Darker shades of grey represent higher $Bn$. The red dashed lines represents steady value of $\overline{r}_y$,  $\overline{r}_{ys}$, at moderate and high yield stresses. (b) Time evolution of velocity norm at different yield stress values. The inset illustrates a magnified view over the range $40<T<50$.}
	\label{fig:unormry}
	\end{figure}

Figure \ref{fig:yieldedTime_Re=50} shows the time evolution of $\bar{r}_y$ at $Re=50$. Upon the start of stirring, there is an immediate rise in $\bar{r}_y$, reflecting the yielding of the fluid. In regime \SE, $\bar{r}_y$ increases to a local maximum as the CW shed vortices travel away from the stirrer (see $T\approx 5$ and $Bn=0.025$ in the figure). The final increase in $\bar{r}_y$ corresponds to the escape of vortices from the central region (e.g., $T\approx 20$ for $Bn=0.025$). Although we expect $\overline{r}_y$ to approach an upper bound for $Bn\ge 0$, numerical estimation of this ultimate threshold in regime \SE~was not feasible within the computational domain used in this study. However, the approach of $\overline{r}_y$ to a steady average value, $\overline{r}_{ys}$, is clear in regimes \ST~and \E~(see red dashed lines in the figure). 

In regime \ST, the relatively slow increase in $\bar{r}_y$ during the early stages (see $Bn=0.4$, $T\lesssim 10$) confirms that the yield stress delays the advection of shed vortices away from the stirrer. Additionally, no further increases in $\bar{r}_y$ are observed, as the vortices remain confined within the central region. Finally, at $Bn=1$, characteristic of regime \E, $\bar{r}_y$ quickly reaches an approximately constant value after stirring begins, as vortex shedding and escape from the central region do not occur. Finally, we note that the small oscillations observed in $\bar{r}_y$ are associated with the interaction of vortices in the flow domain. The oscillations disappear in regime \E~where there is no shedding and, thus, no vortex interaction beyond the two attached eddies.

Figure \ref{fig:unormRe50} shows the time evolution of the kinetic energy. $\ke$ increases rapidly when stirring begins, before reaching an apparently quasi-steady state (see, e.g., $Bn=0.05$, $10\lesssim \newTime \lesssim 25$). This corresponds to the time interval when the shed CW vortices advect within the central region. In regime \SE,  this stage is followed by a second, relatively rapid increase, corresponding to the escape of vortices from the central region (see, e.g., $Bn=0.05$, $30 \lesssim \newTime \lesssim 45$). This transition is not observed in regime \ST, where shed vortices remain trapped in the central region (see, e.g., $Bn=0.4$ in the figure). 

Figure \ref{fig:unormRe50} further illustrates that in regimes \SE~and \ST, the kinetic energy, similar to $\bar{r}_y$, evolves on two distinct timescales: a shorter timescale ($T_s \approx o(1)$), corresponding to small oscillations in $\ke$, and a much longer timescale ($T_l \approx O(10)$), associated with the overall increase in $\ke$. The amplitude of these oscillations diminishes as $Bn$ increases, eventually disappearing entirely in regime \E.

Two critical values of $Bn$ may be identified to distinguish the three regimes.: 
\begin{align*}
	\left\{
	\begin{array}{lll}
		\text{Regime \SE} & \text{if}&  Bn \le Bn_{\ce}  \\
		\text{Regime \ST} & \text{if} & Bn_{\ce} \le Bn \le Bn_\ct\\
		\text{Regime \E} & \text{if} & Bn_{\ct} \le Bn 
	\end{array}
	\right.
\end{align*}
where $Bn_{\ce}$ and $Bn_{\ct}$ are the critical Bingham numbers marking the regime transitions.

\begin{figure}
	\centering
		\includegraphics[trim=0cm 0cm 0cm 0cm, clip=true,height=.38\textwidth]{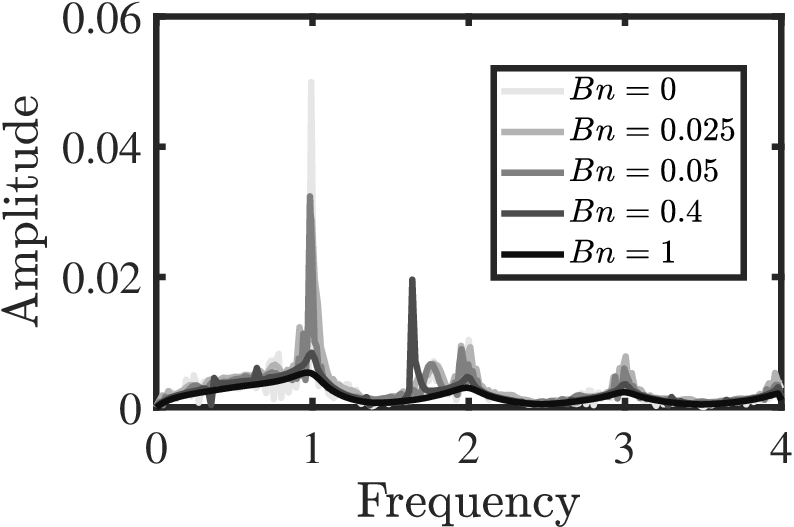}
	\caption{Fourier spectrum of $\ke'$.  $Re=50.$}
	\label{fig:ufftRe50}
\end{figure}

 To differentiate the regimes, we evaluated the Fourier spectrum of the oscillations, $\ke'=\ke-\overline{\ke}$ where $\overline{\ke} = \int_{\newTime}^{\newTime+1} \ke \, d\newTime$. Figure \ref{fig:ufftRe50} shows the frequency spectrum of $\ke'$ at $Re=50$, $0 \le Bn \le 1$. All regimes have peaks at frequency one and its harmonics. This is the fundamental frequency of the system for all $Bn$ considered. Regimes \SE \  and \ST \ have additional distinct peaks  (see $f\approx1.6$ and $1.8$, respectively, in Figure \ref{fig:ufftRe50}). These characteristic peaks vanish as mixing transitions between regimes. In contrast, regime \E~does not display any peaks beyond the fundamental frequency and the harmonics. To estimate the critical values, we evaluated the strength of the characteristic peak for each regime and used the two closest datapoints to find, by extrapolation, the $Bn$ at which the characteristic peak disappears.
 	
\begin{figure}
	\centering
	\subfloat{
		\includegraphics[trim=0cm 0cm 0cm 0cm, clip=true,height=.38\textwidth]{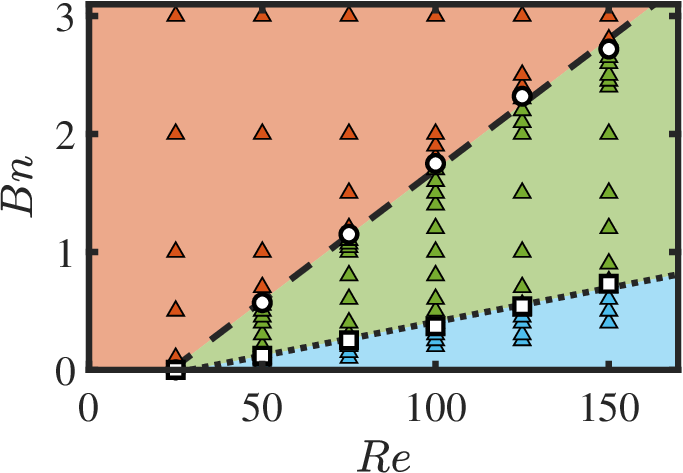}
	}	
			
	\caption{Flow regime map in the ($Re,Bn$) plane. Regimes \E, \ST, and \SE, are indicated in orange, green and blue, respectively. The triangle markers are the data points. Square and circle markers display the estimates of $Bn_{ct}$ and $Bn_{ce}$, respectively. The black dotted and dashed lines are linear fits to $Bn_{ct}$ and $Bn_{ce}$.}
	\label{fig:criticalValues}
\end{figure}

Figure \ref{fig:criticalValues} illustrates the flow regime map in the ($Re,Bn$) plane. The orange, green and blue zones correspond to regimes \E, \ST, and \SE, respectively. Regimes \SE~and \ST~disappear at sufficiently small $Re$ below which there is no shedding even in the Newtonian case. The square and circle markers indicate the estimates of $Bn_\ce$ and $Bn_\ct$, respectively.

%\begin{figure}
%	\centering
%		\subfloat{
%			\includegraphics[trim=0cm 0cm 0cm 0cm, clip=true,height=.35\textwidth]{/effectOfYieldStress/steadyRadiusRe=50.eps}
%		}	
%	\caption{ \color{blue} Steady value of  the averaged radius of the yielded region, $\overline{r}_{ys}$, at different Bingham numbers. The dotted and dashed vertical lines represent $Bn_{ct}$ and $Bn_{ce}$, respectively. The empty and full markers represent regime \ST \  or \E, respectively. $Re=50$.}
%	\label{fig:averagedRadius_Re=100}
%\end{figure}
	
The dashed and dotted lines in Figure \ref{fig:criticalValues} illustrate that the variation of $Bn_\ct$ and $Bn_\ce$ is well-described by the following fits 
 \begin{align}
 	\begin{array}{l}
 	Bn_\ce = 0.0057 Re - 0.17 \\
	Bn_\ct = 0.0222 Re - 0.53
	\end{array}
	\label{eq:bncr}
 \end{align}

The linear variation of the critical Bingham numbers with $Re$ is reminiscent of the effective Reynolds number, $Re_e$, in flows of viscoplastic fluids, 
\begin{align}
	Re_{e} = \frac{\hat{\rho} \hat{U}_o^2}{\hat{\tau}_y+\hat{\mu} \hat{\dot{\gamma}}_c}
	\label{eq:Reeff}
\end{align}
where $\hat{\dot{\gamma}}_c$ is the characteristic strain rate. The primary challenges in identifying the characteristic strain rate in flows of viscoplastic fluids are twofold: firstly,  the physically representative $\hat{\dot{\gamma}}_c$ may vanish as yield stress increases and the fluid becomes quiescent. Secondly, the relationship between $\hat{\dot{\gamma}}_c$ and yield stress is not known a priori; see \citep{thompson2016viscoplastic,ahmadi2022rayleigh} for more details.

In this context, the linear critical equations presented in \ref{eq:bncr} reveal the effective Reynolds numbers at which regime transitions occur. Assume that $\hat{U}_o= \hat{r}_o\hat{\Omega}$ and $ \hat{\dot{\gamma}}_c = a \hat{\Omega}$ where $a$ is a constant, equation \ref{eq:Reeff} yields
\begin{align*}
	& %Bn = \frac{Re}{cRe_{e}} - \frac{a}{c}
	Re_e = \frac{Re}{a+Bn}
\end{align*}

Comparing this with equations \ref{eq:bncr}, we find
\begin{align}
	\begin{array}{lll}
	&\displaystyle Re_{e,\ce} = \frac{Re}{Bn_{\ct} + 0.53} \approx 87 \\
	&\displaystyle Re_{e,\ct} = \frac{Re}{Bn_{\ce} + 0.17} \approx 22
	\end{array}
	\label{eq:Rec}
\end{align}

%and 
%\begin{align*}
%	& \lim_{Bn\rightarrow 0} a = 1\\
%	& \lim_{Bn\rightarrow Bn_{\ct}} a \approx 0.3\\
%	& \lim_{Bn\rightarrow Bn_{\ce}} a \approx 1\\
%\end{align*}

The two distinct definitions of the effective Reynolds number confirm that for a given flow setup, the characteristic strain rate changes with $Bn$. It follows that defining a unique effective Reynolds number that fully  describes the hydrodynamics observed in the ($Re,Bn$) plane, e.g., help distinguish the different regimes at different values of the Bingham number, is not feasible. Interestingly, $Re_{e,\ce}\approx 22$, at which shedding begins, is about half of the threshold Reynolds number beyond which vortices are shed behind a cylinder travelling at a constant velocity in an otherwise quiescent fluid.

\begin{figure}
			\centering
			\subfloat[\label{fig:ryBnAll}]{
				\includegraphics[trim=0cm 0cm 0cm 0cm, clip=true,height=.3\textwidth]{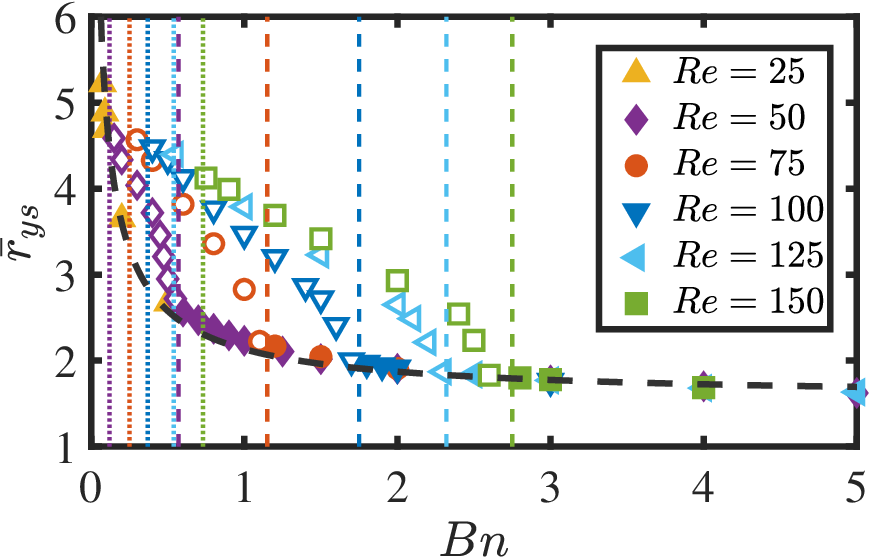}
			}~
			\subfloat[\label{fig:ryBnmAll}]{
				\includegraphics[trim=0cm 0cm 0cm 0cm, clip=true,height=.3\textwidth]{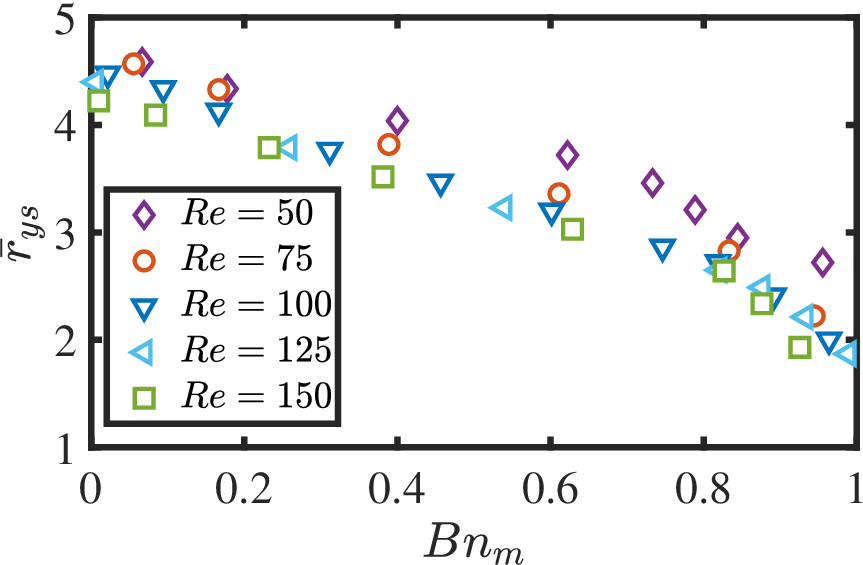}
			}
			\caption{(a) Steady value of the averaged radius of the yielded region, $\bar{r}_{ys}$, at different $Bn$ and $Re$. Each symbol with unique color shows a specific $Re$. The vertical colored dotted and dashed lines represent  $Bn_{ct}$ and $Bn_{ce}$, respectively.  The empty and filled symbols show whether the data point belongs to regime \ST \  or \E, respectively. The black dashed line illustrate the estimation of $\bar{r}_{ys}$ based on Eq \ref{eq:ryFit}. (b) Evolution of $\bar{r}_{ys}$ in regime \ST \  in a modified Bingham number, $\displaystyle Bn_m = \frac{Bn-Bn_{ct}}{Bn_{ce}-Bn_{ct}}$.}
			\label{fig:averagedRadius_Re}
	\end{figure}

To quantitatively compare the extent of localization in different regimes, Figure \ref{fig:ryBnAll} displays the steady average radius of the yielded region, $\overline{r}_{ys}$, in regimes \ST~and \E~for a wide range of $Re$. Distinct colors and marker shapes represent various $Re$ values. Vertical dotted and dashed lines of the same color mark $Bn_\ce$ and $Bn_\ct$ at each $Re$. Empty and filled markers indicate regimes \ST~and \E, respectively. 
	
	In regime $\ST$, $\bar{r}_{ys}$ decreases with decreasing $Re$ and increasing $Bn$, confirming the influence of both yield stress and purely plastic stresses on the flow field. Figure \ref{fig:ryBnmAll} shows that, in this regime, $\bar{r}_{ys}$ values approximately collapse onto the same curve when plotted against a scaled Bingham number, $Bn_m$,
	\begin{align*}	
		Bn_m=\frac{Bn-Bn_\ct}{Bn_\ce-Bn_\ct}
	\end{align*}

Additionally, Figure \ref{fig:ryBnAll} reveals that $\bar{r}_{ys}$ is independent of $Re$ in regime \E. To understand this, we consider the system from the perspective of an observer located at the domain center, rotating with angular velocity $\hat{\Omega}$. In this reference frame, the system is approximately steady: the fluid is yielded within an approximately circular domain of radius of $\bar{r}_{ys}$, the stirrer remains fixed, and the boundary of the yielded region acts like a no-slip boundary rotating at $-\hat{\Omega}$. In this context, the energy balance becomes
\begin{align}
  	 \int_0^{2\pi} (-pu_jn_j + u_i \tau_{ij}n_j) \bar{r}_{ys} d\theta 
	\approx 
	\frac{1}{2} \int_{A_y} Bn \, \dot{\gamma} dA
	+ \frac{1}{2} \int_{A_y} \dot{\gamma}^2 dA 
	\label{eq:EHighBn}
\end{align}	

where $n_j$ is the unit vector normal to the boundary of the yielded region, and $A_y$ is the subdomain where the fluid is yielded. The integrals on the right-hand-side are identical to those across the entire flow domain, as the fluid remains quiescent beyond $\bar{r}_{ys}$. The left-hand side represents the energy input through the domain boundary. Energy input due to pressure is zero due to the no-penetration condition. Assuming $\displaystyle \boldsymbol{\tau}=\tau_y {\bf e}_{r\theta}$, the left-hand side can be estimated as
\begin{align}
	\int_0^{2\pi} (-pu_jn_j + u_i \tau_{ij}n_j) \bar{r}_{ys} d\theta   \approx 2\pi \bar{r}_{ys}^2 \tau_y \Omega
	\label{eq:Einput}
\end{align}

\begin{figure}
			\centering
			\subfloat[]{
				\includegraphics[trim=0cm 0cm 0cm 0cm, clip=true,height=.28\textwidth]{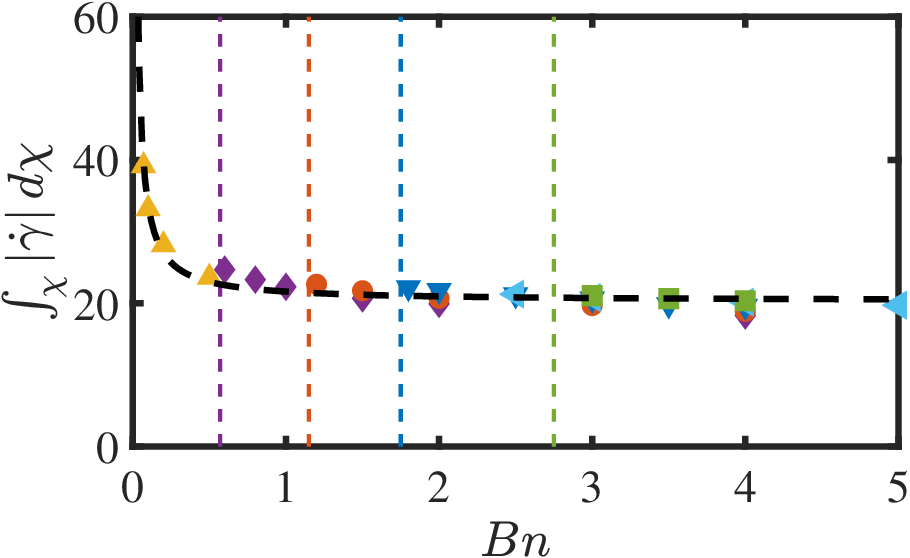}
			}
			\subfloat[]{
				\includegraphics[trim=0cm 0cm 0cm 0cm, clip=true,height=.28\textwidth]{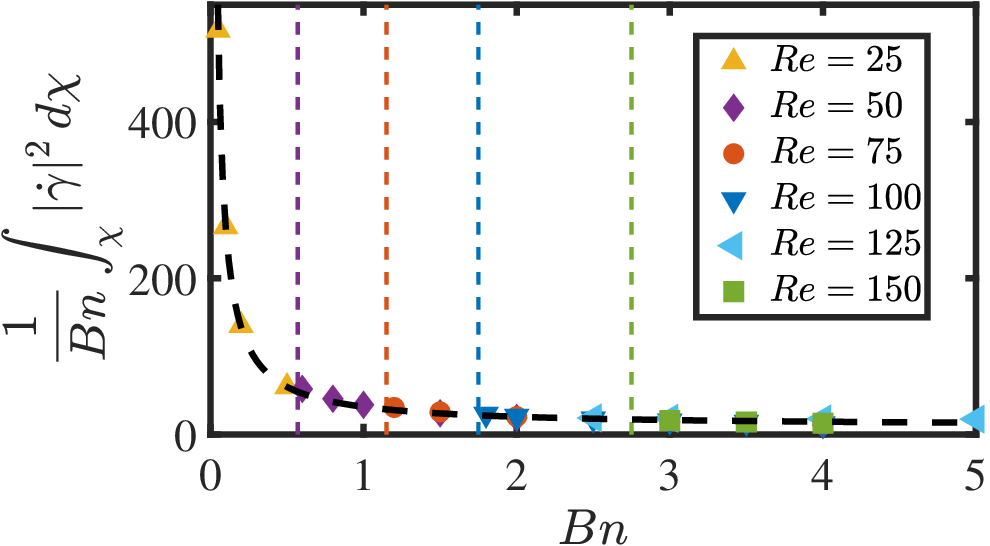}
			}
			\caption{Illustration of the dissipation due to (a) yield stress and (b) purely viscous effects. The vertical colored dashed lines represent $Bn_{ct}$. The black dashed lines show fits of the form $c_{1i}+c_{2i}/Bn$.}
			\label{fig:dissipation}
		\end{figure}

Figure \ref{fig:dissipation} illustrates $\displaystyle \int_{A_y} \dot{\gamma}  dA$ and $\displaystyle\frac{1}{Bn} \int_{A_y} \dot{\gamma}^2 dA$. It shows that energy dissipation due to yield stress and purely viscous forces is independent of $Re$ in regime \E. The black dashed lines display fits of the form $c_{1i}+c_{2i}/Bn$, which closely follow the data points (where $c_{ij}$ are fitted constants). By plugging these fits and equation \ref{eq:Einput} into the approximate energy balance equation \ref{eq:EHighBn}, we find
\begin{align}
	\bar{r}_{ys} \approx \sqrt{2.4+\frac{2.1}{Bn}}
	\label{eq:ryFit}
\end{align}

This result, shown as a black dashed line in Figure \ref{fig:ryBnAll}, closely follows the data points. Overall, this approximate analysis reveals that the boundary of the quiescent yielded region approaches a circle as $Bn$ is increased. Moreover, while dissipation due to purely viscous and yield stresses remain important at high $Bn$, they scale similarly with $Bn$, explaining why the dynamics become independent of the Reynolds number.

\begin{figure}
			\centering
			\subfloat[]{
				\includegraphics[trim=0cm 0cm 0cm 0cm, clip=true,height=.3\textwidth]{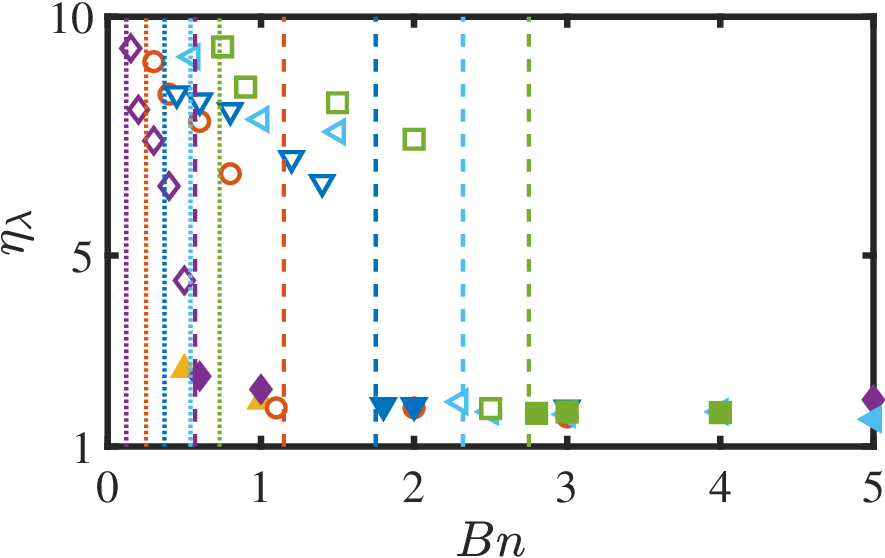}
			}
			\subfloat[]{
				\includegraphics[trim=0cm 0cm 0cm 0cm, clip=true,height=.3\textwidth]{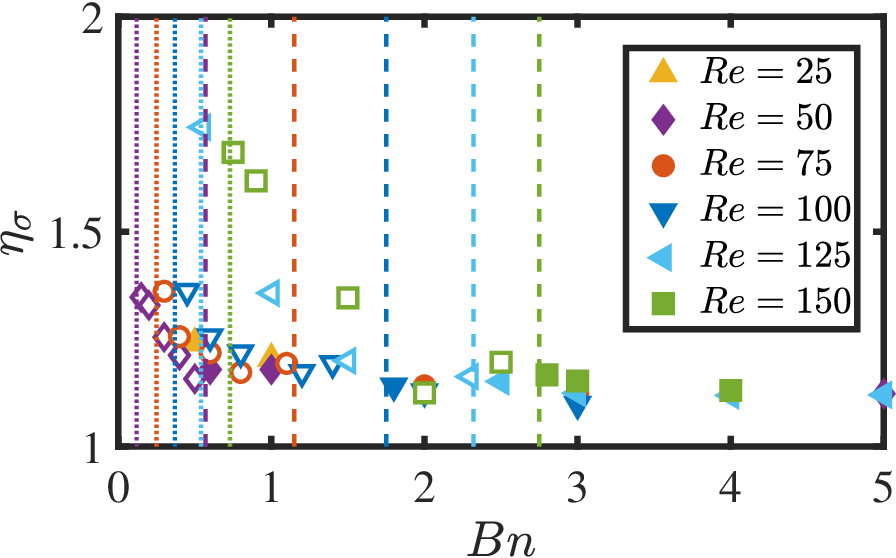}
			}
			\caption{Variation of enhancement factors, (a) $\eta_{\lambda}$ and (b) $\eta_{\sigma}$ with $Bn$ and $Re$. The vertical colored dotted and dashed lines represent  $Bn_{ct}$ and $Bn_{ce}$, respectively.  The empty and filled symbols indicate regime \ST~and \E, respectively.}
			\label{fig:enhancementAll}
		\end{figure}

Finally, figure \ref{fig:enhancementAll} illustrates the variation of enhancement factors with $Bn$ across different $Re$. The initial rapid decrease in $\eta_\lambda$ and $\eta_\sigma$ corresponds to regime \ST. The enhancement factors show little change in regime \E. It is also evident that the influence of a small yield stress value is more pronounced at lower $Re$; for instance, comparing $Re=25$ and $Re=125$ in the figure. At higher $Re$, the influence of $Bn$ on the enhancement factors is less pronounced at low $Bn$, and a very sharp decline in the enhancement factors is observed near $Bn \lesssim Bn_{ct}$. 
			
%-----------------------------------------------------------------
	
	\section{Summary}
	\label{sec:summary}

In this study, we investigated a canonical two-dimensional mixing setup to establish a mechanistic understanding of how fluid mechanics drive transitions in the mixing regimes of yield-stress fluids.

Our numerical simulations model an infinite, two-dimensional domain filled with a quiescent viscoplastic fluid described by the Bingham model. In this setup, a cylinder moves at a constant speed along a circular path, stirring a fluid with uniform density and rheological properties. The bottom half of the domain is initially marked with a passive dye, with the center of the stirrer’s path aligned with the dye interface. This representative model allowed us to explore the fundamental features of yield-stress fluid mixing in two-dimensional settings.

To decouple the influence of flow dynamics from dye concentration, we considered only one-way coupling, where the flow is unaffected by dye concentration changes. This approach simplifies the isolation of causal relationships between fluid mechanics phenomena and mixing events.

In the Newtonian case, we identified three primary mixing mechanisms when stirring a fluid with heterogeneous dye distribution: (1) the stretching and folding of the dye interface within the central region as the stirrer initiates movement, (2) diffusion-dominated mixing when dye distribution becomes approximately uniform along streamlines, and (3) enhanced mixing due to vortex shedding, where shed vortices transport dyed regions into dye-free areas (or vice versa), significantly extending the dye interface.

For the laminar regimes studied, the fluid dynamics evolve on two timescales: the average energy of the system changes on a slow timescale, roughly an order of magnitude slower than the stirrer period, while energy oscillations, related to vortex interactions, occur on a faster timescale. As yield stress increases, the average yielded region size, kinetic energy, and energy oscillations decrease, eventually leading to oscillation suppression. We identified three mechanisms by which yield stress localizes mixing:

\begin{enumerate}
\item {\bf Finite Vortex Advection}: In yield-stress fluids, advection of shed vortices is suppresses with vortices travelling within a finite radius from the stirrer due to energy decay. This defines a maximum distance for dye transport and leads to localized mixing.

\item {\bf Entrapment of Shed Vortices}: At moderate yield stresses, shed vortices cannot escape the stirrer’s vicinity, resulting in periodic interactions between the stirrer and previously shed vortices, which promotes mixing localization.

\item {\bf Suppression of Vortex Shedding}: At high yield stresses, vortex shedding ceases entirely, confining mixing to the interface stretching caused by direct interaction between the stirrer and the dye interface.
\end{enumerate}

We classified mixing regimes based on these mechanisms: in Regime \SE, shed vortices escape the central region, causing dye variance evolution similar to that in Newtonian fluids, with mixing initiated by interface stretching and folding, followed by a diffusion-dominated phase, and then accelerated by escaping eddies. Regime \ST~is defined by the entrapment of vortices near the stirrer, limiting mixing to its immediate vicinity, while Regime \E~shows no vortex shedding, with mixing characterized by initial interface stretching followed by diffusion across streamlines and across the boundaries of the well-mixed region.

Using a fast Fourier transform on energy oscillations, we distinguished these regimes, all featuring a fundamental mode at frequency one. Additional spectral peaks in Regimes \SE~and \ST, introduced by shed vortices, further enabled us to identify critical transition criteria. In the $(Re,Bn)$ plane, we observed distinct separation of the three regimes along lines that relate critical Bingham numbers to Reynolds numbers. This relationship allows us to define two unique, effective Reynolds numbers, each capturing a transition between two regimes at a constant value. This supports the hypothesis that fluid mechanics phenomena underlying mixing regime transitions are closely linked to bluff-body flow dynamics, traditionally described by threshold Reynolds numbers marking stability and flow transition modes.

Comparing the decay of dye concentration variance, $\sigma^2$, across regimes, we hypothesize that, among the various localization mechanisms, vortex entrapment near the stirrer has the most significant impact on mixing. This is supported by the contrasting long-term decay behaviors observed in regimes \SE~and \ST: in regime \ST, where vortices remain entrapped, $\sigma^2$ shows an exponential decay, while in regime \SE, it decays at an accelerating rate due to the influence of vortices that escape into the outer regions. To compare regimes \ST~and \E, we introduced an enhancement factor, $\eta_\lambda$, which quantifies the relative decay rate of $\sigma^2$ during the final, diffusion-dominated stage of mixing, compared to the purely diffusive case. In regime \ST, $\eta_\lambda$ declines sharply with increasing $Bn$, but in regime \E, it becomes largely independent of both $Re$ and $Bn$. This suggests that vortex entrapment is indeed the primary driver of mixing localization. Moreover, once vortex shedding is fully suppressed (in regime \E), increasing stirrer speed has minimal effect on enhancing mixing.

Finally, mixing localization has far reaching effects beyond the initial localization of interface stretching: while the change in the degree of mixing by the end of the advection-dominated phase may not be very significant (e.g. about 10\% for an increase of two orders of magnitude in $Bn$), the decay rate of $\sigma^2$ may decrease dramatically (by more than 80\%).

The fundamental principle of laminar mixing in stirred tanks is the periodic movement of a stirrer along a closed path. By modeling this process in its most simplified form - a stirrer moving on a circular path across a concentration gradient - we observe a gradual transition from attached eddies to vortex shedding, governed by the Reynolds and Bingham numbers. We hypothesize that the mixing modes identified here are archetypes of mixing modes in stirred tanks, while the yield stress-driven localization mechanisms and associated regimes represent canonical modes of mixing in yield-stress fluids. For passive tracers, these transitions correlate with distinct effective Reynolds numbers based on the stirrer size, shape, and speed, alongside Bingham number. Alternatively, the critical Bingham numbers marking these transitions are expected to scale linearly with the Reynolds number based on purely viscous stresses. Given the lack of an a priori definition for effective Reynolds numbers in such flows, we recommend further investigation of the linear critical criteria observed here.

\section*{Acknowledgments}
We gratefully acknowledge financial support from Natural Sciences and Engineering Research Council (NSERC), Canada, through the Discovery Grant program. This research was enabled in part by support provided by Calcul Qu\'ebec (\url{www.calculquebec.ca}) and Compute Canada (\url{www.computecanada.ca}).

%%%%%%%%%%%%%%%%%%%%%%%%%%%%%%%%%%%%%%%%%
\bibliographystyle{jfm}
\bibliography{references}

\end{document}